\documentclass[12pt]{iopart}
\usepackage{graphicx}
\usepackage{bm}
\renewcommand{\thetable}{\Roman{table}}
\renewcommand{\thesection}{\Roman{section}}
\renewcommand{\thesubsection}{\Alph{subsection}}

\newcommand{\water}{H$_2$O}
\newcommand{\pwater}{{\em p}H$_2$O}
\newcommand{\owater}{{\em o}H$_2$O}
\newcommand{\odtwo}{{\em o}D$_2$}

\newcommand{\sotwo}{SO$_2$}

\newcommand{\hydrogen}{H$_2$}
\newcommand{\wno}{cm$^{-1}$}
\newcommand{\phtwo}{{\em p}H$_2$}
\newcommand{\ohtwo}{{\em o}H$_2$}
\newcommand{\pdtwo}{{\em p}D$_2$}

\begin{document}

\title{Microscopic molecular superfluid response: theory and simulations}

\author{Tao Zeng and Pierre-Nicholas Roy}

\address{Department of Chemistry, University of Waterloo, Waterloo, Ontario N2L 3G1, Canada}

\begin{abstract}
Since its discovery in 1938, superfluidity has been the subject of much investigation because  it provides a unique example of a macroscopic manifestation of quantum mechanics. About sixty years later, scientists successfully observed this phenomenon in the microscopic world though the spectroscopic Andronikashvili experiment in helium nano-droplets. 
This reduction of scale suggests that not only helium but also {\em para}-\hydrogen (\phtwo) can also be a candidate for superfluidity.
This expectation is based on the fact that the smaller number of neighbours and surface effects of a finite-size cluster may hinder solidification and promote a liquid-like phase. The first prediction of superfluidity in \phtwo~clusters was reported in 1991 based on quantum Monte Carlo simulations.
The possible superfluidity of  \phtwo was later indirectly observed in a spectroscopic Andronikashvili experiment in 2000. 
Since then, a growing number of studies have appeared, and theoretical simulations have been playing a special role because they help guide and interpret  experiments. 
In this review, we go over the theoretical studies of \phtwo~superfluid clusters since the experiment of 2000. 
We provide a historical perspective and introduce the basic theoretical formalism along with key experimental advances.
We then present illustrative results of the theoretical studies and comment on the possible future developments in the field. 
We include sufficient theoretical details such that the review can serve as a guide for newcomers to the field.
\end{abstract}

\maketitle

\section{Introduction} \label{sec:intro}

From a phenomenological point of view, a superfluid is a substance in a state where flow occurs without viscosity. 
This behaviour was first discovered for $^4$He liquid at low temperature in 1938 by Kapitsa~\cite{superfluid_discovery_1} and Allen and Misener~\cite{superfluid_discovery_2}. 
Being a very rare manifestation of large-scale quantum coherence, superfluidity has been considered 
as the ``jewel in the crown of low temperature physics''~\cite{basic_superflu}.
As a matter of fact, another important physical phenomenon, superconductivity, can be considered as a special case of superfluidity, since the Cooper electron pairs flow without energy dissipation (viscosity).~\cite{tilley_book}
Several Nobel Prizes in physics (1962, 1972, 1973, 1978, 1987, 1996, and 2003) were awarded for achievements related to superfluidity and
superconductivity. 
Superfluidity is intrinsically related to the permutation symmetry of bosons and Bose-Einstein condensation~\cite{london_superflu_bec,feynman_sf1,feynman_sf2,feynman_sf3}.
Our fundamental understanding of superfluidity in bulk liquid helium is due to Landau~\cite{landau41,landau47}.
He proposed the dispersion relation for elementary excitations in liquid helium that explained the lack of dissipation (viscosity) between the liquid and an object flowing inside the liquid with a velocity below a threshold, the Landau velocity. 
As a subject of almost eighty years of age, superfluidity has been discussed in many classic monographs~\cite{basic_superflu,tilley_book,khalatnikov_book, clark_derrick, nozieres_pines,dalfovo_bec_rmp,untracold_qf_stoof,annett_super_book}, where macroscopic systems form the focus. 
In this report, we turn our attention to a more recently developed discipline, the study of superfluid-like response that occurs on the microscopic scale. 
More specifically, we focus on superfluidity in {\em molecular} systems rather than the traditional {\em atomic} superfluidity of helium. 
We limit the scope of our review to the fundamental ideas necessary to explain molecular scale superfluid response and refer the reader
to the aforementioned references for a more general discussion.

Two subjects that provide background for our exposition of molecular superfluidity are the two-fluid model and the Andronikashvili experiment. 
The two-fluid model was first proposed by Tizsa~\cite{tisza_2fluid, tisza_2flud_2} but was also independently later developed  by Landau~\cite{landau41}. 
In this model,  the low temperature ($T < 2.17$~K) $^4$He liquid is viewed as of two inseparable components, the normal and the superfluid components. The normal component contains all the entropy and viscosity of the helium while the superfluid component contains none. 
The two components flow with different velocities and interact differently with an object placed in contact with the liquid. 
When an object is flowing through such a liquid, the normal component exerts a drag on it as in the case of a common Newtonian fluid, 
whereas the superfluid part slips. 
The two-fluid model can be summarized as follows:
\begin{enumerate}
\item Density: $\rho=\rho_N + \rho_S$;
\item Velocity: $\rho \vec{v} = \rho_S \vec{v}_s+ \rho_N \vec{v}_N$;
\item Viscosity: $\eta_S=0$ and $\eta_N\ne 0$;
\item Entropy: $\rho S=\rho_N S_N$.
\end{enumerate}
This summary is adapted from Table 2.1 of Ref.~\cite{basic_superflu}.

\begin{figure}[h]
\centerline {\includegraphics[width=1.0\columnwidth]{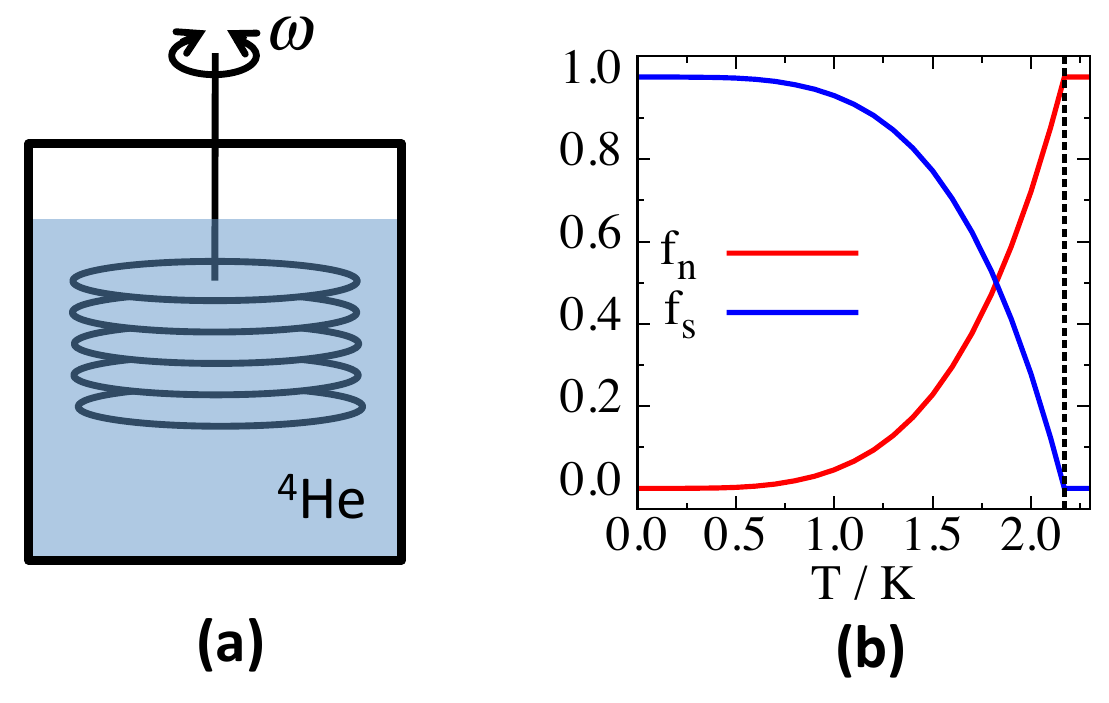}}
\caption{Sketches of (a) Andronikashvili experiment and (b) the measured superfluid ($f_s$) and normal ($f_n$) fractions. The result in (b) is not to scale and it is adapted from the original publication~\cite{andron_bucket_1}. }
\label{fig:bucket}
\end{figure}

Based on the two-fluid model, Andronikashvili conducted an experiment to measure the superfluid fraction of $^4$He liquid~\cite{andron_bucket_1,andron_bucket_2}.
His experiment is sketched in Fig.~\ref{fig:bucket}(a). 
A stack of suspended and closely packed disks were immersed in liquid $^4$He and were allowed to oscillate about a pivot axis. 
The viscous liquid entrained in the inter-disc space would adiabatically follow the oscillation and contribute to the total moment of inertia ($I^{\rm total}$) 
of the oscillator. 
By measuring the angular frequency of oscillation $\omega$, Andronikashvili obtained $I^{\rm total}$ through the relation $\omega=\sqrt{\frac{k}{I^{\rm total}}}$, where $k$ is the torsional constant of the suspending fibre. 
The effective moment of inertia of $^4$He can be calculated as $I^{\rm eff}=I^{\rm total}-I^{\rm disc}$.  
He found that below the characteristic $\lambda$~temperature ($T_{\lambda}=2.17$~K) of $^4$He, 
the contribution of the liquid to $I^{\rm eff}$ decreased and reached zero at $T=0$~K, i.e., the liquid becomes completely superfluid. 
Based on the third point in the above summary, Andronikashvili equated the normal fraction, $f_n=\frac{\rho_N}{\rho}$, to the quotient of 
\begin{eqnarray}
\frac{I^{\rm eff}(T < 2.17~{\rm K})}{I^{\rm eff} (T=2.17~{\rm K})},
\end{eqnarray}
and obtained the normal and superfluid fractions as illustrated in Fig.~\ref{fig:bucket}(b). 
The superfluid fraction is defined to be $f_s=1-f_n$.

Although superfluidity was discovered in the context of a macroscopic quantum phenomenon, people have asked the question: 
how small a system can be and still exhibit superfluid-like features? 
This question would allow us to bridge the gap between the macroscopic quantum mechanical world and the atomic scale quantum motion of particles.
The answer to this question is partly revealed by the microscopic Andronikashvili experiment (also called spectroscopic Andronikashvili experiment). Theoretical simulations and explanations of this type of experiment form the focus of the present review. 
Another equally important question is whether there is any other superfluid substance, other than helium. 
The answer to this question is also rooted in the microscopic Andronikashvili experiment. 
We will commence  our discussion by introducing this crucial experiment.

\subsection{Molecular scale superfluidity versus molecular superfluids} \label{subsec:vs}

Theoretical predictions of microscopic scale superfluid systems were first reported around 1989 and 1990. 
The superfluidity of nano-scale He systems was first suggested by Lewart~\etal, who carried out variational Monte Carlo simulation for He clusters containing $20$ to $240$ atoms.~\cite{lewart_liquid-helium}
Based on the results of  Feynmann path-integral Monte Carlo (PIMC)~\cite{ceperley_rmp_1995} simulations and other theoretical arguments, Sindzingre \etal~\cite{ceperley_area_estim} and Pitaevskii \etal~\cite{He_cluster_superflu} concluded that the pure $^4$He cluster with only 64 atoms demonstrated superfluidity through weak anomalies in heat capacity and reduced effective moment of inertia. 
The first microscopic Andronikashvili experiment can be traced back to the high-resolution infrared (IR) spectroscopy of an SF$_6$ molecule doped inside a liquid helium droplet consisting of about 4000 atoms. 
This spectrum was measured by Hartmann \etal~\cite{vilesov_sf6} in 1995, following the pioneering works of Goyal \etal~\cite{goyal_sf6_he_prl,goyal_sf6_he}.
It is the first spectroscopic experiment that showed a clear rotational structure of the IR spectrum of  doped species in a liquid helium droplet. 
While focussing on the determination of the temperature of the helium droplet, the authors did not realize that they had conducted a microscopic Andronikashvili experiment. 
However, they did notice that the rotational constant of SF$_6$ was reduced from 2730 to 1019(30) MHz, indicating a substantial increase of moment of inertia. 
The authors explained this increase of moment of inertia as a result of the rigid attachment of He atoms to the SF$_6$ molecule. 
However, they pointed out that this oversimplified rigid model is ``somewhat of an oxymoron'', and considered the origin of this increase as ``unclear''.  
In a follow-up letter, it was concluded that the above SF$_6$ spectrum can be analyzed using the same Hamiltonian as that of a free molecule, but with different parameters~\cite{vilesov_SF6_theory}.
This suggested that  the symmetry of the molecule was not disturbed by the helium. 
In a later study of both SF$_6$ and SF$_6$-rare gas complexes in $^4$He droplets~\cite{hartmann_sf6-rg}, the same group of researchers started to associate the sharp rotational lines in the spectra, i.e., the phenomenological free rotation of the dopants, to the superfluidity of $^4$He.

\begin{figure}[h] 
\centerline {\includegraphics[width=0.7\columnwidth]{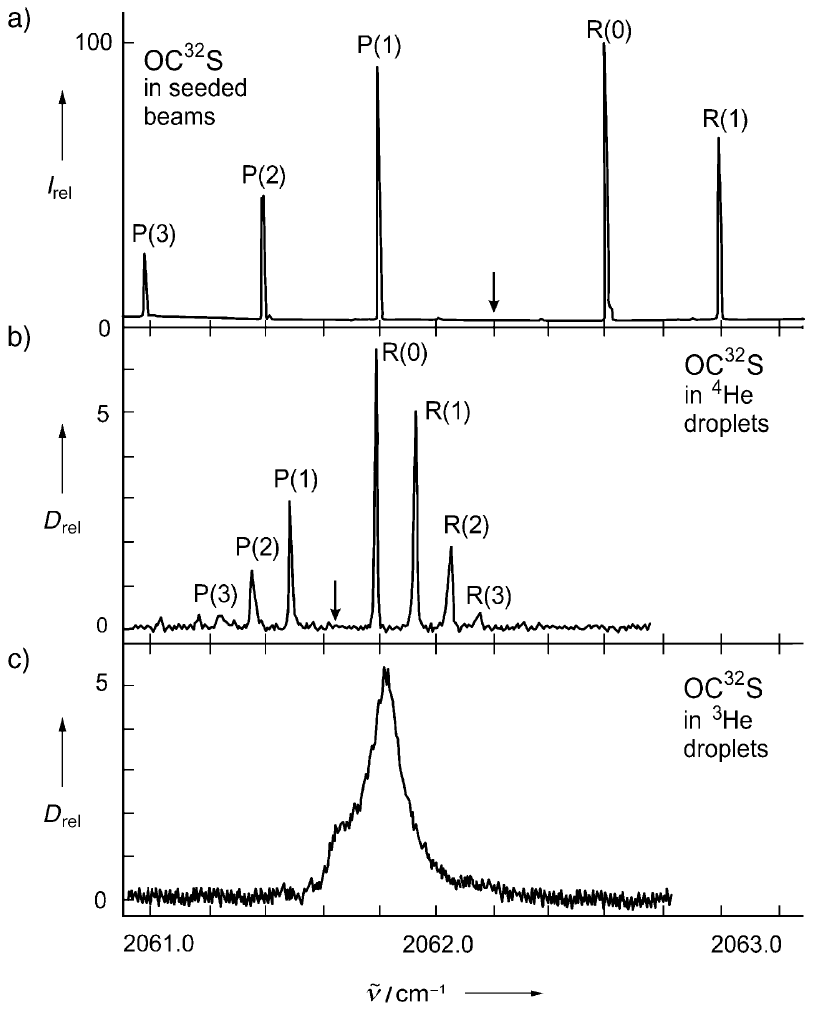}}
\caption{Comparison of the rotational spectra of OCS: a) free molecule; b) in pure $^4$He droplet; c) in pure $^3$He droplet. The P- and R-lines are labeled by the initial $j$ value and correspond to transitions with $j\rightarrow j-1$ and $j\rightarrow j+1$, where $j$ is the angular momentum quantum number. The Q-branch is missing (position indicated by arrows) in a) and b). The reduced rotational spacing (effective rotational constant) is obvious in b). This figure is taken from Fig. 12 of Ref.~\cite{vilesov_acie}. Reproduced with the permission from~\cite{vilesov_acie}. Copyright \copyright~2004 WILEY-VCH Verlag GmbH \& Co. KGaA, Weinheim.}
\label{fig:ocs_he_grebenev}
\end{figure}

In 1998, the spectroscopic study coined as the microscopic Andronikashvili experiment was reported by Grebenev, Toennies, and Vilesov~\cite{grebenev_science}, the latter two having participated in the above studies of SF$_6$. 
This time, they doped an OCS molecule in $^4$He and $^3$He droplets and compared their IR spectra, which are reproduced in Fig.~\ref{fig:ocs_he_grebenev}. 
The $^3$He droplet has a temperature of 0.15~K, which is too high for the fermionic $^3$He atoms to be superfluid. 
They found that when doped in the $^4$He droplet, OCS exhibited narrow rotational lines in its IR spectrum with a reduced rotational constant. 
In contrast, a broad peak was observed when OCS was doped in the $^3$He droplet. 
This finding unambiguously connects the phenomenological free rotation of the doped rotor to the bosonic exchange of its surrounding. 
The authors interpreted this striking difference as follows. The normal component of the $^4$He droplet is coupled to OCS and adiabatically follows its rotation, forming an effective rotor. This adiabatic following increases the effective moment of inertia and reduces the effective rotational constant and rotational line spacings. The superfluid component of the $^4$He droplet is completely decoupled from the rotation of the effective rotor as a result of its phenomenological null viscosity and promotes coherent  quantum rotation of the effective rotor, yielding  narrow rotational lines in the IR spectrum. 
This explanation is an adaptation of the two-fluid model to the microscopic scale. 
On the other hand, the 100\% non-superfluid $^3$He hinders the rotation of OCS through collisions and yields incoherent quantum rotations and a blurring of the rotational structure in the IR spectrum.
the $^4$He droplet plays the role of the bulk liquid helium while the OCS dopant represents the stack of discs.
The rotation of the dopant is then analogous to the elastic oscillation of the discs.

Another salient feature of this work was that about 60 $^4$He atoms were enough to provide a superfluid environment for the OCS dopant.
This observation is consistent with the theoretical predictions mentioned above. 
To do so, the authors also measured the OCS IR spectra doped in mixed  $^3$He/$^4$He  droplets and  controlled the number of $^4$He atoms. 
They found that with about 60 $^4$He atoms, the same sharp rotational lines as OCS in pure $^4$He droplets appear, suggesting that the OCS has been fully coated with a superfluid shell of $^4$He. 
As discussed below, this record of minimum number of atoms (or molecules) for superfluidity has been broken many times in the subsequent years. 
The authors termed this microscopic scale superfluidity  ``molecular superfluidity''. 
However, since atoms ($^4$He), not molecules, constitute this superfluid droplet, we prefer to name this phenomenon molecular-scale superfluidity and reserve the quoted term for the superfluidity displayed by actual molecular systems. 
There have been many spectroscopic studies on $^4$He nano-droplets. Interested readers should refer to the excellent review chapters of Refs.~\cite{vilesov_arrc, toennies_he_review,vilesov_acie,choi_he_review,stienkemeier_he_review,toennies_0.37K,callegari_nanocryostats,szalewicz_helium_review} for a comprehensive introduction to the advances in this field.

\begin{figure}[h] 
\centerline {\includegraphics[width=0.7\columnwidth]{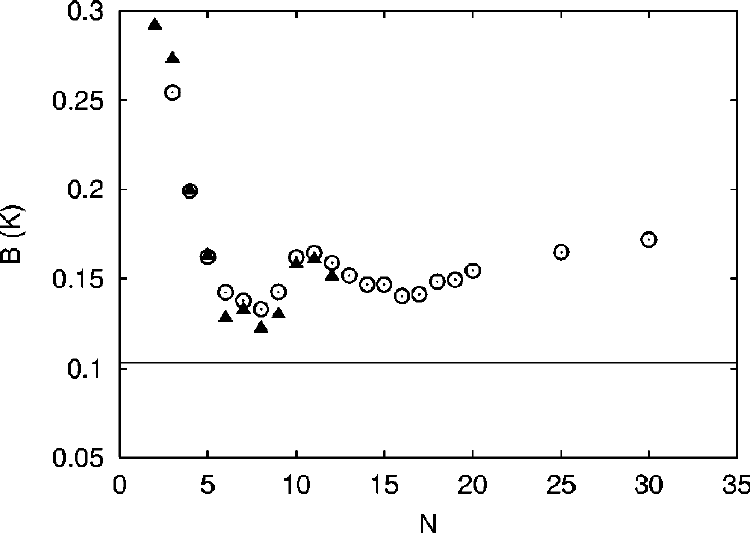}}
\caption{Evolution of effective rotational constant $B$ of N$_2$O(He)$_N$ clusters with $N$. Circles and triangles indicate theoretical~\cite{moroni_pn_pimc} and experimental~\cite{xu_turnaround_1} values. The horizontal line is the nano-droplet limit~\cite{nauta_n2o_he}. 
This figure is taken from Fig. 3 of Ref.~\cite{moroni_pn_pimc}. Reproduced with the permission from~\cite{moroni_pn_pimc}. Copyright \copyright~2004 American Institute of Physics.}
\label{fig:n2o_he}
\end{figure}

Can even smaller systems exhibit superfluidity? 
Researchers have tried to answer this question by performing microscopic Andronikashvili experiment with small $^4$He clusters rather than droplets. 
In 2002, Tang \etal~measured high-resolution IR and microwave (MW) spectra for OCS($^4$He)$_N$ van der Waals complexes, with $N$ ranging from 2 to 8~\cite{tang_OCS_He}. 
They observed that the moment of inertia of the effective rotor keeps increasing and becomes larger than that of the $^4$He droplet limit. 
They concluded that there must be a turnaround in the moment of inertia as $N$ increases further. 
In a follow-up study of N$_2$O($^4$He)$_N$~\cite{xu_turnaround_1}, they did observe this experimental turnaround at $N=7$ (Fig.~\ref{fig:n2o_he}). 
Corroborated with theoretical simulations~\cite{moroni_pn_pimc}, Xu \etal~studied N$_2$O($^4$He)$_N$ with $N$ up to 19, completing the first solvation shell~\cite{xu_turnaround_2}.
They concluded that the first turnaround in effective rotational constant (moment of inertia) should be regarded as the onset of superfluidity at the microscopic level. Therefore, even a few $^4$He atoms can exhibit superfluid features. 
The connection between the turnaround of the moment of inertia and superfluidity will be further discussed in Sec.~\ref{sec:formu}\ref{subsec:noncl_I}. 
For reviews and theoretical models of superfluid helium clusters, readers can refer to Refs.~\cite{barnett_review_he_cluster,toennies_helium_today,surin_review,paesani_he_review,he_overview,sf_model_ovchinnikov,novikov_diagram,alexey_gs_BEC,lehmann_model,kwon_superfluid_helium,yang_ellis_helium_droplet,surin_heco_2009}.

Before closing this section, we would like to caution readers on two issues. First, the similarity between the microscopic and macroscopic Andronikashvili experiments is only of a phenomenological sense, and different mechanisms are behind the two phenomena. The adiabatic following in the macroscopic case stems from thermal excitations of the bosons, while the microscopic counterpart is a manifestation of quantum hydrodynamic effects.~\cite{lehmann_sf_hydro,lehmann_hydro_erratum,lehmann_impurity_he,lehman_sf6_hydro} Detailed knowledge of the quantum hydrodynamics model is beyond the scope of this paper and interested readers may refer to the cited references. Strictly speaking, the term ``normal component'' should only be used for the macroscopic Andronikashvili experiment, and it may be more appropriate to call the microscopic counterpart a ``hydrodynamic component''. However, since similar terms such as ``normal fluid density''~\cite{ceperley_rmp_1995} and ``normal fraction''~\cite{sindzingre_pH2_superfluid} have been extensively used to describe the non-superfluid portion of microscopic fluids in the past two decades, we follow the same convention and use ``normal component'' for microscopic systems throughout this report. 
Second, strictly speaking, the term of phase transition is only well defined in the thermodynamic limit, i.e. in the bulk phase. On the contrary, for finite size systems such as nano-droplets and clusters, the transition from a non-superfluid to a superfluid phase is a continuous process as the temperature decreases. Therefore, it is hard to describe a finite size system as completely superfluid or completely normal, as the superfluid fraction may still be asymptotically reaching the lower (0) or upper (1) limit as the temperature changes. In the text below, when terms like ``completely superfluid'' or ``completely non-superfluid'' are encountered, people should be aware that they mean the superfluid fractions are close to $1$ or $0$.

\subsection{{\em Para}-hydrogen as a molecular superfluid candidate}

\begin{figure}[h]
\centerline {\includegraphics[width=1.0\columnwidth]{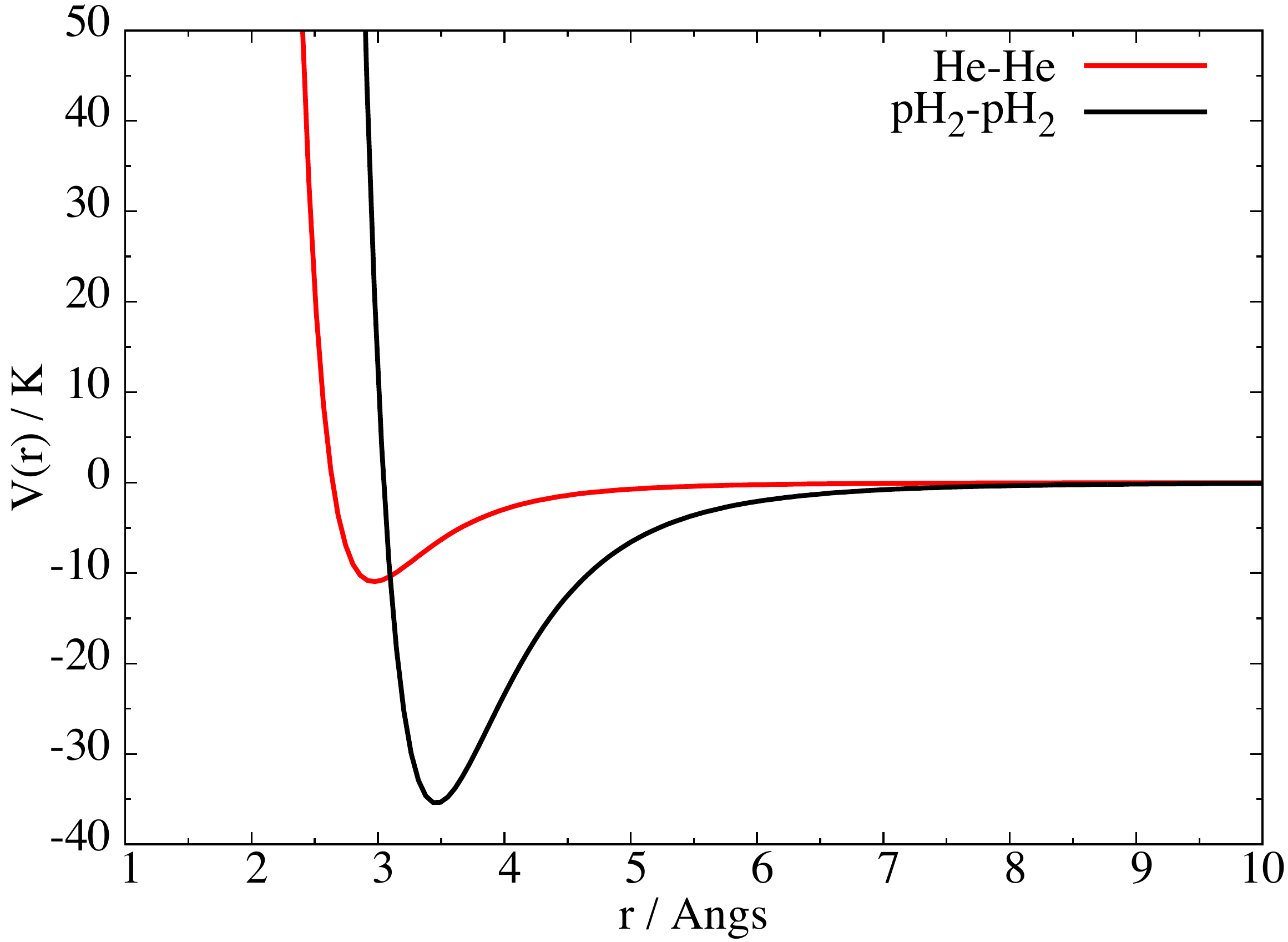}}
\caption{Potential energy curves of He-He and \phtwo-\phtwo~interactions.}
\label{fig:heh2}
\end{figure}

As to the second question of whether or not other substances could be superfluid, the most likely candidate is {\em para}-hydrogen (\phtwo). Molecular hydrogen is a boson, satisfying the superfluid requirement of Bose-Einstein statistics. Furthermore, \phtwo~has zero total nuclear spin and a ground rotational state of $J=0$. This makes it a spherical particle like $^4$He. In 1972, Ginzburg and Sobyanin pointed out the possibility of \phtwo~being a superfluid~\cite{ginzburg_pH2_superfluid}. Using a modified London formula of $\lambda$-temperature for an ideal bose gas~\cite{modified_london_form},
\begin{eqnarray}
T_{\lambda,0}&=&\frac{3.31\hbar^2}{Mk_B}\left(\frac{n}{g}\right)^{2/3}, \label{eqn:t_lambda}
\end{eqnarray}
where $M$ is the atomic or molecular mass, $n$ the concentration, $k_B$ the Boltzmann constant, and $g$ the nuclear spin degeneracy, they calculated $T_{\lambda,0}\approx 6$~K for \phtwo, the onset temperature of \phtwo~superfluid. This temperature is higher than the predicted $T_{\lambda,0}$ (3~K) and the actual $T_{\lambda}$ (2.17~K) of helium, suggesting superfluid phase of \phtwo~would emerge more easily. 
This is due to the lighter mass of \phtwo~that makes it more quantum than helium. Nevertheless, this temperature is far too low compared to the triple point (13.8~K) of \phtwo~and therefore, \phtwo~solidifies and prohibits its potential superfluidity under normal condition. Such a contradiction to helium, which remains liquid under normal pressure till $T=0$~K, is due to the much stronger interaction between \phtwo~molecules. 
A comparison between the \phtwo-\phtwo~and He-He interaction potential energy curves is showed in Fig.~\ref{fig:heh2}, where the \phtwo-\phtwo~potential is calculated by isotropically averaging the Patkowski 4-D potential for H$_2$-H$_2$ interaction~\cite{patkowski_h2_pot} and the He-He one takes the Aziz potential~\cite{aziz_he_pot}. 
These potentials are chosen for comparison purposes here.
The comparison shows a much deeper and wider potential well of \phtwo-\phtwo~interaction. The possibility of superfluid {\em ortho}-hydrogen ({\em o}\hydrogen) was excluded as a potential candidate for superfluidity. This is because of its $J=1$ ground rotational state, which provides more orientational degrees of freedom and therefore the molecule retains the quadrupole moment. This quadrupole moment is averaged to be zero by the spherical ground rotational state of \phtwo. Therefore, a  stronger interaction between {\em o}\hydrogen~molecules and a higher melting point are expected. {\em Ortho}-deuterium ({\em o}D$_2$) has similar a ground rotational state as \phtwo,
and substituting its molecular mass into Eq.~\ref{eqn:t_lambda} leads to the superfluid transition temperature about 0.9~K.
However, {\em o}D$_2$molecules behave as distinguishable particles, because they can be in either a state with total nuclear spin of 0 or 2~\cite{grebenev_pH2_5ring}, and therefore, the {\em o}D$_2$ superfluidity is suppressed.

In their original paper~\cite{ginzburg_pH2_superfluid}, Ginzburg and Sobyanin proposed several ways to observe superfluidity in \phtwo, including the  formation of \phtwo~films on different substrates and the introduction of impurities or vacancies in bulk \phtwo. 
Inspired by this pioneering work, Maris, Siedel, and coworkers attempted to obtain superfluid hydrogen by supercooling liquid hydrogen~\cite{maris_sf_h2_1,maris_sf_h2_2,maris_sf_h2_3}.
The liquid hydrogen was supercooled down to 10.6~K, which is still too high, and they failed in obtaining superfluid hydrogen. Knuth \etal~\cite{knuth_h2_supercool} further pushed the temperature down to about 5~K but it could not be determined whether the resultant \phtwo~droplet was superfluidic or not. More recently, Grisenti \etal~\cite{grisenti_supercool_h2} claimed to have supercooled \phtwo~liquid to 1.3~K, but the authors did not report evidence of superfluidity at that time.

Nineteen years after the prediction of Ginzburg and Sobyanin, a theoretical study revived interest in superfluid hydrogen. 
In 1991, Sindzingre \etal~performed a simulation of small \phtwo~clusters with PIMC method~\cite{sindzingre_pH2_superfluid}. 
They calculated the superfluid fractions for (\phtwo)$_N$ clusters with $N=13$, 18, and 33, and observed that below 2~K, (\phtwo)$_{13,18}$ become  superfluid whereas (\phtwo)$_{33}$ remains non-superfluid. 
Long permutation cycles of \phtwo~particles were observed in (\phtwo)$_{13,18}$ and this was deemed  responsible for superfluidity. This study pointed out that experiments with nano-clusters would be a promising direction for the seeking of superfluid hydrogen.

An experimental breakthrough was reported in 2000 by Grebenev \etal~\cite{grebenev_OCS_pH2} The authors investigated the IR spectra of OCS(\phtwo)$_{14,15,16}$ and OCS({\em o}D$_2$)$_{15,16,17}$ clusters embedded in pure $^4$He and in mixed $^3$He/$^4$He droplets. The pure droplets provided an environmental temperature of 0.38~K for the clusters while the mixed droplets had a temperature of 0.15~K. The spectra of OCS({\em o}D$_2$)$_{15,16,17}$ clusters displayed Q-branches, the transitions with $\Delta J=0$. This was also observed for OCS(\phtwo)$_{14,15,16}$ in the $^4$He droplets. However, no Q-branch was observed for OCS(\phtwo)$_{14,15,16}$ in $^3$He/$^4$He droplets.
The axial symmetry of \phtwo~or {\em o}D$_2$ distributions in the OCS(\phtwo)$_{14,15,16}$ and OCS({\em o}D$_2$)$_{15,16,17}$ clusters along the OCS axis makes them resemble symmetric top molecules. For symmetric tops, the observation of a Q-branch in the rovibrational spectrum stems from excitation of the angular momentum along the dipole moment vector (OCS axis in this case).
Therefore, the presence of Q-branch requires a moment of inertia about the OCS axis. The disappearance of Q-branch for OCS(\phtwo)$_{14,15,16}$ when embedded in the $^3$He/$^4$He droplets indicates that as the temperature is reduced from 0.38 to 0.15~K, the contribution of \phtwo~to the moment of inertia about the OCS axis decreases to zero. This observation matched the reduction in the moment of inertia in the macroscopic Andronikashvili experiment. 
Therefore, Grebenev \etal~considered the results to be the first evidence for \phtwo~superfluidity. 
This revolutionary discovery stimulated a large number of follow-up studies on \phtwo~superfluidity, both experimental and theoretical. 
The main objective of this report is to give a comprehensive account of the scientific work dedicated to the particular case of molecular superfluidity with special emphasis on theoretical studies.

The rest of the report is arranged as follows. In Sec.~\ref{sec:exp}, we give a brief introduction to the experimental observations and interpretations of \phtwo~superfluidity in the last twelve years. In Sec.~\ref{sec:formu}, we cover some background information regarding  the theoretical formalisms  used to  simulate and interpret superfluid \phtwo. Details of the simulation algorithms based on the specific PIMC method are covered in Sec.~\ref{sec:algor}. 
We chose the PIMC method because of its particular ability to directly calculate the superfluid fraction of a microscopic system. 
In Sec.~\ref{sec:results}, illustrative results of theoretical simulations are presented and in Sec.~\ref{sec:challenge}, current theoretical challenges are discussed. 
In Sec.~\ref{sec:summary}, we conclude the report and provide an outlook for the future of this field.

\section{Experimental observations and interpretations} \label{sec:exp}

Since the aforementioned breakthrough of 2000, there have been two main streams in the study of \phtwo~superfluidity, 
and both are based on spectroscopic studies of \phtwo~clusters.
One avenue is to surround a rotor with \phtwo~molecules to form a cluster, and subsequently embed the cluster in a helium droplet.
One then measures the spectrum of this system. 
These studies employ a similar experimental setup as that used in the Grebenev \etal experiment~\cite{grebenev_OCS_pH2},
and we term these studies {\em nano-droplet experiments}. 
Another direction is to place a dopant inside a \phtwo-cluster and then directly measure its spectrum. 
We will call such studies {\em cluster experiments}. 
For technical details of the helium droplet experiment, which was formally called helium nano-droplet isolation (HENDI) spectroscopy, please refer to Refs.~\cite{callegari_hendi,electronic_he_droplet,glyoxal_sf_he,hendi_selection,hendi_selection_2,hendi_selection_3,hendi_kupper,toennies_beams,makarov_hendi,hendi_vilesov,hendi_stienkemeier,paesani_hcccn.hcn,mckellar_he_cluster}.

Few experiments have been carried out to study the superfluidity of pure \phtwo~clusters because of the lack of a permanent dipole moment and the small quadrupole moment of the molecule. Despite these experimental difficulties, Tejeda \etal~managed to obtain the Raman spectrum of small (\phtwo)$_N$ clusters formed in cryogenic free jets, and successfully identified clusters with $2 \le N \le 8$~\cite{tejeda_raman_ph2_cluster}. 
The authors could also track the transition of clusters from liquid to solid during the jet expansion. 
More recently, Kuyanov-Prozument and Vilesov performed Raman spectroscopic studies on pure large \phtwo~clusters with about $10^4$ molecules~\cite{vilesov_raman_large_h2}. 
They found that the $S_0\left( 0\right)$ line of  clusters formed of highly dilute \phtwo~in helium is not split, 
and this is  strong evidence that those clusters remain liquid-like at $T=1-2$~K, which is low enough for \phtwo~to be superfluid. 
In light of these advances, the direct experimental observation of superfluidity in pure \phtwo~clusters is within reach.

\subsection{Nano-droplet experiments} \label{subsec:droplet_exp}

Grebenev \etal~conducted a series of studies for the similar OCS(\phtwo)$_N$ and OCS(\odtwo)$_N$ clusters doped in helium droplets following their pioneering work of 2000. 
They first looked into the OCS-H$_2$, OCS-D$_2$, and OCS-HD van der Waals complexes doped in helium droplets~\cite{grebenev_ocs-h2,grebenev_ocs-hd}, but these studies are not directly related to \phtwo~superfluidity because there is no \phtwo~exchange involved. 
In 2002, they reported a study of OCS(\phtwo)$_N$ and OCS(\odtwo)$_N$ clusters doped in the 0.15~K $^3$He/$^4$He droplet with $N=2-8$~\cite{grebenev_pH2_5ring,vilesov_ocs_ph2_he}. 
Again, they observed Q-branches for OCS(\odtwo)$_{5,6}$, but not for OCS(\phtwo)$_{5,6}$. 
They assumed a symmetric top spectroscopic model for OCS(\phtwo)$_{5,6}$ and tried several structural models for the two. 
They found that the structures with a rigid \phtwo~pentagon around the OCS axis best account for the spectrum of OCS(\phtwo)$_{5}$, and so does a rigid \phtwo~hexagon around the OCS axis for OCS(\phtwo)$_{6}$. 
They called this arrangement of \phtwo~the {\em donut model}. 
They analyzed the absence of a Q-branch as follows. 
The H\"onl-London factor for a Q-branch transition is $A_{KJ}=K^2[J(J+1)]^{-1}$, where $K$ is the angular momentum projected onto the dipole axis and $J$ the total angular momentum. Therefore, Q-branch transitions require non-zero $K$ of the initial rovibrational state of the cluster. 
In order to satisfy the bosonic permutation symmetry of \phtwo, the $K$ value of the \phtwo~ring has to be a multiple of the number of \phtwo~on the ring.  Therefore, the lowest nonzero $K$ is 5 or 6 for OCS(\phtwo)$_{5,6}$, which corresponds to $J$ larger than the two respective values, making the levels too high in energy to be thermally excited at $T=0.15$~K. This has the effect of  quenching the Q-branch transitions. 
On the other hand, any $K$ value is allowed for the essentially distinguishable \odtwo~and therefore, 
low energy levels with small values of $J$ and $K$ are thermally occupied and give rise to the observed Q-branch transitions of OCS(\odtwo)$_{5,6}$.

The same group pushed their investigation further and studied the IR spectra of OCS(\phtwo)$_N$ and OCS(\odtwo)$_N$ clusters with $N=8-16$ and embedded in both $^4$He and $^3$He/$^4$He droplets~\cite{grebenev_2008}. 
The OCS(\odtwo)$_N$ clusters exhibited Q-branch transitions in both droplets, while for OCS(\phtwo)$_N$ with $N \ge 11$, except 12, Q-branches appeared at 0.38~K and disappeared at 0.15~K. For OCS(\phtwo)$_{12}$, there were no Q-branches at both temperatures. 
They could not explain this phenomenon using the similar rigid structure model as in Ref.~\cite{grebenev_pH2_5ring}, and therefore, they proposed a floppy model. 
In this model, two or more floppy \phtwo~rings with internal degrees of freedom surround the OCS axis, and the mutual hindrance of two adjacent rings leads to \phtwo~tunnelling through the periodic potential and enhances \phtwo~permutation within each ring. 
This \phtwo~exchange induces superfluidity about the OCS axis and the disappearance of the Q-branch at 0.15~K. 
At the higher temperature of 0.38~K, the authors speculated that some internal degrees of freedom within each \phtwo~ring would be thermally excited and account for the emergence of Q-branches. 
The reason for the special spectra of OCS(\phtwo)$_{12}$ remained unexplained in that work.

In 2010, Grebenev \etal~revisited their results for OCS(\phtwo)$_N$ and OCS(\odtwo)$_N$ clusters embedded in helium droplets from the previous decade, and reanalyzed their spectroscopic data~\cite{grebenev_2010}. 
They also investigated  clusters that were doped in the pure $^3$He droplets. 
They found that in the latter case, the rotational structure of their high-resolution IR spectra of the clusters was blurred. 
This lack of rotational coherence can readily be attributed to the non-superfluid character of $^3$He in the 0.15 to 0.38~K temperature range.
They also observed that just as OCS(\phtwo)$_{12}$, OCS(\phtwo)$_{5,6}$ displayed no Q-branch when doped in both $^4$He and $^3$He/$^4$He droplets (at both 0.38 and 0.15~K). 
In that work, they introduced internal vibrations, internal rotations, and inter-ring tunnelling in a multi-ring system to their floppy model. 
They concluded that the large amplitude zero point motion and the coupling between the rings can enhance  large-scale \phtwo~permutations and in turn superfluidity. 
They explained the absence of a Q-branch for OCS(\phtwo)$_{12}$ at both temperatures as a result of the high symmetry ($C_6$) of the two interlocked six-membered \phtwo~rings. 
With this structure, the first excited rotational state ($J=6$ and $K=6$) would be too high to be thermally occupied at both temperatures. 
The cluster structures proposed by Grebenev \etal~do not always match those obtained from theoretical simulations for the clusters without helium droplets. For example, a six-membered \phtwo~ring around the OCS axis has not been reported based on theoretical studies. 
They attributed this discrepancy to the presence of the surrounding helium atoms. 
Therefore,  helium droplets may not just act as thermostats with varying temperatures, and their influence on the cluster structure requires further study.

Other experimental groups also contributed to our understanding in this area. Moore and Miller studied hydrogen-HF complexes doped in helium droplets using high-resolution IR Laser spectroscopy~\cite{more_miller_h2_hf, moore_miller_hd_hf,moore_HF_H2}. 
They observed free rotation of HF when there were twelve HD molecules surrounding it~\cite{moore_miller_hd_hf}. 
This free rotation cannot be due to hydrogen superfluidity because the HD molecule is not a boson. 
Instead, they attributed it to the isotropic solvent cage formed by the twelve HD. 
This interpretation requires caution when associating the observed phenomenological free rotation to superfluidity. 
In their first attempt~\cite{moore_HF_H2}, they ``unfortunately'' could not have \phtwo~with high enough purity to form pure HF(\phtwo)$_N$ clusters in helium droplets, and therefore, they could not study the possible superfluid phenomenon of \phtwo. 
This problem was solved later~\cite{moore_HF_H2_2} and they concluded that the structures of HF(\phtwo)$_N$ with $N=4-6$ are highly symmetric. 
These symmetric distributions most likely stem from the large degree of delocalization of the highly quantum \phtwo~molecules. 
In conclusion, their research did not directly point to \phtwo~superfluidity in the HF(\phtwo)$_N$ clusters.

More recently, there was an attempt to use larger molecules to form clusters with \hydrogen. Kuma \etal~successfully doped a tetracene molecule in (Ar)$_N$, (Ne)$_N$ and (\hydrogen)$_N$ clusters with $N=1-2000$ and subsequently embedded these clusters in $^4$He droplet~\cite{tetracene_h2}. 
They used Laser Induced Fluorescence Spectroscopy to study these systems. 
Unlike the rigid behaviour of the (Ar)$_N$ and (Ne)$_N$ clusters, they found that the (\hydrogen)$_N$ clusters remained fluxional. 
This behaviour could be a manifestation of the \phtwo~superfluidity. Nevertheless, no definite conclusion has been drawn yet.

\subsection{Cluster experiments} \label{subsec:exp_cluster}

Following their initial investigation of the OCS-\hydrogen~dimer~\cite{tang_ocs_h2_1st}, 
Tang and McKellar conducted the first study of doped \hydrogen~clusters without the use of helium droplets~\cite{tang_h2_ocs_2nd}. 
They reported the infrared spectra of OCS(\phtwo)$_N$, OCS(\ohtwo)$_N$, and OCS(HD)$_N$, with $N=2-7$. 
The rotational structure of these ``bare'' clusters resembles that of a symmetric top rotor. 
From the intensities of the $P(1)$ and $R(0)$ transitions, they determined that the temperature of the clusters ranged from $0.15-0.20$~K, close to the temperature of the $^3$He/$^4$He mixed droplets. 
As in the case of clusters embedded in $^3$He/$^4$He droplets, Q-branches were observed for the OCS(\ohtwo)$_N$ and OCS(HD)$_N$, but not for  OCS(\phtwo)$_N$ clusters. 
However, the disappearance of the Q-branch is not clearly connected to the superfluidity of \phtwo~because it already occurs for $N=1$ without any bosonic exchange. 
The evolution of the vibrational shift for the bare OCS(\phtwo)$_N$ clusters differs from that of the clusters doped in helium droplets for $N=5-7$.
This indicates that  a six-membered \phtwo~ring is formed under the influence of helium, while a five-membered ring is formed in the bare case. 
This finding clearly suggests that one should not treat the clusters embedded in helium as surrogates for the bare clusters. 
This work established the feasibility of direct spectroscopic studies of hydrogen clusters without the need for helium droplets. 
The authors considered this as the most significant aspect of their work.

Tang and McKellar substituted the OCS dopant with an N$_2$O molecule. 
In 2005, they published a high-resolution IR spectroscopic study of N$_2$O(\phtwo)$_N$ and N$_2$O(\ohtwo)$_N$ clusters, 
with $N=2-13$~\cite{tang_n2o_h2}. 
Again, a simple symmetric top-type rotational structure was observed in their spectra. 
While the \ohtwo~clusters exhibited prominent Q-branch features, the \phtwo~ones did not. 
The authors did not associate this absence of a Q-branch to \phtwo~superfluidity, and concluded that there is no obvious indication of superfluid effects for  these \phtwo~clusters.

In addition to IR spectroscopy, Microwave (MW) spectra can also be used to investigate the superfluid response of  \phtwo. 
Efforts in this direction were pioneered by Yu \etal~\cite{yu_ocs_h2_1,yu_ocs_h2_2} and Michaud \etal~\cite{michaud_ocs_h2_1}, who measured the Fourier-transform MW spectra for the van der Waals complexes of OCS-\phtwo, OCS-\ohtwo, OCS-\odtwo, OCS-\pdtwo, and OCS-HD. 
Although these studies were not dedicated to \phtwo~superfluidity, they provided technical preparation for follow-up work. 
In 2008, Michaud and J\"ager published the first MW study of doped \hydrogen~clusters~\cite{michaud_ocs_h2_2}. 
They used a pulsed-jet Fourier transform MW spectrometer to measure rotational spectra of OCS(\phtwo)$_N$ and OCS(\ohtwo)$_N$, with $N=2-7$. 
They only obtained the $J=1-0$ transitions and calculated the effective rotational constants ($B_{\rm eff}$) as half of these transition energies, 
following a rigid rotor model. 
Their $B_{\rm eff}$ for both OCS(\phtwo)$_N$ and OCS(\ohtwo)$_N$ clusters monotonically decrease with the increase of $N$, and they speculated that the onset of \phtwo~superfluidity, which is signified by a turnaround of $B_{\rm eff}$, would occur at a larger cluster size.

Such a turnaround for \phtwo~clusters was first observed by Li \etal~In 2010 in a combined theoretical and experimental study of CO$_2$(\phtwo)$_N$ clusters with $N=1-18$~\cite{huili_prl}. 
They measured the IR spectra of the clusters and observed a monotonic decrease of $B_{\rm eff}$ for $N=1-8$. 
The $B_{\rm eff}$ then has a turnaround at $N=9$ and reaches its maximum at $N=12$, and then steadily decreases for larger  clusters. 
This experimental observation is consistent  with the results of PIMC simulations. 
Unlike the disappearance of a Q-branch, a $B_{\rm eff}$ turnaround was observed for doped $^4$He clusters and was solidly considered to be the onset of $^4$He superfluidity~\cite{xu_turnaround_1,moroni_pn_pimc,xu_turnaround_2}. 
Therefore, this work provided a stronger evidence of \phtwo~superfluidity than previous studies. 
In a follow-up publication~\cite{mckellar_co2_h2}, McKellar investigated the high-resolution IR spectra of CO$_2$(\ohtwo)$_N$ clusters with $N=1-7$ and CO$_2$(\phtwo)$_N$ clusters with $N=1-17$. 
As in the case of the aforementioned findings for OCS(\hydrogen)$_N$ and N$_2$O(\hydrogen)$_N$ clusters, a simple symmetric top-type rotational structure was observed for the CO$_2$(\hydrogen)$_N$ clusters. 
Also, the \ohtwo~clusters exhibited prominent Q-branch features while the \phtwo~clusters did not.

In addition to the heavy dopants mentioned above, researchers started to look into the possibility of using lighter rotors to probe the superfluid response of \phtwo. 
In 2012, Raston \etal~reported results of a joint experimental and theoretical study using CO as a chromophore~\cite{raston_coh2_superfluid}. 
They measured MW spectra for CO(\phtwo)$_N$ clusters with $N=2-8$ and investigated the $J=1-0$ transitions. 
Two types of transitions were observed: a-type transitions associated with the end-over-end rotation of the whole cluster, and b-type transitions attributed to the hindered rotation of the chromophore. 
They proposed a novel theory, which is introduced in Sec.~\ref{sec:results}\ref{subsec:linear_dopant}, to extract $B_{\rm eff}$ from these two types of transitions. 
They concluded that unlike CO$_2$(\phtwo)$_N$ clusters, the superfluid response to dopant rotation persists in CO(\phtwo)$_N$ even for large $N$ values.

Despite the many experimental breakthroughs introduced above, 
one has to admit that the superfluid information extracted from IR and MW spectra is somewhat limited. 
For example, one cannot readily associate the absence of a Q-branch in the bare \phtwo~clusters to  superfluidity. 
To gain a deeper understanding of this subject, theoretical analysis and simulations are required. 
Theoretical studies are not simply a complement to experiments. 
They can actually guide novel experiments and provide new interpretations. 
In the next section, we introduce the fundamental theoretical formalisms that have been used in investigating the superfluid response of \phtwo.

\section{Theoretical formalism} \label{sec:formu}

Superfluidity is a many-body phenomenon due to bosonic exchange.
The exact solution to such a  problem is a formidable task. 
Practical simulations of superfluid systems have to involve some approximations. 
In this section, we first introduce a quantum many-body method that describes thermal equilibrium states at finite temperature, 
the {\em path-integral formulation}. 
This method has been most extensively used in simulating microscopic bosonic systems. 
We also discuss several methods that approximately describe the ground state of many-body quantum systems. 
These methods provide information at absolute zero temperature. 
After introducing the basic formalisms, we turn our attention to the calculation of the non-classical effective moment of inertia. 
Since the study of \phtwo~superfluidity rests on the microscopic Andronikashvili experiment, 
accurate evaluation of the effective moment of inertia is  crucial. 
At the end of this section, we discuss the difference between superfluid responses with respect to the molecular dopant rotation and with respect to an external field. 
Understanding this difference is essential in oder to connect  superfluid response to actual experimental measurements.

\subsection{Path-integral formulation for finite temperature} \label{subsec:pi}
We begin our account of the path-integral (PI) method by defining the thermal density operator of distinguishable particles,
which follow Boltzmann distribution at their equilibrium states and are therefore also called boltzmannons,
\begin{equation}
\hat{\rho}\left( \beta \right)=e^{-\beta \hat{H}}.
\end{equation}
$\beta=\frac{1}{k_BT}$, $k_B$ is the Boltzmann constant, $T$ temperature, and $\hat{H}$ 
the Hamiltonian operator of the whole system. 
Note that we do not make specific reference to particle indistinguishability and quantum statistics here.
We will introduce those concepts later in the discussion.
Exploiting its exponential form, $\hat{\rho}\left(\beta\right)$ can be expressed as
\begin{eqnarray}
\hat{\rho}\left(\beta\right)&=&\left( e^{-\tau\hat{H}} \right)^M=\hat{\rho}\left(\tau \right)^M; \nonumber\\
\tau&=&\frac{\beta}{M}~.
\end{eqnarray}
$M$ is an integer and $\tau$ will be referred to as the imaginary time step in the discussion that follows. It has a unit of reciprocal energy.

With the position representation $\left\{ \left| {\bf R} \right> \right\}$ of all degrees of freedom, 
the partition function of the system can be expressed as
\begin{eqnarray}
Z&=&{\rm Tr}\left(\hat{\rho}\left(\beta \right) \right)=\int d {\bf R}\left< {\bf R} \right|\hat{\rho}\left(\beta\right)\left| {\bf R} \right>=\int d {\bf R}\left< {\bf R} \right|\hat{\rho}\left(\tau\right)^M\left| {\bf R} \right> \nonumber\\
&=&\int d {\bf R} \left(\prod_{i=1}^{M-1}\int d{\bf R}_i\right)\left< {\bf R} \right|\hat{\rho}\left(\tau\right)\left| {\bf R}_1 \right>\left< {\bf R}_1 \right|\hat{\rho}\left(\tau\right)\left| {\bf R}_2 \right> \cdots \nonumber\\
&&\cdots \left< {\bf R}_{M-2} \right|\hat{\rho}\left(\tau\right)\left| {\bf R}_{M-1} \right>\left< {\bf R}_{M-1} \right|\hat{\rho}\left(\tau\right)\left| {\bf R} \right>\nonumber\\
&=&\prod_{i=1}^{M}\int d{\bf R}_i \left< {\bf R}_{i} \right|\hat{\rho}\left(\tau\right)\left| {\bf R}_{i+1} \right> \label{eqn:z1}
\end{eqnarray}
with $\left| {\bf R}_{M+1} \right>=\left| {\bf R}_{1} \right>$. ``Tr'' labels the trace operation and ${\bf R}$ the ``collective'' coordinates of all particles in the system. 
The resolution of the identity, $\int d {\bf R}_i \left| {\bf R}_i \right>\left< {\bf R}_i \right|$ has been inserted between each  adjacent $\hat{\rho}\left(\tau \right)$ factor above. 
In this report, whenever the boundaries of an integral are not given, integration over all space is assumed.
The matrix element $\left< {\bf R}_{i} \right|\hat{\rho}\left(\tau\right)\left| {\bf R}_{i+1} \right>$ is called the high-temperature density matrix or imaginary time propagator for an imaginary time step $\tau$. 
That time step $\tau$ is often called a time-slice.
The ${\bf R}_i$ positions are  referred-to as the beads. 

In practice, the Trotter factorization approximation~\cite{trotter_factor,simon_trotter,trotter_suzuki} is often introduced in order to obtain an expression for the high temperature density operator
\begin{equation}
\rho\left(\tau\right)=e^{ -\tau\hat{H} }\approx e^{-\tau\hat{T} }e^{ -\tau\hat{V} }, \label{eqn:trotter_prim}
\end{equation}
where $\hat{T}$ and $\hat{V}$ respectively stand for kinetic and potential energy operators of the system.
This corresponds to an approximate description of the density operator since 
\begin{equation}
\left[ \hat{T},\hat{V} \right]\ne0~.
\end{equation}
It has been shown that the error is of order  $\tau^2$~\cite{schulman_pi}. 
The error vanishes in the limit of an infinite number of factors (or slices) ($M\rightarrow \infty$ and $\tau \rightarrow 0$). 
With this approximation and the locality of $\hat{V}$, we can obtain the following expression for a matrix element of the high temperature density matrix,
\begin{eqnarray}
\left< {\bf R}_{i} \right|\hat{\rho}\left(\tau\right)\left| {\bf R}_{i+1} \right>&\approx&\left< {\bf R}_{i} \right| e^{-\tau\hat{T}} \left| {\bf R}_{i+1} \right> e^{ -\tau V\left( {\bf R}_{i+1}  \right)} \label{eqn:linka}
\end{eqnarray}
One can then obtain an explicit expression for the matrix elements of the kinetic energy contribution to the propagator such that,
\begin{eqnarray}
\left< {\bf R}_{i} \right| e^{-\tau\hat{T}} \left| {\bf R}_{i+1} \right> e^{ -\tau V\left( {\bf R}_{i+1}  \right)} 
&=&\left(\prod_{j=1}^n \lambda_{j,\tau}^{-3} e^{ -\frac{\pi}{\lambda_{j,\tau}^2} \left( {\bf r}_{j,i} -{\bf r}_{j,i+1} \right)^2 }\right) e^{ -\tau V\left( {\bf R}_{i+1} \right) }, \label{eqn:linkb}
\end{eqnarray}
where $j$ is the index and $n$ the total number of particles and ${\bf r}_{j,i}$ is the position vector of particle $j$ at slice $i$, i.e.
\begin{equation}
{\bf R}_i=\left({\bf r}_{1,i},{\bf r}_{2,i},\ldots, {\bf r}_{n,i}\right).
\end{equation}
$\lambda_{j,\tau}$ stands for the thermal wavelength of particle $j$ at a temperature corresponding to $\tau$,
\begin{equation}
\lambda_{j,\tau}=\left( \frac{2\pi \hbar^2 \tau}{m_j} \right)^{\frac{1}{2}}, \label{eqn:thermal_wl}
\end{equation}
and $m_j$ is the mass of particle $j$. 

To go from Eq.~\ref{eqn:linka} to Eq.~\ref{eqn:linkb}, an identity operator in momentum representation, $\int d{\bf P} \left| {\bf P} \right>\left< {\bf P} \right|$, has been inserted to replace $\hat{T}$ by its eigenvalues and the standard Gaussian integral has been used to integrate the momentum. 
${\bf P}$ labels the ``collective'' momenta of all the particles in the system.
Substituting Eq.~\ref{eqn:linkb} into Eq.~\ref{eqn:z1}, we have
\begin{eqnarray}
Z&=&\left(\prod_{j=1}^n \lambda_{j,\tau}^{-3} \right)^M \int d {\bf R}_1\int d {\bf R}_2\cdots\int d {\bf R}_M \nonumber\\
&&\times e^{ -\sum_{j=1}^n \frac{\pi}{\lambda_{j,\tau}^2} \sum_{i=1}^M  \left( {\bf r}_{j,i}- {\bf r}_{j,i+1}\right)^2 -\tau \sum_{i=1}^MV\left( {\bf R}_i \right) }. \label{eqn:z2}
\end{eqnarray}
The above is an integral over all possible  configurations of the ${\bf R}_i$ beads.
It can be viewed as integrating over the spatial configurations of closed cyclic paths.
In particular, each path corresponds to one particle and contains $M$ beads, and each bead corresponds to the position of a particle at a (imaginary) time slice. 
Adjacent beads in one path interact through a spring-like potential proportional to $\frac{mr^2}{\tau}$, where $m$ is the mass of the particle represented by the path and $r$ is the distance between the two beads. 
This spring term is the path representation of the kinetic energy and describes the quantum motion as a classical diffusion process. 
Beads at the same time slice (i.e. with the same index) interact with each other through the potential $V\left( {\bf R} \right)$.
A quantum particle has therefore been mapped onto a ring polymer and each bead of the polymer feels a fraction of the potential energy.
This path or ring-polymer picture is sketched in Fig.~\ref{fig:path}(a). 
The essence of the PI method is therefore to map a many-body quantum system to a set of configurational classical ring-polymers. 
By integrating over all paths (configurations of all polymers), we can in principle obtain $Z$ and all thermodynamic properties as ensemble averages.

In reality, it is difficult to integrate over all the path variables. 
A practical approach is to sample the paths using the Monte Carlo method.
This is  the path-integral Monte Carlo (PIMC) simulation method \cite{ceperley_rmp_1995} mentioned above. 
Path configurations are randomly sampled using the integrand in Eq.~\ref{eqn:z2} as a distribution and Metropolis algorithm \cite{metropolis_random}.
The approach is used to obtain ensemble averaged properties such as the internal energy,
\begin{eqnarray}
E&=&-\frac{1}{Z}\frac{\partial Z}{\partial \beta}~.
\end{eqnarray}
Properties that are not normalized by the partition function such as the Helmholtz free energy ($A=-\frac{ \ln Z}{\beta}$) cannot be calculated by PIMC directly since Metropolis sampling yields ensemble averages as the ratio of integrals.
However, one may employ the technique of thermodynamic integration~\cite{kirk_tdi,briels_tdi,schlitter_tdi,sprik_tdi,kastner_thermo_int} to calculate the free energy difference by running several PIMC simulations.

In order to use PIMC to study superfluidity, one needs to account for Bose-Einstein statistics of indistinguishable particles. 
The Bose-Einstein distribution only includes states that are symmetric upon particle permutation.
This property must be included in the trace operation in Eq.~\ref{eqn:z1}. 
In the position representation, the following symmetrization operator, $\hat{S}=\frac{1}{n_{B}!}\sum_{i=1}^{n_{B}!}\hat{p}_i$, acts on
each basis state $\left|{\bf R} \right>$. 
$n_{B}$ stands for the number of bosons and the index $i$ runs through all $n_B!$ particle permutations with exchange operator
$\hat{p}_i$. 
For systems with more than one type of bosons, 
the total symmetrizer is a product of symmetrizers for each type of bosons. 
Therefore, the Bose-Einstein thermal density matrix in the position representation is
\begin{eqnarray}
\left< {\bf R} \right| e^{ -\beta\hat{H} }\hat{S}\left| {\bf R'}\right>. \label{eqn:be_rho}
\end{eqnarray}
Obviously, Eq.~\ref{eqn:be_rho} is invariant with respect to any permutation of particles, satisfying the requirement of Bose-Einstein statistics. With this density matrix and following the similar derivation to obtain Eq.~\ref{eqn:z2}, one obtains the Bose-Einstein partition function
\begin{eqnarray}
Z^{\rm BE}&=&\frac{1}{N_{p}}\left(\prod_{j=1}^n \lambda_{j,\tau}^{-3} \right)^M \int d {\bf R}_1\int d {\bf R}_2\cdots\int d {\bf R}_M \nonumber\\
&&\times \sum_{p=1}^{N_p} e^{ -\sum_{j=1}^n \frac{\pi}{\lambda_{j,\tau}^2} \left( \sum_{i=1}^{M-1}  \left( {\bf r}_{j,i}- {\bf r}_{j,i+1}\right)^2 + \left( \left( {\bf r}_{j,M}- {\bf r}_{p_j,1}\right)^2 \right) \right) -\tau \sum_{i=1}^MV\left( {\bf R}_i \right)}  \label{eqn:zbe}
\end{eqnarray}
Here $p$ stands for a certain permutation, $p_j$ the new particle index for particle $j$ after the permutation $p$, $N_p$ the total number of permutation and
\begin{eqnarray}
N_p&=&\prod_{i=1}^{M_B} n_i!,
\end{eqnarray}
where $M_B$ is the number of boson types in the system and $n_i$ is the number of type $i$ bosons.

Comparing Eqs.~\ref{eqn:z2} and~\ref{eqn:zbe}, we see that the path representation of bosonic systems is similar to that of boltzmannons, except that the last bead of particle $j$ should be connected to the first bead of particle $p_j$.
This modification yields  different values of the spring-like kinetic term. 
Such an example for two bosons is illustrated in Fig.~\ref{fig:path}(b). 
PIMC simulations for bosonic systems  require the sampling of both permutations and path configurations.

We would like to reemphasize the fact that the PIMC method involves only one approximation, the Trotter factorization, and this approximation is exact if the number of slices reaches infinity. 
In practise, one can always carry out a convergence study to determine the necessary number of slices to maintain the balance between accuracy and efficiency. 
Here we only present a very brief introduction to the PIMC method to provide the necessary background for the discussion of \phtwo~superfluidity. 
For the theoretical foundations of the path-integral formulation, readers should refer to the excellent textbooks and classic papers~\cite{chandler_pimc1,feynman_pi_rmp,ceperley_rmp_1995,schulman_pi,feynman_hibbs,kleinert_pi,kashiwa_pi,Roepstorff_pi,weigel_pi,khandekar_pi,kristen_pi}. 
More specific introductions to PIMC and its application to quantum fluids such as helium can be found in the works of Ceperley and co-workers\cite{ceperley_rmp_1995,pimc_quantum_fluid,ceperley_pimc_quantum_fluid,ceperley_pimc_3He, ceperley_pimc_he_h2,ceperley_pimc_he_mix, galli_pimc_he,mcmahon_rmp_2012} and  references therein. 
Specific PIMC algorithms required for the study of microscopic superfluidity will be discussed further in Sec.~\ref{sec:algor}. 
We conclude this section by quoting Boninsegni and coworkers who stated that \begin{quote}
{\em At least for Bose systems, PIMC is the only presently known method capable of furnishing in principle exact numerical estimates of physical quantities, including the superfluid density, and the condensate fraction}~\cite{boninsegni_worm_1}.
\end{quote}

\subsection{Zero temperature (ground state) description}
The  PIMC method  introduced above is based on the thermal density operator of a many-body quantum system. Another way to extract information from these systems is to study their eigenstates. 
Because of the computational complexity associated with the solution of the many-body Schr\"odinger equation, one needs to rely on stochastic sampling methods to obtain ground state properties. 
We  describe such methods below.

A well developed approach is the {\em Diffusion Monte Carlo} (DMC) method.
The approach has been widely used in the study of microscopic superfluids composed of $^4$He and \phtwo particles.
DMC is a stochastic projection method based on the similarity between the Schr\"odinger equation and the diffusion equation. A state that satisfies the time-dependent Schr\"odinger equation
\begin{eqnarray}
i\hbar\frac{\partial \left|\Psi \left(t \right)\right>}{\partial t}&=&\left( \hat{H} - E_T \right)\left|\Psi \left(t \right)\right>,
\end{eqnarray}
also satisfies the imaginary time Schr\"odinger equation
\begin{eqnarray}
-\frac{\partial \left|\Psi \left(\tau \right)\right>}{\partial \tau}&=&\left( \hat{H} - E_T \right)\left|\Psi \left(\tau \right)\right>, \label{eqn:itse}
\end{eqnarray}
where $\tau$ is real and associated with $\frac{it}{\hbar}$ with $t$ being imaginary.
Here $E_T$ is an energy offset and it is a useful parameter in DMC simulations. 
The formal solution of Eq.~\ref{eqn:itse} is
\begin{eqnarray}
\left|\Psi \left(\tau \right)\right>&=&e^{ -\tau\left(\hat{H} - E_T \right)}\left|\Psi \left(0 \right)\right>. \label{eqn:formal_solution}
\end{eqnarray}
The initial state can be written as a linear combination of eigenstates of $\hat{H}$
\begin{eqnarray}
\left|\Psi \left(0 \right)\right>&=&c_0\left|\Phi_0\right>+\sum_i c_i \left|\Phi_i\right>. \label{eqn:expansion}
\end{eqnarray}
The subscript ``0'' labels the eigenstate with the lowest energy in the expansion and this state is singled out. Substitution of Eq.~\ref{eqn:expansion} in Eq.~\ref{eqn:formal_solution} results in
\begin{eqnarray}
\left|\Psi \left(\tau \right)\right>&=&c_0 e^{ -\tau \left(E_0 - E_T \right) }\left|\Phi_0\right>+ \sum_i c_i e^{ -\tau \left(E_i - E_T \right) }\left|\Phi_i\right>.
\end{eqnarray}
Obviously, in the $\tau\rightarrow\infty$ limit, 
the first term on the right-hand-side of this equation dominates and we have
\begin{eqnarray}
\left|\Psi \left(\tau\rightarrow\infty \right)\right>&\propto&\left|\Phi_0\right>.
\end{eqnarray}
In words, operating with the propagator 
$e^{ -\tau\left(\hat{H} - E_T \right)}$
on an initial state long enough would yield the ground state of a complex many-body system, as long as the ground state has a non-zero overlap with the initial state.
Note that $\tau$ here has the similar meaning of imaginary time as in Sec.~\ref{sec:formu}\ref{subsec:pi}. However, in Sec.~\ref{sec:formu}\ref{subsec:pi}, $\tau$ takes the limit of approaching zero, while here approaching infinity.

In the position representation, Eq.~\ref{eqn:formal_solution} becomes
\begin{eqnarray}
\Psi\left({\bf R},\tau \right)&=&\left< {\bf R}\right| e^{ -\tau\left(\hat{H} - E_T \right)}\left|\Psi \left(0 \right)\right> \nonumber\\
&=&\int {\bf R'}\left< {\bf R}\right| e^{ -\tau\left(\hat{H} - E_T \right)}\left| {\bf R'} \right>\left<{\bf R'}\right.\left|\Psi \left(0 \right)\right> \nonumber \\
&=&\int {\bf R'}\left< {\bf R}\right| e^{ -\tau\left(\hat{H} - E_T \right)}\left| {\bf R'} \right>\Psi\left({\bf R'},0 \right).
\end{eqnarray}
The
$\left< {\bf R}\right|e^{ -\tau\left(\hat{H} - E_T \right)}\left| {\bf R'} \right>$
factor is called the imaginary time Green's function, $G\left( {\bf R} \leftarrow {\bf R'}, \tau \right)$. Obviously, it is similar to the $\left< {\bf R}_i \right| \hat{\rho}\left( \tau \right) \left| {\bf R}_{i+1} \right>$ factors in Eq.~\ref{eqn:z1} and its exact form is unknown except for some simple cases such as the harmonic oscillator. 
Its short time (small $\tau$) approximation is similar to the one used in the PIMC derivation above.
But this time, using the Hermitian factorization formula~\cite{hermitian_trotter_suzuki}
\begin{eqnarray}
e^{-\tau\left(\hat{A} +\hat{B} \right)}&\approx&e^{ -\frac{\tau\hat{B}}{2} } e^{ -\tau\hat{A} } e^{ -\frac{\tau\hat{B}}{2}},
\end{eqnarray}
instead of the primitive Trotter approximation in Eq.~\ref{eqn:trotter_prim}, is more convenient. 
The error of this Hermitian approximation is of  order  $\tau^3$. 
Following a derivation similar to that  for Eq.~\ref{eqn:linkb}, 
we have the short imaginary time approximate Green's function
\begin{eqnarray}
G\left( {\bf R} \leftarrow {\bf R'}, \tau\rightarrow 0 \right)&\approx&\left(\prod_{j=1}^n \lambda_{j,\tau}^{-3} e^{ -\frac{\pi}{\lambda_{j,\tau}^2} \left( {\bf r}_{j} -{\bf r}_{j}' \right)^2 }\right)\nonumber\\
&&\times e^{ -\tau \frac{\left( V\left( {\bf R}\right)+V\left( {\bf R'}\right) \right) }{2}+\tau E_T}. \label{eqn:green_short}
\end{eqnarray}
The initial function $\Psi\left({\bf R'},0 \right)$ can be represented as a set of discrete sampling points that are called ``walkers'' in the realm of DMC. In a DMC simulation, these walkers undergo Brownian motion following the propagator of $G\left( {\bf R} \leftarrow {\bf R'}, \tau\rightarrow0 \right)$ for each short time step $\tau$. 
A large number of such steps are equivalent to a long time propagation of the initial state and 
the resultant walkers distribution reflects the amplitude of the ground state wave function. 
This approximate wave function can then be used for the evaluation of properties ~\cite{dmc_descendent_weight1,dmc_descent_weight2}.

The term in the first parentheses on the right-hand-side of Eq.~\ref{eqn:green_short} is actually the propagator of a diffusion process~\cite{second_course_diffusion}. 
The second exponential term in Eq.~\ref{eqn:green_short} subjects the diffusion process to a potential and concentrates (depletes) the walkers in (from) the low (high) potential region. 
In practise, this gathering process is realized through the {\em birth/death algorithm}~\cite{reynolds_birth-death}, which kills the walkers if they are in regions of high potential or otherwise causes them to proliferate. 
The energy offset $E_T$, which appears in the potential exponential term, can be adjusted wisely to control the total number of walkers to balance the efficiency and accuracy of a simulation.

With the fixed-node approximation~\cite{anderson_fixed_node_75,anderson_fixed_node_76,moskowitz_fixed_node_82}, DMC can also be used to obtain approximate excited state wave functions. 
This method requires one to introduce trial nodal surfaces of the excited state wave function of interest and independent DMC simulations are performed in each region surrounded by the nodes.
It becomes an exact method if the nodal surfaces are determined {\it a priori} by symmetry.
Walkers attempting to cross the nodal surfaces would be eliminated. i.e., an infinitely high potential barrier is placed at the nodes.
This constraint is equivalent to imposing Dirichlet boundary conditions for the excited state wave function. The walkers distribution within each region will mimic the absolute value of amplitude of the wave function. The accuracy of this approximation crucially depends on the quality of the trial nodal surfaces.

The sampling efficiency of DMC can be largely increased using importance sampling~\cite{grimm_isdmc,ceperley_isdmc}. 
In this method, a {\em trial} or {\em guiding} wave function that is a (hopefully) good approximation to the wave function of interest is introduced to the simulation. 
The effect of this trial function is not as simple as providing a good initial distribution of walkers. 
Its inclusion leads to an effective short time Green's function, which enhances the density of walkers in the regions with large trial function amplitude and diminishes the population fluctuation of walkers in the course of simulation. Walkers that are close to the nodal surface of the trial function will be carried away and the fixed-node approximation is naturally included. The efficiency of this sampling scheme largely relies on how closely the trial function resembles the wave function of interest.

A method called rigid body diffusion Monte Carlo (RBDMC) was proposed and developed to study the problem of quantum rotation~\cite{buch_rbdmc}. This method ignores the non-commutators between different components of angular momentum and treats rotations about different axes as translations along different axes. Therefore, it employs a Green's function similar to Eq.~\ref{eqn:green_short}, but its arguments include angles that specify orientation of rigid molecules. 
For a small enough time step, this is a good approximation, and it can be combined with the fixed-node approximation and importance sampling~\cite{niyaz_arnhf,whaley_SF6,viel_isrbdmc}. Because of its ability to describe quantum rotation, RBDMC has been used in simulating rotors doped inside $^4$He and \phtwo~clusters.

The development and application of DMC is a very broad and deep subject. 
This methodology has been developed to handle an extensive range of problems including electronic structure, solid state physics, and large amplitude rovibrational motion of weakly interacting complexes.
In the present report, we only provide a brief overview and cover some necessary background for forthcoming discussion. Interested readers should refer to Refs.~\cite{barnett_dmc_93,foulkes_qmc_review,anderson_dmc,hammond_mc_in_qc,mccoy_dmc_irpc, mccoy_dmc_2006,needs_dmc_2010,mccoy_dmc_2012} for more details.

A methodology that is closely connected to DMC was developed to obtain excitation energies without solving the Schr\"odinger equation. It is called the projection operator imaginary time spectral evolution (POITSE) method \cite{poitse1}. With this method, one can obtain excitation energies of systems such as doped He and \hydrogen~clusters. The $B_{\rm eff}$ can be extracted from these states and the effective inertia can be studied. Because of its importance, a brief account of this method is given here.
The essential quantity of POITSE is the spectral function
\begin{eqnarray}
\kappa\left(E\right)&=&\sum_n \left| \left< \Phi_0 \right|\hat{A}\left| \Phi_n \right> \right|^2 \delta\left( E-E_n+E_0 \right),
\end{eqnarray} 
where $\left\{ \left| \Phi_n \right> \right\}$ and $\left\{ E_n \right\}$ are eigenstates and eigenenergies of the system, and the subscript $0$ denotes the ground state. The operator $\hat{A}$ couples the ground state to a certain set of excited states. The Laplace transform of the spectral function 
yields an imaginary time ($\tau$) correlation function  
\begin{eqnarray}
\tilde{\kappa}\left( \tau \right)&=&\sum_n \left| \left< \Phi_0 \right|\hat{A}\left| \Phi_n \right> \right|^2 e^{ -\tau\left(E_n - E_0\right) } \nonumber\\
&=&\left< \Phi_0\right| \hat{A} e^{-\tau\left( \hat{H}-E_0 \right) } \hat{A}^\dagger \left| \Phi_0 \right>.
\end{eqnarray}
The POITSE method consists of two steps: (i) evaluate $\tilde{\kappa}\left( \tau \right)$ by Monte Carlo  and (ii) evaluate $\kappa\left(E\right)$ through an inverse Laplace transform. 
Readers should be reminded that the inverse Laplace transform is an unstable procedure. It may lead to biased results.
In general the exact $\left| \Phi_0 \right>$ and $E_0$ are unknown and one needs to use their approximate counterparts, $\left| \Phi_T \right>$ and $E_{ref}$, which may come from a DMC simulation. 
Here the subscript ``T'' and ``ref'' stand for trial function and reference energy. 
The use of an approximate ground state introduces a systematic bias in excitation energies. A way to eliminate this bias is to renormalize the approximate imaginary time correlation function by the factor~\cite{poitse1}
\begin{eqnarray}
\left< \Phi_T\right| e^{ -\tau\left( \hat{H}-E_{ref} \right)  } \left| \Phi_T \right>,
\end{eqnarray}
i.e.
\begin{eqnarray}
\tilde{\kappa}\left( \tau \right)&\approx&\frac{\left< \Phi_T\right| \hat{A} e^{ -\tau\left( \hat{H}-E_{ref} \right) }\hat{A}^\dagger \left| \Phi_T \right>}{\left< \Phi_T\right| e^{ -\tau\left( \hat{H}-E_{ref} \right) } \left| \Phi_T \right>}.
\end{eqnarray}

Apparently, the same propagator
$e^{ -\tau\left( \hat{H}-E_{ref} \right)}$
as in the DMC simulation is involved in POITSE and a similar propagation technique is employed. 
The evaluation of $\tilde{\kappa}\left( \tau \right)$ involves propagating walkers whose initial distribution follows the probability density of $\left| \Phi_T \left( {\bf R} \right) \right|^2$. 
The resultant $\tilde{\kappa}\left( \tau \right)$ will then be inverse Laplace transformed to $\kappa\left(E\right)$ with the Maximum Entropy Method (MEM)~\cite{mem1} based on Bayesian statistics~\cite{baysesian_statistics1}. 
From $\kappa\left( E \right)$, the energies of the excitations induced by $\hat{A}$ would be readily obtained. 
Detailed algorithms for the POITSE method are beyond the scope of the present report and readers should refer to Refs.~\cite{poitse1,poitse2} and the references therein to learn more about this approach. 
One last comment on POITSE is that by choosing an appropriate coupling operator $\hat{A}$,
e.g., the orientation of the dopant molecule,
one can single out a set of excitations that are most useful in the study of  superfluid clusters.

Another quantum Monte Carlo method that has been used to investigate ground states of superfluid \phtwo~clusters is the Reptation Quantum Monte Carlo (RQMC). This method is introduced in Refs.~\cite{rqmc1,rqmc2} and here we only summarize its features without giving any derivation. The method does not propagate walkers in a sequence of imaginary time steps as DMC does. It samples a segment of walkers that is called {\em reptile} by Moroni and Baroni, the two proposers of this method, and the sampling is based on the Langevin equation. The length of the reptile represents the length of imaginary time, and by sampling the reptile, imaginary time correlation functions can be obtained. This is the most important feature of RQMC. Also, RQMC has an advantage that the ground state local properties are evaluated without mixed estimates and population control bias.
However, this method may be subject to serious ergodicity problems and a long projection may be needed to have converged results.
The path-integral ground state (PIGS) approach \cite{Hetenyi1999,Sarsa2000} is closely related to RQMC.
The PIGS is also referred to as the Variational Path Integral method \cite{ceperley_rmp_1995,Sarsa2000,Hinde2006}.
One advantage of the PIGS method is that it does not  suffer from the population bias problem \cite{Boninsegni:2012cn} of DMC. The approach has been successfully used to simulate condensed helium\cite{cuervo2005} and weakly bound   parahydrogen clusters \cite{cuervo2006,cuervo2008,cuervo2009}. A new Langevin equation Path integral Ground State (LePIGS) approach has recently been proposed \cite{constable2013}. The advantage of the method is that since the sampling is performed using equations of motion, the design of Monte Carlo moves is not required. The methods has recently been applied to simulate hydrogen clusters\cite{lepigsH2} and to predict the vibrational Raman shifts of \phtwo clusters \cite{lepigsH2raman} with very good agreement with the results of Ref. \cite{tejeda_raman_ph2_cluster}.

Beside  quantum Monte Carlo methods, one could in principle handle many-body bosonic problems by expanding the Hamiltonian operator with basis functions followed by diagonalisation.
Recently, de Lara-Castells and Mitrushchenkov used such an approach to study  doped \phtwo~clusters~\cite{ph2_ocs_ci}. 
They developed a method that closely resembles the treatment of electron correlation  in quantum chemistry using a full-configuration-interaction nuclear orbital approach. 
Ideally, the only error of this type of method comes from the limit of basis-set size and a Born-Oppenheimer type approximation. 
It describes low-lying excited states as well as the ground state. 
The unfavourable basis-set scaling with respect to system size, however, restricts its usage. 
The further development and usage of this method will be dependent on the advances of quantum chemistry methodologies to treat systems with a large number of electrons.

\subsection{Non-classical effective moment of inertia} \label{subsec:noncl_I}

The non-classical reduction of the moment of inertia of $^4$He and \phtwo~clusters
discussed in Secs.~\ref{sec:intro} and~\ref{sec:exp} provides a window into  the 
nature of microscopic superfluidity.
As the size of clusters increases and more bosons surround the dopant rotor, 
the classical moment of inertia ($I^{\rm cl}$) calculated from the mass density distribution can only increase since more mass has been added. 
Therefore, the non-increase of the effective moment of inertia ($I^{\rm eff}$) can only come from a dynamical quantum effect.
Phenomenologically, this non-classical behaviour is attributed to the coherent decoupling between the superfluid and the normal components of the bosons. 
For a macroscopic rotor like the disks in the classic Andronikashvili experiment, 
the non-classical behaviour of helium $I^{\rm eff}$ has been discussed in Sec.~\ref{sec:intro} and illustrated in Fig.~\ref{fig:bucket}(b). 
For a microscopic superfluid, an additional finite size effect is present. As the number of bosons ($N$) increases, the original normal component can be converted to a superfluid.
Therefore, $I^{\rm eff}$ not only ceases to increase, but starts to decrease, displaying a downward turnaround (upward turnaround of $B_{\rm eff}$). 
The first turnaround in the $B_{\rm eff}$ vs. $N$ diagram is naturally considered to signal the onset of superfluidity of the cluster while the later turnarounds are manifestations of the non-monotonic dependance of the superfluid and normal components as a function of $N$.

In spectroscopic Andronikashvili experiments, effective rotational constants can be extracted by fitting the energy levels of a model Hamiltonian to the spectra. 
For example, for a spectrum with the dominant characters of a prolate symmetric top, 
effective rotational energy levels
\begin{eqnarray}
B_{\rm eff}J\left(J+1\right)+\left(A_{\rm eff}-B_{\rm eff} \right)K_a^2
\end{eqnarray}
can be used to fit the spectra to obtain $B_{\rm eff}$ and $A_{\rm eff}$. Throughout this report, we follow the convention of $A \ge B \ge C$ for rotational constants. Effective moments of inertia can be calculated as
\begin{eqnarray}
I_b^{\rm eff}=\frac{\hbar^2}{2{B_{\rm eff}}}
\end{eqnarray}
and similarly for the other components. The contribution from the bosons can be readily calculated as $I_b^{\rm B}=I_b^{\rm eff} - I_b^{\rm rot}$, where $I_b^{\rm rot}$ is the moment of inertia of the rotor itself. It is straightforward to generalize this fitting procedure to spectra with different characteristic model Hamiltonian. This procedure was widely used in the IR experiments introduced in Sec.~\ref{sec:exp}. In the MW experiments introduced in Sec.~\ref{sec:exp}\ref{subsec:exp_cluster}, a linear molecule model was employed for fitting and the $J=1-0$ transition was directly used to calculate $B_{\rm eff}$ through
\begin{eqnarray}
\Delta E_{J=1-0}&=&2B_{\rm eff}.
\end{eqnarray}

In theoretical studies, a good way to calculate $I^{\rm eff}$ is using linear response theory and  the following definition,
\begin{eqnarray}
I^{\rm eff}_n&=&\left.\frac{\partial\left< \hat{L}_n \right>}{\partial \omega}\right|_{\omega=0},
\end{eqnarray}
where $\hat{L}_n$ stands for the total angular momentum operator of the bosons along the $n$ direction and $\left< \right>$ stands for the thermal average. 
This derivative measures the linear response of the bosons to a rotational field with an infinitesimally small angular velocity $\omega$ along the $n$ direction~\cite{rotating_bucket_1,rotating_bucket_2,Ieff_stringari}. 
The Andronikashvili experiment measures the contribution from the liquid to the rotor. 
Therefore, $I^{\rm eff}$ should be calculated in a frame rotating with the rotor. 
This requires that the Hamiltonian operator in the average $\left< \right>$  be written 
in a rotor-fixed frame (RFF).

Let us momentarily assume that the rotor is rotating about the space-fixed $z$-axis. 
Then the $x'$ and $y'$ coordinates in the RFF are connected to the space-fixed frame (SFF) 
coordinates $x$ and $y$ through
\begin{eqnarray}
x'&=&x\cos \left( \omega t \right)+y\sin\left( \omega t \right); \nonumber \\
y'&=&-x\sin\left( \omega t \right)+y\cos\left( \omega t \right). \label{eqn:x'y'}
\end{eqnarray}
For a one-particle state function $\Psi\left(x,y,z,t\right)$, this coordinate transform modifies the time-derivative side of its Schr\"odinger equation as
\begin{eqnarray}
i\hbar\frac{\partial \Psi\left(x,y,z,t \right)}{\partial t}&=&i\hbar\left( \frac{\partial \Psi'\left(x',y',z,t \right)}{\partial t} + \frac{\partial \Psi'\left(x',y',z,t \right)}{\partial x'}\frac{\partial x'}{\partial t}\right.\nonumber\\
&&\left. +\frac{\partial \Psi'\left(x',y',z,t \right)}{\partial y'}\frac{\partial y'}{\partial t} \right) \nonumber\\
&=&i\hbar\left( \frac{\partial \Psi'\left(x',y',z,t \right)}{\partial t}\right. \nonumber\\
&&\left. -\omega \left( x'\frac{\partial}{\partial y'} -y'\frac{\partial}{\partial x'}\right)\Psi'\left(x',y',z,t \right) \right) \nonumber\\
&=&i\hbar\frac{\partial \Psi'\left(x',y',z,t \right)}{\partial t} + \omega\hat{L}_z'\Psi'\left(x',y',z,t \right).
\end{eqnarray}
The prime denotes the function, operator, and coordinates in the RFF. 
The chain rule and Eq.~\ref{eqn:x'y'} are employed to obtain the second equality on the right hand side.
Since the Laplacian operator is invariant to frame rotation, the kinetic energy operator is unchanged in this coordinate transform. Usually, the potential energy operator depends on the physical position and orientation of a particle and therefore, it is also invariant to the transformation. Finally, the time-dependent Schr\"odinger equation of this one-particle state in the RFF becomes
\begin{eqnarray}
i\hbar\frac{\partial \Psi'\left( {\bf r}',t \right)}{\partial t}&=&\left(\hat{T}'+\hat{V}'-\omega\hat{L}_z'\right)\Psi'\left({\bf r}',t \right) \nonumber\\
&=&\left(\hat{H}-\omega\hat{L}_z' \right)\Psi'\left( {\bf r}',t \right).
\end{eqnarray}
Thus, the Hamiltonian operators in the RFF and SFF are connected by
\begin{eqnarray}
\hat{H}'&=&\hat{H}-\omega\hat{L}_z'.
\end{eqnarray}

The derivation above is given in page 260 of Ref.~\cite{untracold_qf_stoof}. With the inverse transform
\begin{eqnarray}
x&=&x'\cos \left( \omega t \right)-y'\sin\left( \omega t \right); \nonumber \\
y&=&x'\sin\left( \omega t \right)+y'\cos\left( \omega t \right). \label{eqn:xy}
\end{eqnarray}
and using the chain rule, one can easily show that
\begin{eqnarray}
x'\frac{\partial}{\partial y'}-y'\frac{\partial}{\partial x'}&=&x\frac{\partial}{\partial y}-y\frac{\partial}{\partial x}; \nonumber\\
\hat{L}_z'&=&\hat{L}_z.
\end{eqnarray}
Therefore,
\begin{eqnarray}
\hat{H}'&=&\hat{H}-\omega\hat{L}_z,
\end{eqnarray}
where all operators on the right-hand-side are in the SFF. This connection between Hamiltonians can be generalized to the case of many particles and a frame rotating with an angular velocity $\omega$ along an arbitrary axis specified by a unit vector $\hat{n}$ as
\begin{eqnarray}
\hat{H}'&=&\hat{H}-\omega \hat{n} \cdot{\bf \hat{ L}},\label{eqn:h_rff_sff}
\end{eqnarray}
where ${\bf \hat{ L}}$ is  the total angular momentum operator of all particles. We would like to reemphasize that the dot product $\hat{n} \cdot{\bf \hat{ L}}$ is invariant  with respect to the change between the RFF and SFF and all the operators on the right-hand-side of Eq.~\ref{eqn:h_rff_sff} are conventionally chosen to be in the SFF.

With the above formulas, we see that the contribution of a given type of bosons to the inertia
of the whole system is
\begin{eqnarray}
I^{\rm eff}_n&=&\left.\frac{\partial\left< \hat{L}_n \right>}{\partial \omega}\right|_{\omega=0} \nonumber\\
&=&\left(\frac{\partial}{\partial \omega} \frac{{\rm Tr} \left(e^{  -\beta \hat{H}' }\hat{L}_n\right)}{Z' }\right)_{\omega=0} \nonumber\\
&=&\left(\frac{\partial}{\partial \omega} \frac{{\rm Tr} \left(e^{ -\beta \hat{H}' }\hat{L}_n\right)}{{\rm Tr} \left(e^{ -\beta \hat{H}' }\right)}\right)_{\omega=0}, \label{eqn:ieff1}
\end{eqnarray}
where $\hat{L}_n$ is the total angular momentum operator of that given type of bosons. 
In the further derivation below, we will use an important formula for taking the derivative of an exponential operator:
\begin{eqnarray}
\frac{\partial }{\partial \lambda}e^{ -\beta \hat{H} }&=&-\int_{0}^{\beta} e^{ -\left(\beta-\tau \right)\hat{H} }\frac{\partial \hat{H}}{\partial \lambda} e^{-\tau\hat{H} }d\tau.
\end{eqnarray}
This formula is introduced in Refs.~\cite{wilcox_derivative,sinder_derivative}. 
With this formula, we have
\begin{eqnarray}
\frac{\partial e^{ -\beta\hat{H}' }}{\partial \omega}&=&\int_{0}^{\beta} e^{ -\left(\beta-\tau \right)\hat{H} }\hat{L}_n e^{ -\tau\hat{H} }d\tau,
\end{eqnarray}
and we can continue the derivation of Eq.~\ref{eqn:ieff1} as
\begin{eqnarray}
I^{\rm eff}_n&=&\left[\frac{{\rm Tr}\left( \int_0^\beta e^{ -\left( \beta-\tau\right)\hat{H}' }\hat{L}_n e^{ -\tau\hat{H}' }\hat{L}_n d\tau\right)}{Z'}\right]_{\omega=0} -\nonumber\\
&&\left[\frac{{\rm Tr}\left( \int_0^\beta e^{\left( - \beta-\tau\right)\hat{H}' }\hat{L}_n e^{ -\tau\hat{H}' } \right)Tr\left( e^{ -\beta\hat{H}' } \hat{L}_n\right)}{Z'^2}\right]_{\omega=0}.
\end{eqnarray}
Using the cyclic invariance of the trace operation and the commutation between the integration and the trace, the second term on the right-hand-side can be easily shown to be
\begin{eqnarray}
-\beta\left< \hat{L}_n\right>^2.
\end{eqnarray}
Following a similar procedure, the first term becomes
\begin{eqnarray}
\int_0^\beta \left< \hat{L}_n\left( \tau \right) \hat{L}_n \right> d\tau,
\end{eqnarray}
with the definition of the imaginary time evolved angular momentum operator in the Heisenberg picture
\begin{eqnarray}
\hat{L}_n\left( \tau \right)&=&e^{ \tau\hat{H}' }\hat{L}_n e^{ -\tau\hat{H}' }.
\end{eqnarray}
Therefore,
\begin{eqnarray}
I^{\rm eff}_n&=&\int_0^\beta \left< \hat{L}_n\left( \tau \right) \hat{L}_n \right> d\tau-\beta\left< \hat{L}_n\right>^2. \label{eqn:Ieff2}
\end{eqnarray}
In the $\omega=0$ limit, the averages in Eq.~\ref{eqn:Ieff2} should be carried out with the density operator in the SFF
\begin{eqnarray}
e^{-\beta\hat{H}}.
\end{eqnarray} 

In the usual case where time-reversal symmetry is obeyed,
\begin{eqnarray}
\left< \hat{L}_n \right>&=&0,
\end{eqnarray}
and 
\begin{eqnarray}
I^{\rm eff}_n&=&\int_0^\beta \left< \hat{L}_n\left( \tau \right) \hat{L}_n \right> d\tau. \label{eqn:Ieff3}
\end{eqnarray}
$I^{\rm eff}$ is thus obtained from integrating the imaginary time angular momentum correlation function and this connection is reasonable. 
For an Andronikashvili experiment, if the bosons adiabatically follow the rotation the the rotor (a stack of disks or a molecular dopant), their total angular momentum should be maximally correlated, giving a large $I^{\rm eff}$. 
On the other hand, if the bosons are superfluid and not dragged by the rotor, their total angular momentum should be least correlated, giving a negligible $I^{\rm eff}$.

Given a set of eigenstates ($\left\{ \left| m \right> \right\}$) of $\hat{H}$,
\begin{eqnarray}
\hat{H}\left| m\right>&=&E_m\left| m\right>,
\end{eqnarray}
 Eq.~\ref{eqn:Ieff3} can be further simplified as
 \begin{eqnarray}
 I^{\rm eff}_n&=&\int_0^\beta \left< \hat{L}_n\left( \tau \right) \hat{L}_n \right> d\tau \nonumber\\
 &=&\frac{1}{Z}{\int_0^\beta \sum_m\left<m\right| e^{ -\left(\beta-\tau\right)\hat{H} }\hat{L}_n e^{ -\tau \hat{H} }\hat{L}_n\left|m\right>} \nonumber\\
 &=&\frac{1}{Z}{\int_0^\beta \sum_{m,k}\left<m\right| e^{ -\left(\beta-\tau\right)\hat{H} }\hat{L}_n e^{ -\tau \hat{H} }\left| k\right>\left<k\right|\hat{L}_n\left|m\right>} \nonumber\\
 &=&\frac{1}{Z}{\int_0^\beta \sum_{m,k} e^{-\left(\beta-\tau\right)E_m }\left<m\right|\hat{L}_n\left| k\right> e^{ -\tau E_k }\left<k\right|\hat{L}_n\left|m\right>}\nonumber\\
 &=&\frac{1}{Z}{\sum_{m,k} e^{ -\beta E_m }\left| \left< m \right|\hat{L}_n\left| k \right> \right|^2\int_0^\beta  e^{ \left(E_m-E_k \right)\tau }d\tau}\nonumber\\
 &=&\frac{1}{Z}\sum_{m,k} e^{ -\beta E_m }\left| \left< m \right|\hat{L}_n\left| k \right> \right|^2\frac{ e^{ \beta\left(E_m - E_k \right) }-1}{E_m - E_k}\nonumber\\
 &=&\frac{1}{Z}\sum_{m,k}\frac{ e^{ -\beta E_k }-e^{ -\beta E_m }}{E_m-E_k}\left| \left< m \right|\hat{L}_n\left| k \right> \right|^2  \nonumber\\
 &=&\frac{2}{Z}\sum_{m,k}e^{ -\beta E_k }\frac{\left|\left<m \right|\hat{L}_n\left|k \right> \right|^2}{E_m-E_k}, \label{eqn:Ieff4}
 \end{eqnarray}
 which is Eq. 3 of Ref.~\cite{Ieff_stringari}. Here we provide a derivation of this equation, which is skipped in the reference. For most realistic applications, it is impossible to obtain the set of eigenstates and therefore, impossible to calculte $I^{\rm eff}$ by a straightforward application of Eq.~\ref{eqn:Ieff4}. Instead, one usually calculates $I^{\rm eff}$ through Eq.~\ref{eqn:Ieff3} in PIMC simulations. Discussion of this approach is deferred to Sec.~\ref{sec:algor}, in which review specific algorithms used in PIMC studies of microscopic superfluid systems.
 
\subsection{Response to an external field versus response to molecular dopant rotation} \label{subsec:SFFvsMFF}

In a macroscopic Andronikashvili experiment, the rotor (the stack of disks in Fig.~\ref{fig:bucket}(a)) is only allowed to rotate about a fixed axis. On the other hand, due to the large difference between the period of the macroscopic rotor and the relaxation time of the helium to its rotation, adiabatic separation between the quantum liquid and the classical rotor is strict and the limit of $\omega=0$ in Eq.~\ref{eqn:ieff1} is truly satisfied. 
The linear response theory calculation of $I^{\rm eff}$ introduced in Sec.~\ref{sec:formu}\ref{subsec:noncl_I} is exact for this kind of systems. 
Applying this theory to a microscopic cluster, however, deserves further discussion.

For a microscopic cluster, a true analogue of the macroscopic Andronikashvili experiment is to have a rotor that rotates about a fixed axis in the SFF, grips the cluster tightly to force its normal component to rotate with it, and rotates infinitesimally slowly to satisfy the linear response condition. 
This thought experiment is illustrated in Fig.~\ref{fig:sff_mff}(a). 
With such a rotor, Eq.~\ref{eqn:Ieff3} can be applied to calculate the true $I^{\rm eff}$ of a microscopic cluster. 
This {\em true} $I^{\rm eff}$ measures the response of the cluster to an external rotating field exerted by the rotor. 
Because of the isotropy of the SFF, any axis can be chosen to be the pivot axis of the field. 
However, such a rotor does not exist. 
At the scale of a microscopic cluster, this external rotor would have to follow the laws of quantum mechanics. 
The requirement for this {\em quantum} rotor to rotate about a fixed axis with infinitesimally small angular momentum is a violation of the uncertainty principle.

In an actual spectroscopic Andronikashvili experiment, one measures the response of a cluster to the rotation of a molecular dopant, and this is illustrated in Fig.~\ref{fig:sff_mff}(b). 
One needs to rely on a Born-Oppenheimer type adiabatic separation between the quantum rotor and the bosons to fulfill the two requirements of fixed axis and infinitesimally slow rotation. 
The Born-Oppenheimer approximation states that nuclei are far heavier than electrons and therefore, 
their motion is much slower and can be considered as stationary in the view of electrons. 
Consequently, molecular electronic structure studies can be conducted in the nuclear-fixed frame. 
This approximation forms the foundation of quantum chemistry~\cite{szabo_book,mol_ele_stru_theo,cramer_book,levine_qc5,jensen_book,piela_book}. 
Analogously, if the rotor is heavy enough and its characteristic rotational energy level spacing is much smaller than that of the characteristic boson energy level spacing in the field of the rotor, one can consider the relaxation time of the bosons to the orientation of the rotor is much shorter than the period of the rotor rotation. 
Therefore, the superfluid response of the bosons can be calculated in the molecular-fixed frame (MFF) of the rotor. 
From this point on, we use the term ``MFF'' to denote the frame rigidly attached to a quantum rotor (molecule), and ``RFF'' is reserved for the frame attached to a macroscopic classical rotor. 
In the MFF, the principal axes of the molecule are fixed and a heavy enough molecule would rotate about these axes slowly, making a good approximation to the ``stack of disks''. 
On the contrary, the rotation of a light molecule is highly quantum, i.e., its orientation fluctuates with large amplitude, preventing the bosons to follow. 
The coupling between this kind of rotors and the bosons is analogous to the vibronic coupling between nuclei and electrons in some molecules~\cite{bersuker_1989,fischer_vc,bersuker_2006,bersuker_2008}.

Besides being slow (heavy), 
another requirement for a good molecular superfluid probe is that its interaction 
with the bosons should be anisotropic enough to drag the bosons along with its rotation~\cite{tang_briding_bap,mak_ch4_he,renorm_nh3_he}. 
For obvious reasons, if the rotor-boson interaction is highly isotropic, 
the bosons will not respond to the orientation of the rotor and there will be no analogy with the entrainment of helium by the ``stack of discs'' in Fig.~\ref{fig:bucket}(a). 
Under these circumstances, superfluidity will be overestimated since the decoupling does not solely come from bosonic exchange. 
The last requirement, which is also obvious, is that the rotor-boson interaction needs to be attractive enough such that the dopant rotor molecule will inside in the centre of the cluster.

The preceding discussion illustrates how a molecular dopant can, under certain conditions, approximate an infinitesimally slowly rotating external field that drags the bosonic particles. The extent of this drag in turn reveals the genuine superfluid response.
In short, a good superfluid probe in spectroscopic Andronikashvili experiments needs to be less quantum than the bosons themselves, and have highly anisotropic and strong enough interactions with the bosons. 
These requirements should be considered as three general guidelines for choosing a good dopant for spectroscopic Andronikashvili experiments. 
A good way to judge the quality of a molecular dopant is to compare the superfluid fractions calculated in the MFF and SFF. In Sec.~\ref{sec:results}\ref{subsec:nonlinear}, we introduce such an example.

\section{Simulation algorithms for Path-Integral Monte Carlo} \label{sec:algor}

Among all theoretical methods, PIMC has been the most extensively employed in the study of microscopic superfluids. 
In this section, we introduce some algorithms that are especially useful for simulating and analyzing superfluid clusters and droplets. 
We first look at algorithms for sampling rigid-body rotations. 
These algorithms are extremely important because, as pointed out above, the superfluid information is stored in the hindered rotation of the molecular dopant. 
We then introduce a very efficient algorithm for sampling bosonic exchange that has been adapted from lattice models to continuous space microscopic superfluidity research in the last decade. 
Lastly, we go back to the effective moment of inertia and introduce an algorithm for evaluating this property.

\subsection{Sampling rigid-body rotation}

The PIMC scheme introduced in Sec.~\ref{sec:formu}\ref{subsec:pi} is based on a system of point-like particles. As a result, the spring-like kinetic energy term in Eq.~\ref{eqn:z2} only depends on the position of one particle at two adjacent imaginary time slices, not its orientation. 
An extension to treat rigid-body rotation is straightforward. 
One needs to include the orientation representation in the resolution of the identity and the rotational kinetic energy in the Hamiltonian operator. 
We first focus on the case of one rotor and note that the generalization to more rotors is straightforward.

For the case of one rotor, the identity operator should be written as a direct product of the resolutions of identity in the position and orientation representations
\begin{eqnarray}
{\bf \hat{1}}=\int d {\bf r} \left|{\bf r} \right>\left<{\bf r} \right| \int d{\bf \Omega}\left|{\bf \Omega}\right>\left<{\bf \Omega}\right|,
\end{eqnarray}
where ${\bf r}$ labels the position of the centre of mass of the rotor and ${\bf \Omega}$ the angles that specify its orientation. For a linear rotor, ${\bf \Omega}$ is usually chosen to be its polar and azimuthal angles in the SFF. For a non-linear rotor, the three Euler angles~\cite{zare_1988} are the most natural candidates. Quaternions can also be used and for spherical top molecules their usage increases simulation efficiency~\cite{mak_ch4_he}. The integration range of $\int d{\bf \Omega}$ should include the whole volume of all the angles under consideration. With this identity operator, the partition function in Eq.~\ref{eqn:z1} becomes
\begin{eqnarray}
Z=\prod_{i=1}^{M}\int d{\bf r}_i \int d{\bf \Omega}_i \left< {\bf r}_i \right|\left<{\bf \Omega}_i \right|\hat{\rho}\left(\tau\right)\left| {\bf \Omega}_{i+1} \right>\left| {\bf r}_{i+1} \right>,
\end{eqnarray}
with
\begin{eqnarray}
\hat{\rho}\left(\tau \right)\approx e^{ -\tau\hat{T}_t } e^{ -\tau\hat{T}_r } e^{ -\tau\hat{V} }. \label{eqn:rhotr}
\end{eqnarray}
Here the restriction of $M+1=1$ applies again for the slice index $i$ and the subscripts ``t'' and ``r'' are used to denote the translational and rotational kinetic operators. Note that for a rigid-body
\begin{eqnarray}
\left[\hat{T}_r , \hat{T}_t \right]=0
\end{eqnarray}
and the approximation in Eq.~\ref{eqn:rhotr} comes from
\begin{eqnarray}
\left[\hat{T}_r , \hat{V} \right]\ne0;\left[\hat{T}_t , \hat{V} \right]\ne0.
\end{eqnarray}
Sandwiching the propagator with the position and orientation bases results in
\begin{eqnarray}
\left< {\bf r}_i \right|\left<{\bf \Omega}_i \right|\hat{\rho}\left(\tau\right)\left| {\bf \Omega}_{i+1} \right>\left| {\bf r}_{i+1} \right>&\approx&\left<{\bf r}_i\right| e^{ -\tau\hat{T}_t }\left| {\bf r}_{i+1} \right>\nonumber\\
&&\times\left<{\bf \Omega}_i\right| e^{ -\tau\hat{T}_r }\left| {\bf \Omega}_{i+1} \right> e^{ -\tau V\left({\bf r}_{i+1}, {\bf \Omega}_{i+1} \right) } \nonumber\\
&=&\frac{1}{\lambda^3} e^{ -\frac{\pi}{\lambda^2}\left({\bf r}_i-{\bf r}_{i+1} \right)^2 } e^{-\tau V\left({\bf r}_{i+1}, {\bf \Omega}_{i+1} \right)}\nonumber\\
&&\times\left<{\bf \Omega}_i\right| e^{ -\tau\hat{T}_r }\left| {\bf \Omega}_{i+1} \right>. \label{eqn:link2}
\end{eqnarray}
The comparison of Eqs.~\ref{eqn:link2} and~\ref{eqn:linkb} reveals that adding the angular degrees of freedom of the rigid rotor leads to an angular dependence of the potential and a rotational kinetic propagator connecting orientations in adjacent slices. Therefore, besides positions, one also needs to sample orientations in a way similar to that introduced in Sec.~\ref{sec:formu}\ref{subsec:pi}.

While handling the angular dependence of the potential propagator is easy, the rotational kinetic propagator requires more discussion. For a linear rotor
\begin{eqnarray}
\hat{T}_r=B\hat{{\bf j}}^2,
\end{eqnarray}
where $B$ is the rotational constant and ${\hat{\bf j}}$ the angular momentum operator of the rotor. 
An approach to handle the rotational kinetic propagator is to insert the resolution of the identity based on the complete eigenstates of ${\hat{\bf j}}^2$, $\left\{ \left|jm\right> \right\}$, into the propagator:
\begin{eqnarray}
\left<{\bf \Omega}_i\right| e^{ -\tau\hat{T}_r }\left| {\bf \Omega}_{i+1} \right>&=&\left<{\bf \Omega}_i\right| e^{ -\tau B \hat{{\bf j}}^2 }\left| {\bf \Omega}_{i+1} \right>\nonumber\\
&=&\sum_{jm}\left<{\bf \Omega}_i\right| e^{ -\tau B \hat{{\bf j}}^2 }\left|jm \right>\left< jm \right|\left| {\bf \Omega}_{i+1} \right>\nonumber\\
&=&\sum_{jm}\left<{\bf \Omega}_i\right|\left.jm\right> e^{ -\tau B j\left( j+1\right) }\left< jm\right.\left| {\bf \Omega}_{i+1} \right>\nonumber\\
&=&\sum_{jm}Y_{jm}\left({\bf \Omega}_i\right)e^{-\tau B j\left( j+1\right)}Y_{jm}^*\left({\bf \Omega}_i\right)\nonumber\\
&=&\sum_{j}\frac{2j+1}{4\pi}P_j\left(x_{i,i+1}\right) e^{ -\tau B j\left( j+1\right) }, \label{eqn:rotpro_linear}
\end{eqnarray}
where $j$ and $m$ are the famous orbital and magnetic quantum numbers, $Y_{jm}\left({\bf \Omega} \right)$ the spherical harmonic function, $P_l\left( x\right)$ the Legendre polynomial, and $x_{i,i+1}$ represents the dot product (overlap) of the two unit vectors specifying the orientations of the rotor in adjacent slices. 
As in the relative displacement $\left({\bf r}_i - {\bf r}_{i+1}\right)$ dependence of the translational propagator in Eq.~\ref{eqn:link2}, the rotational propagator only depends on the relative orientation. To get to the last equality of Eq.~\ref{eqn:rotpro_linear}, the following well-known equalities have been used:
\begin{eqnarray}
\hat{{\bf j}}^2\left|jm \right>&=&j\left(j+1\right)\left|jm\right>; \\
\left<{\bf \Omega}\right|\left.jm\right>&=&Y_{jm}\left({\bf \Omega}\right);\\
\sum_m Y_{jm}\left({\bf \Omega}\right)Y_{jm}^*\left({\bf \Omega}'\right)&=&\frac{2j+1}{4\pi}P_j\left(x \right).
\end{eqnarray}
The last equality is the famous addition theorem of the spherical harmonics and $x$ is again the aforementioned dot product. The formula in the last line of Eq.~\ref{eqn:rotpro_linear} is used in actual PIMC simulations involving linear rigid rotor. The infinite summation over $j$ has to be truncated to a large value $j_{max}$, which is determined by a convergence study. It is highly time-consuming to let the computer carry out the large summation over $j$ during a simulation. In reality, one can calculate the rotational propagator for a dense grid of $x\in \left[ -1,1 \right]$ in advance and use an interpolation method to obtain  the propagator value for the $x_{i,i+1}$ calculated from the ${\bf \Omega}_i$ and ${\bf \Omega}_{i+1}$, which are randomly generated on the fly.

The nuclear spin statistics of a rotor can be incorporated into the summation over $j$. 
For example, rotating a \hydrogen~molecule by $\pi$ about any axis passing through its centre of mass, and perpendicular to the molecule's axis, is equivalent to exchanging two identical protons (Fermions) and the \hydrogen~total wave function will need to be anti-symmetric with respect to this rotation. The singlet (triplet) nuclear spin state of \phtwo~(\ohtwo) is symmetric (anti-symmetric) with respect to this rotation and the corresponding rotational state would have to be symmetric (anti-symmetric) to satisfy the anti-symmetry requirement of the total wave function. Therefore, \phtwo~(\ohtwo) can only have rotational states with even (odd) $j$, and this restriction is imposed in the summation over $j$ in Eq.~\ref{eqn:rotpro_linear}. For molecules whose nuclear spin state is not coupled to the rotational state, there is no restriction on $j$.

Fig.~\ref{fig:h2_rho} shows the rotational propagators of \phtwo, \ohtwo, and the fictitious spinless \hydrogen~with distinguishable protons. The fictitious \hydrogen~corresponds to a rotor without coupling between nuclear spin and rotational states, e.g., HD, and as expected, its propagator decays as the relative orientation increases (the overlap decreases). 
On the other hand, the symmetry and anti-symmetry of the \phtwo~and \ohtwo~propagators with respect to the relative orientation is evident. 
In an actual PIMC simulation, there is no technical difficulty in sampling orientation for the fictitious \hydrogen-type and \phtwo-type rotors, because their rotational propagators are positive definite. However, the negative propagator of the \ohtwo-type rotor leads to the notorious sign problem in PIMC: the contributions from the paths with positive and negative weights largely cancel each other, resulting in very low sampling efficiency~\cite{ceperley_pimc_fermions,Mak_pimc_sign}.

To solve this sign problem for rotational PIMC, Ceperley proposed a method called restricted PIMC~\cite{ceperley_pimc_fermions}. This method is similar to the fixed-node approximation in DMC excited state simulations. It takes an arbitrary imaginary time slice as the reference point and samples orientations in all the other slices, with the restriction that the relative orientations between the sampled and the fixed reference orientations are in the positive region of the propagator, i.e., the $x>0$ region in Fig.~\ref{fig:h2_rho}. 
Any moves that lead to
\begin{eqnarray}
\left<{\bf \Omega}_{i_{\rm ref}}\right| e^{ -k\tau B \hat{{\bf j}}^2 }\left| {\bf \Omega}_{i_{\rm ref}+k} \right><0
\end{eqnarray}
are rejected. 
One needs to perform such restricted simulations for arbitrarily chosen reference beads and average to obtain the final result. 
Apparently, much more computational steps are needed in this scheme. 
Maybe due to this reason, to the best of our knowledge, there has been no theoretical study of  microscopic superfluids employing a rotor that needs this treatment. 
For an additional discussion of the linear rotor PIMC propagator, the reader is referred to publications of Marx and M\"user~\cite{marx_line_rot_pimc,muser_line_rot_pimc,marx_muser_rot_pimc}. 
It is noteworthy that in practice, \phtwo~can be treated as a structureless point-like particle, i.e., there is no need to sample its orientation. 
This is because of its spherically symmetric ground rotational state ($j=0$) and the fairly large gap (about 350 \wno) between the ground and the first rotational excited state ($j=2$). 
These states can hardly be coupled by the weak van der Waals interaction in \phtwo~clusters. For the theoretical studies discussed in Sec.~\ref{sec:results}, this treatment is implied unless further specified.

The above approach  for linear rotors can be readily extended to treat non-linear rotors. 
The orientation of a non-linear rotor is parametrized by three Euler angles and in this report, 
we use the angles $\left(\phi,\theta,\chi \right)$ defined in Fig. 3.2 of Ref.~\cite{zare_1988}. 
The rotational kinetic operator in the MFF (in this report, MFF is always chosen to be the principal axes frame of the rotor) is
\begin{eqnarray}
\hat{T}_r&=&A\hat{j}_a^2+B\hat{j}_b^2+C\hat{j}_c^2, \label{eqn:asym_tr}
\end{eqnarray}
where $A$, $B$, and $C$ are rotational constants along the principal-axes $a$, $b$, and $c$ and $\hat{j}_a$ etc. are angular momentum operators of the rotor projected along those axes. 
The most natural basis set to describe a non-linear rotor is the Wigner basis $\left\{\left|jkm \right>\right\}$~\cite{bunker_jensen_book}. $j$ is the total angular momentum quantum number, $m$ the projection of angular momentum along a SFF axis, and $k$ the projection along a principal axis in MFF. 
For each $j$, the ranges of $m$ and $k$ are from $-j$ to $j$, with an increment of $1$. 
The Wigner bases are eigenstates of spherical top ($A=B=C)$ and symmetric top (two of the rotational constants equal) rotors, and the eigenstates for the most general asymmetric tops $(A\ne B\ne C)$ can be expanded in the Wigner basis:
\begin{eqnarray}
\left|j\hat{k}m \right>=\sum_k C_{\hat{k},k}^{j}\left|jkm \right>.
\end{eqnarray}
$\hat{k}$ is used to label different asymmetric top states with the same $j$ and $m$. Obviously, $j$ and $m$ are conserved by $\hat{T}_{r}$ due to its isotropy in the SFF.  
Below, we limit our discussion to the case of asymmetric top rotors, of which the spherical and symmetric tops are just special cases.

Because of the isotropy of  space, the rotational propagator are invariant to any rotational operation on the rigid rotor, i.e.,
\begin{eqnarray}
\hat{R}\left( {\bf \Omega} \right)\left<{\bf \Omega}_i\right| e^{ -\tau\hat{T}_r }\left| {\bf \Omega}_{i+1} \right>&=&\left<{\bf \Omega}_i+{\bf \Omega}\right| e^{ -\tau\hat{T}_r }\left| {\bf \Omega}_{i+1} +{\bf \Omega}\right>\nonumber\\
&=&\left<{\bf \Omega}_i\right| e^{ -\tau\hat{T}_r }\left| {\bf \Omega}_{i+1} \right>. \label{eqn:rot_act_rho}
\end{eqnarray}
$\hat{R}\left({\bf  \Omega} \right)$ is the rotational operator that acts on the rotor and is parametrized by the Euler angles ${\bf \Omega}$. Note that the notation 
${\bf \Omega} +{\bf \Omega}' \label{eqn:ang_sum}$ does not represent numerical vector addition of two sets of Euler angles, but rather represents the resultant Euler angles between the MFF and SFF after two consecutive rotational operations, $\hat{R}\left({\bf \Omega}' \right)\hat{R}\left({\bf \Omega} \right)$, have been applied to the rotor. 
Because of the non-commutation between rotational operators, the addition of $\Omega+\Omega'$ in Eq.~\ref{eqn:rot_act_rho} is not commutative in contrast to  the usual numerical addition.

Now if we replace the rotational operator in Eq.~\ref{eqn:rot_act_rho} by $\hat{R}\left({{\bf \Omega}_i}\right)^{-1}$, we then  have
\begin{eqnarray}
\hat{R}\left({{\bf \Omega}_i}\right)^{-1}\left<{\bf \Omega}_i\right| e^{ -\tau\hat{T}_r }\left| {\bf \Omega}_{i+1} \right>&=&\left<{\bf 0}\right| e^{ -\tau\hat{T}_r }\left| {\bf \tilde{\Omega}}_{i,i+1} \right>\nonumber\\
&=&\left<{\bf \Omega}_i\right| e^{ -\tau\hat{T}_r }\left| {\bf \Omega}_{i+1} \right>,
\end{eqnarray}
where ${\bf \tilde{\Omega}}_{i,i+1}$ are the Euler angles that specify the $\left(i+1\right)$th orientation of the rotor in the $i$th MFF. 
As in the case of linear rotors, the non-linear rotor PIMC rotational propagator also only depends on the relative orientation between two slices, and ${\bf \tilde{\Omega}}_{i,i+1}$ represent the relative Euler angles between slices $i$ and $i+1$. One can similarly calculate
$\left<{\bf 0}\right| e^{ -\tau\hat{T}_r }\left| {\bf \tilde{\Omega}} \right>$
on a grid of ${\bf \tilde{\Omega}}$ and use interpolation to extract the propagator value for any sampled ${\bf \Omega}_i$ and ${\bf \Omega}_{i+1}$.

The actual value of
$\left<{\bf 0}\right| e^{ -\tau\hat{T}_r }\left| {\bf \tilde{\Omega}} \right>$
is calculated as follows:
\begin{eqnarray}
\left<{\bf 0}\right| e^{ -\tau\hat{T}_r }\left| {\bf \tilde{\Omega}} \right>&=&\sum_{j\hat{k}m}\left<{\bf 0}\right| e^{ -\tau\hat{T}_r }\left|j\hat{k}m \right>\left<j\hat{k}m \right.\left| {\bf \tilde{\Omega}} \right>\nonumber\\
&=&\sum_{j\hat{k}m}\left<{\bf 0} \left| j\hat{k}m \right.\right> e^{ -\tau E_{j\hat{k}} }\left<j\hat{k}m \right.\left| {\bf \tilde{\Omega}} \right>\nonumber\\
&=&\sum_{j\hat{k}m}\left( \frac{2j+1}{8\pi^2} \right)C_{\hat{k'},m}^j e^{-\tau E_{j\hat{k}} }\nonumber\\
&&\times\sum_{k=-j}^j C_{\hat{k},k}^j d^{j}_{mk}\left( \tilde{\theta} \right)\cos\left( m\tilde{\phi}+k\tilde{\chi} \right), \label{eqn:noya_rho}
\end{eqnarray}
where $d_{mk}^j\left( {\theta} \right)$ is the famous Wigner small-$d$ function~\cite{zare_1988}. 
This is essentially Eq.~15 of Ref.~\cite{noya_rotden} except that different labels are used. The long derivation to obtain the last line in Eq.~\ref{eqn:noya_rho} is skipped and the interested readers should consult Ref.~\cite{noya_rotden} for details. 
In Ref.~\cite{noya_rotden}, Noya \etal~used the sentence 
``It can be seen that the propagator is also real (and positive)'' to justify their replacement of the complex phase factor 
\begin{eqnarray}
e^{-i\left(m\tilde{\phi}+k\tilde{\chi}  \right)} \label{eqn:rotden_phase}
\end{eqnarray}
by
\begin{eqnarray}
\cos\left( m\tilde{\phi}+k\tilde{\chi} \right)
\end{eqnarray}
to obtain the final real formula of the propagator. However, this may not be easily seen by other readers and we hereby supplement the replacement with a clearer justification. In the Appendix,
we show that the rotation of a non-linear rotor can be represented by a real basis. Replacing the $\sum_{j\hat{k}m}\left| j\hat{k}m \right>\left<j\hat{k}m \right|$ by the resolution of the identity in this real basis in Eq.~\ref{eqn:noya_rho} would not change the value of the propagator, but there is no source of imaginary unit any more. Therefore, the propagator must be real and the imaginary part of the phase factor Eq.~\ref{eqn:rotden_phase} can be dropped.

The nuclear spin states can also be coupled to the rotational states of the non-linear molecule. For example, the two protons of \water~can be coupled to have singlet or triplet nuclear spin states, forming the {\em para}-\water~(\pwater) and {\em ortho}-\water~(\owater) water molecules. 
Following the same symmetry argument as for \phtwo~and \ohtwo, we note that any rotation of \pwater (\owater) that leads to an exchange of the two protons will have to be symmetric (anti-symmetric). 
Hence, \pwater~(\owater) only takes the $\left\{\left|jkm \right> \right\}$ basis with even (odd) $k$ if $k$ denotes the angular momentum along the $C_2$ axis of the molecule. 
Consequently, only certain $\hat{k}$ values should be included in the $\sum_{j\hat{k}m}$ in the first line of Eq.~\ref{eqn:noya_rho}. 
As a result, the summation over $m$ and $k$ in the last line of Eq.~\ref{eqn:noya_rho} are also restricted to be even (odd) for \pwater~(\owater). 
For rotors whose rotational states are decoupled from their nuclear spin states, e.g., a fictitious spinless  \water~with two distinguishable protons, there is no restriction on the summation.

For simplicity, we drop the tilde on the relative Euler angles for now on. 
We compare the rotational propagators of \pwater, \owater, and the fictitious spineless \water~in Fig.~\ref{fig:h2o_rotden}. 
We chose the MFF $z$-axis to be the $C_2$ axis of \water. 
With this choice of axis, the exchange of the two protons is only related to the first and the third relative Euler angles $\phi$ and $\chi$, i.e., either $\phi$ or $\chi$ is increased by an odd multiple of $\pi$ but not both. 
The symmetry (anti-symmetry) of proton exchange in \pwater~(\owater) is illustrated as the invariance (multiplication of $-1$) of the propagator if $\phi$ or $\chi$ is increased by odd multiples of $\pi$ in Fig.~\ref{fig:h2o_rotden}(a) ((b)). Increasing $\phi$ or $\chi$ by an even multiple of $\pi$ brings the \water~to an indistinguishable orientation and consequently, the propagator is invariant. 
This invariance is evident in all panels of Fig.~\ref{fig:h2o_rotden}. 
Noya \etal~did not consider the coupling between nuclear spin and rotational state and they obtained propagators as the one in Fig.~\ref{fig:h2o_rotden}(c)~\cite{noya_rotden}. 
Therefore, they concluded that the propagator is always positive. Here we supplement their statement by showing an example, \owater, whose propagator can be negative.

Readers who are interested in the development of the PIMC propagator of non-linear rotors should refer to Refs.~\cite{rossky_semi_class1,rossky_semi_class2,marx_muser_rot_pimc,muser_berne_rotden,blinov_RCF,noya_rotden,noya_molphys}. 
It is noteworthy that  M\"user and Berne were the first to develop the propagator using the scheme introduced here in 1996~\cite{muser_berne_rotden}. 
However, they made a mistake in their derivation and their propagator (Eq. 5 of Ref.~\cite{muser_berne_rotden}) is only correct for symmetric and spherical tops, 
not for general asymmetric tops as they claimed. 
This mistake was finally rectified by Noya \etal~in 2011~\cite{noya_rotden}, and the correct propagator for general non-linear rotors (Eq. 15 of Ref.~\cite{noya_rotden}) was reported for the first time. 
Because of this late advance, there had been no report of theoretical studies on microscopic superfluids using asymmetric top dopants until our research group published a study of \pwater(\phtwo)$_N$ clusters in 2013~\cite{tobywater}.

\subsection{Sampling bosonic exchanges: the worm algorithm} \label{subsec:worm}

\begin{figure}[h]
\centerline {\includegraphics[width=0.8\columnwidth]{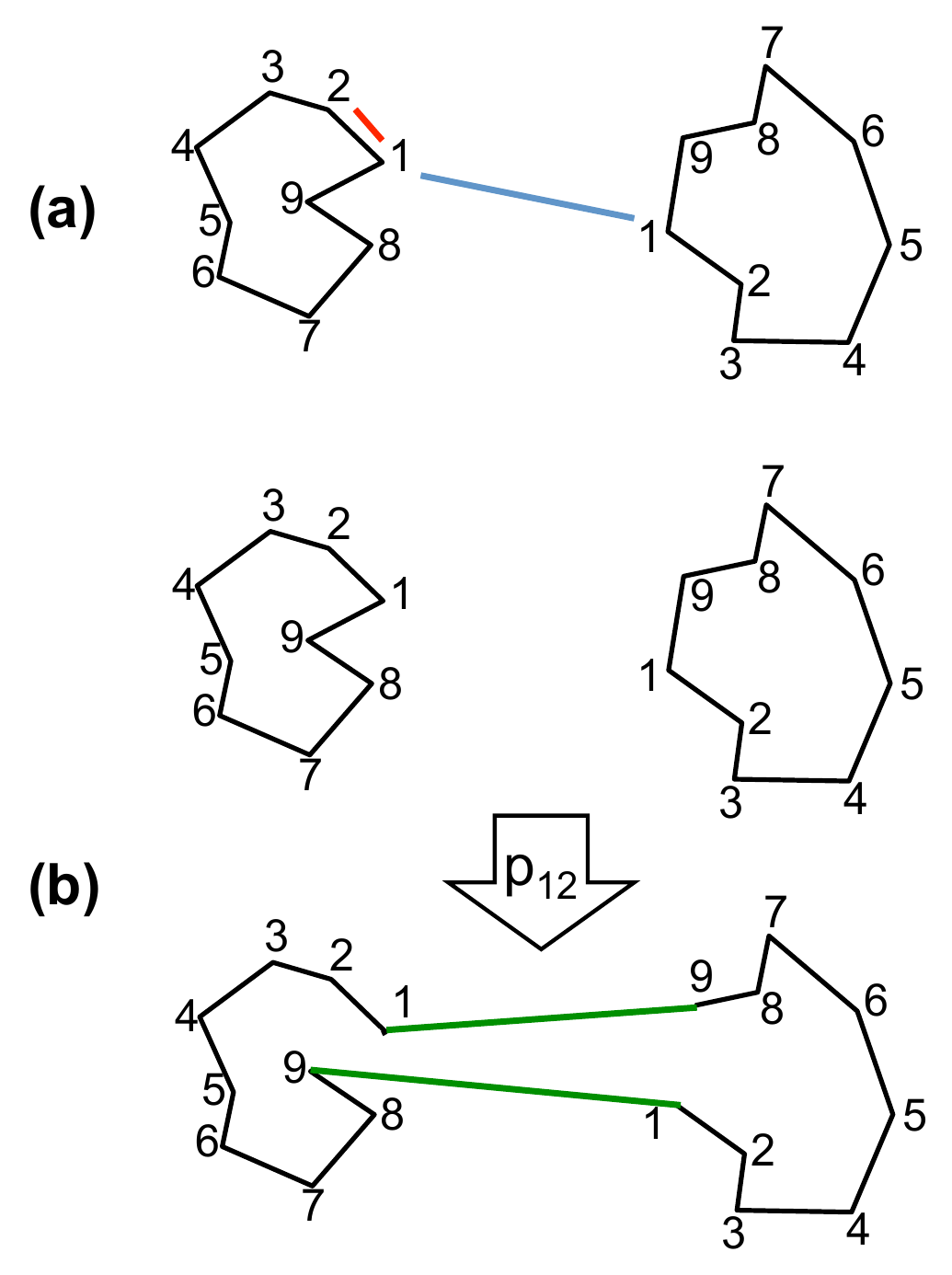}}
\caption{(a) Path representation of two boltzmannon particles with nine slices. The red line represents the spring-like interaction between beads at two adjacent slices and the blue line the potential between beads at the same slice. Note that although only one red and one blue lines are shown, every pair of adjacent beads interact with the spring-like interaction and at every slice beads interact with potential based on their positions. The numbers are slice indices. (b) Path representation of the permutation operation $\hat{p}_{12}$ on two bosons. The green lines represent the new connections and spring-like interactions after the permutation introduces new path connectivity}
\label{fig:path}
\end{figure}

\begin{figure}[h]
\centerline {\includegraphics[width=0.8\columnwidth]{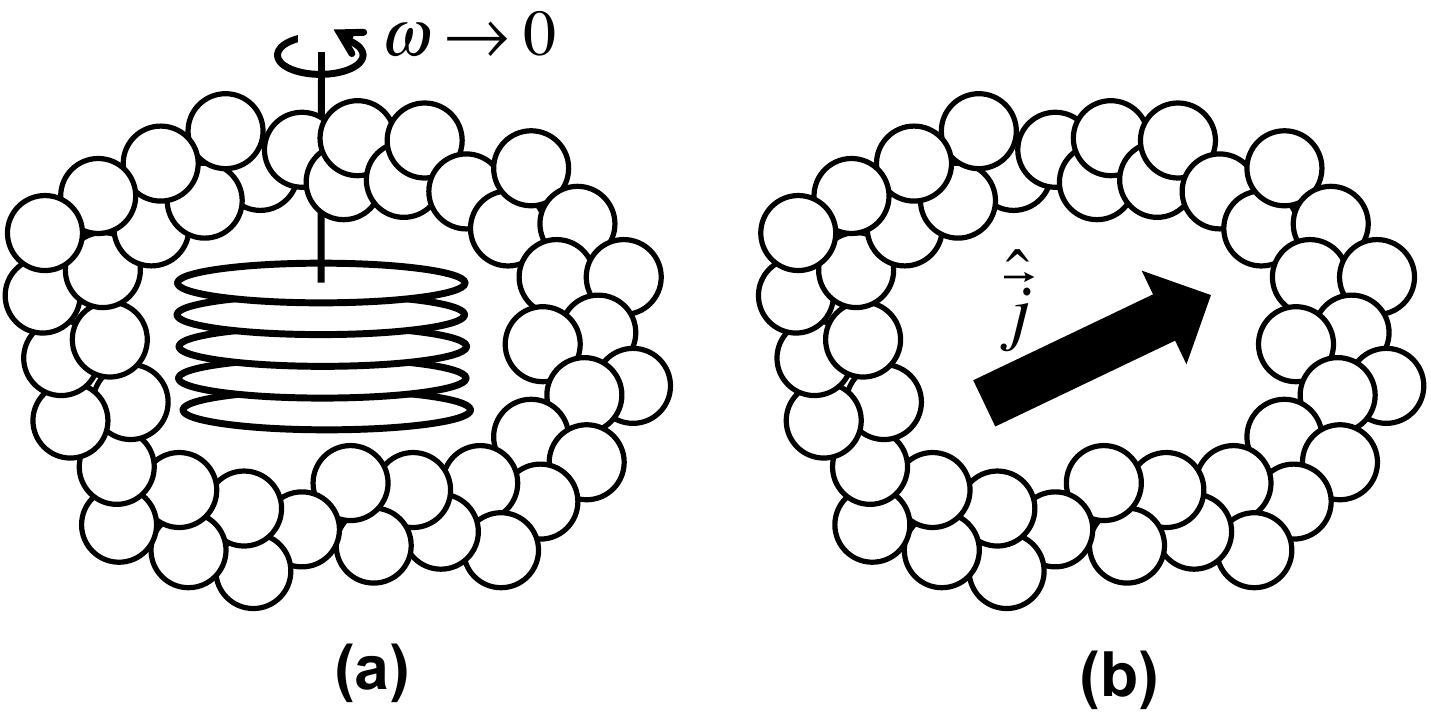}}
\caption{(a) Conducting an Andronikashvili experiment for a microscopic quantum cluster with a fictitious rotor that follows classical mechanics. The cluster is represented by a group of spheres. The rotor is a representation of a rotating external field to the bosons in the cluster and we image it as a stack of disks as in the macroscopic Andronikashvili experiment. (b) Conducting a spectroscopic Andronikashvili experiment with a molecular rotor represented by the arrow. This is a quantum rotor, as indicated by the angular momentum operator $\hat{\vec{j}}$}
\label{fig:sff_mff}
\end{figure}

\begin{figure}[h]
\centerline {\includegraphics[width=0.8\columnwidth]{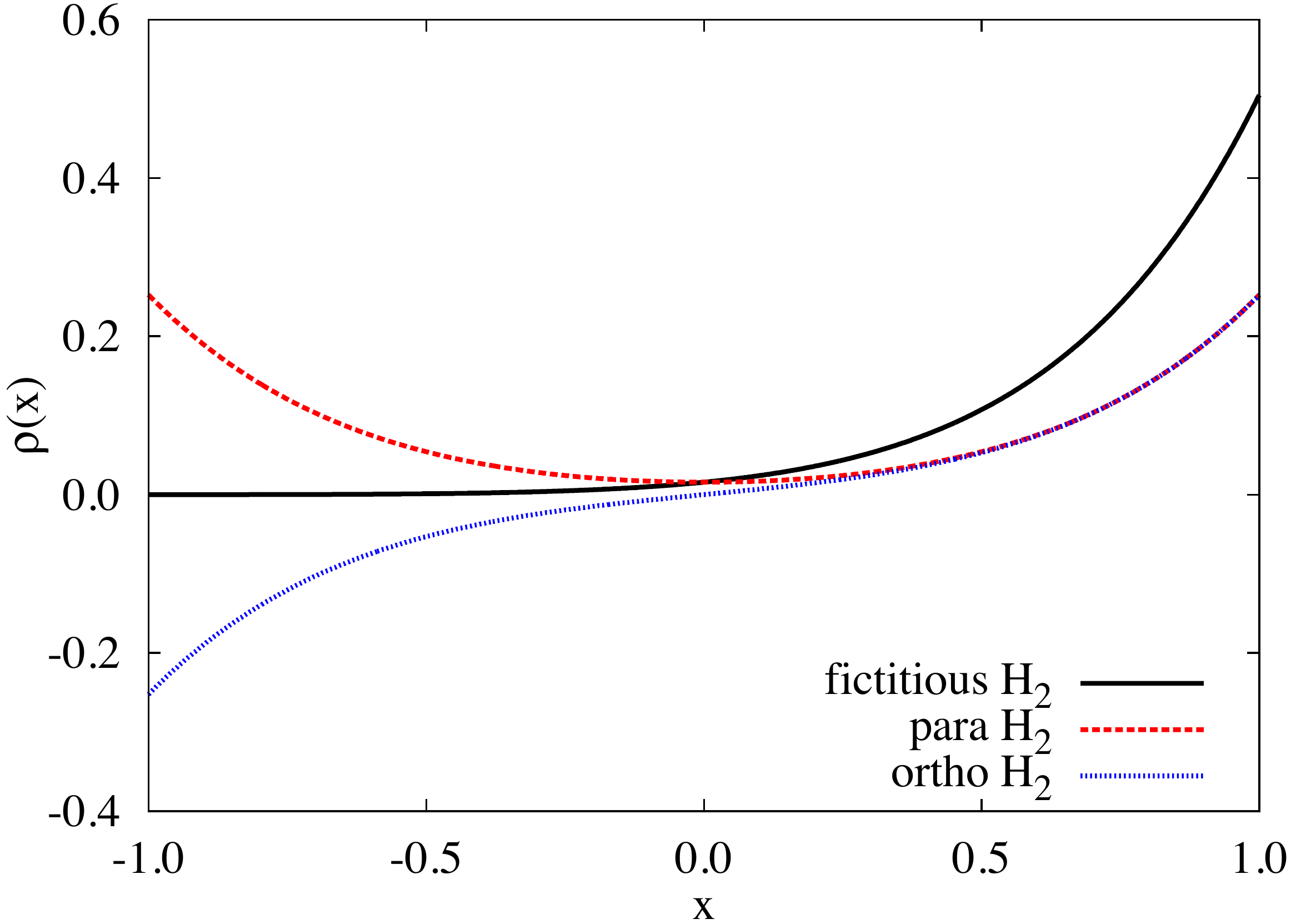}}
\caption{PIMC rotational propagator as a function of relative orientation between two imaginary time slices (the dot product $x$) for \phtwo, \ohtwo, and the fictitious \hydrogen~with two distinguishable protons. The propagators have the same $\tau$ corresponding to temperature of 512~K and the rotational constant of \hydrogen~is chosen to be 59.322~\wno~\cite{h2_b_59.3}.}
\label{fig:h2_rho}
\end{figure}

\begin{figure}[h]
\centerline {\includegraphics[width=0.7\columnwidth]{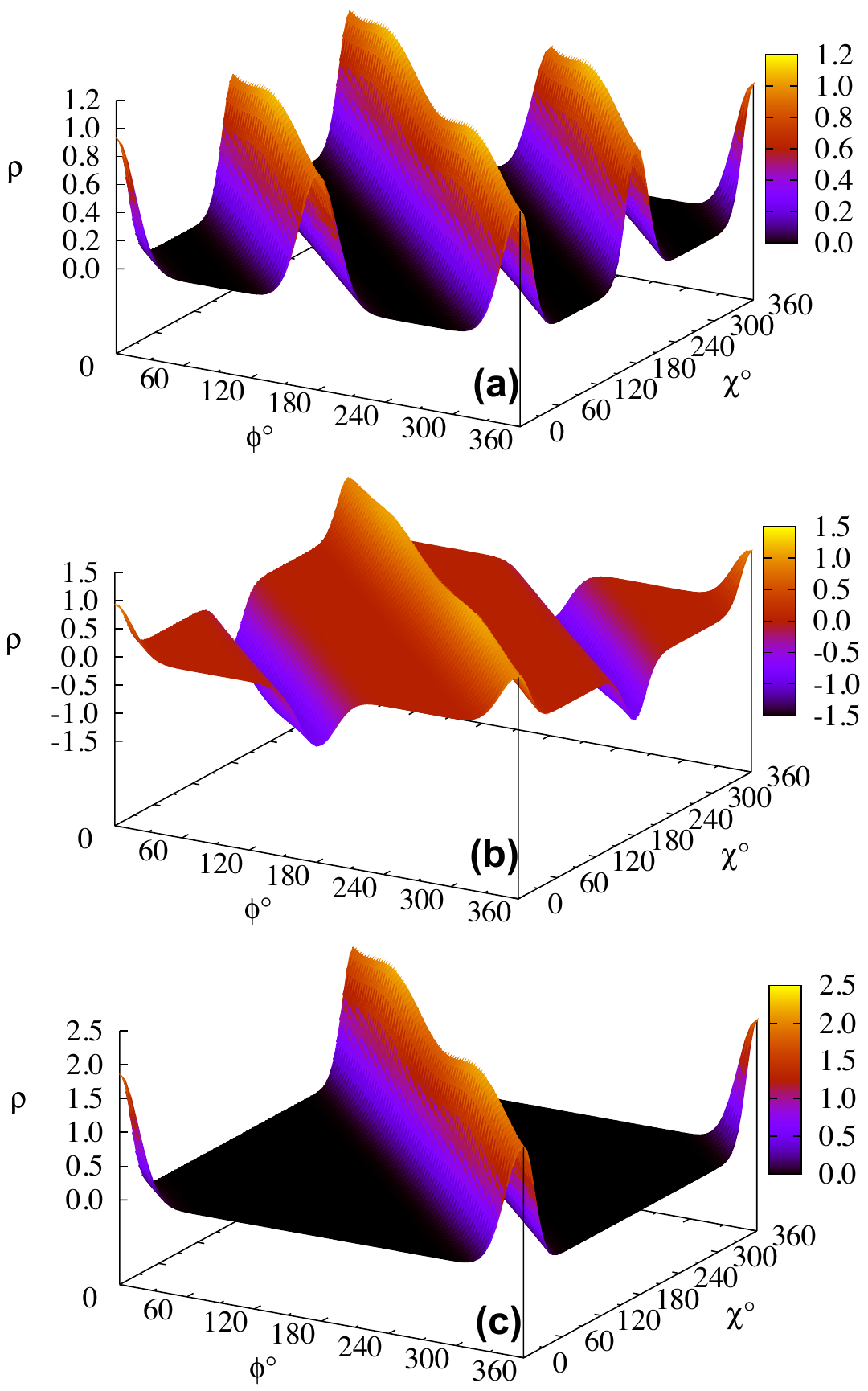}}
\caption{PIMC rotational propagator as a function of the relative Euler angles $\phi$ and $\chi$ for (a) \pwater, (b) \owater, and (c) the fictitious \water~with two distinguishable protons. The second relative Euler angle is fixed at $10^\circ$. The propagators have the same $\tau$ corresponding to temperature of 512~K.The rotaitonal constants of \water~is chosen to be $A=27.8806$~\wno, $B=14.5216$~\wno, and $C=9.2778$~\wno~\cite{vanderAvoird:2011gn}. The MFF $z$-axis is chosen to be the $C_2$ axis of \hydrogen. To reproduce the graphs, one will need to follow the same choice of parameters and axis.} 
\label{fig:h2o_rotden}
\end{figure}

As mentioned in Sec.~\ref{sec:formu}\ref{subsec:pi}, permutation sampling is a key step in the PIMC simulation of bosons. 
Although it is formally correct to separate the sampling of path configurations and permutations as implied in Eq.~\ref{eqn:zbe} and illustrated in Fig.~\ref{fig:path}(b), such an algorithm would be extremely inefficient. 
This is because the end beads to be reconnected by the new spring-like kinetic term generated by permutation (the long green lines in Fig.~\ref{fig:path}(b)) are usually far away, making the permutation unfavourable. A joint sampling of configurations and permutations is highly desirable.

\begin{figure}[h] 
\centerline {\includegraphics[width=0.7\columnwidth]{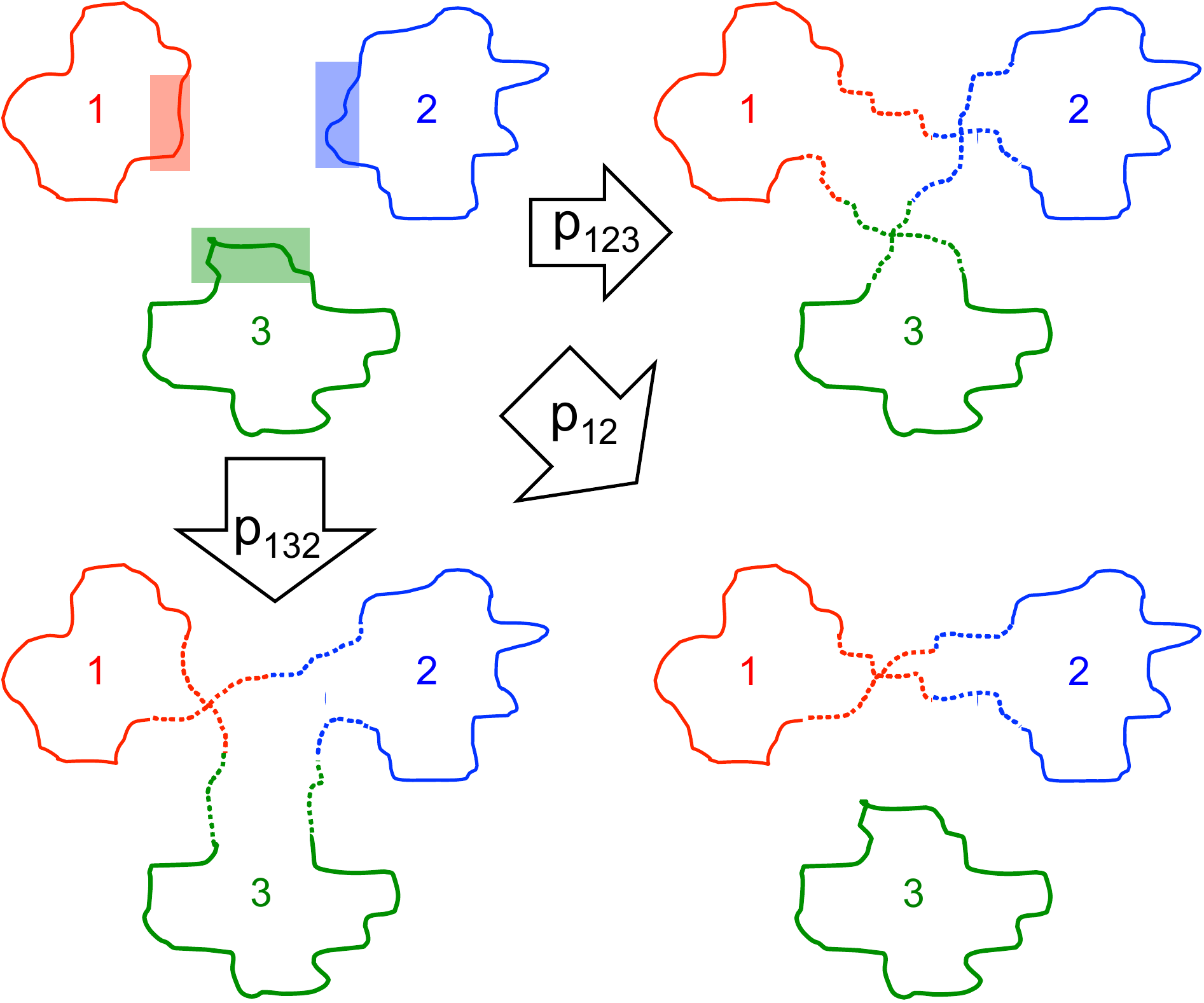}}
\caption{Sketch of the traditional permutation sampling for three bosons. See the main text for discussion. } 
\label{fig:sampath1}
\end{figure}

Traditional methods for this joint sampling scheme are introduced in Sec. 5I and 5J of Ceperley's seminal review (Ref.~\cite{ceperley_rmp_1995}) and Ref.~\cite{boninsegni_pimc_perm}. 
Bosons that are close enough to each other, judged by their distances compared to their thermal wavelength, are considered to be in  exchange range.
A permutation table of the number of the bosons is generated based on this criterion. 
The three particles on the top left of Fig.~\ref{fig:sampath1} are examples of bosons in exchange range as their distances are comparable to the size of their rings, an estimate of their thermal wavelength. 
In this figure, we use different colours to denote paths belonging to different particles. 
A few beads from each path are selected (shaded in Fig.~\ref{fig:sampath1}) and they correspond to the same imaginary time slices in each path. 
These segments are replaced by  new ones (the dotted curves in Fig.~\ref{fig:sampath1}) 
that have the path connectivity corresponding to a sampled permutation from the permutation table. 
The new configuration and connectivity are accepted or rejected based on a similar Metropolis scheme 
as for the path sampling without permutation. 
Three such examples of joint sampling are shown in Fig.~\ref{fig:sampath1}. 
Comparing  Fig.~\ref{fig:path}(b) and Fig.~\ref{fig:sampath1} indicates that the long green lines in the former is replaced by the dotted curves in the latter, i.e., the sharp increase in kinetic energy between the two end beads in the separated sampling scheme is attenuated by a segment of connected beads in the joint sampling scheme. 
That sampling scheme benefits from a higher acceptance ratio and a more efficient exploration of the joint phase space of path configuration and permutation. 
One disadvantage of this method is that the construction of the explicit permutation table for the nearby particles can be very time-consuming, as the number of particles increases. 

Another sampling scheme that does not involve the construction of the permutation table is the worm algorithm. 
The worm algorithm (WA) was originally proposed to solve the problem of unfavourable size scaling of computer time in quantum Monte Carlo (QMC) simulations of lattice models~\cite{worm_lattice1,pimc_lattice2,worm_lattice3}. 
Its application to PIMC is an extension from discrete to continuous space. 
Refs.~\cite{boninsegni_worm_1,boninsegni_worm_2} provide comprehensive accounts of its implementation in PIMC and application to the study of microscopic superfluids. 
Here we only present a brief overview on the features and algorithms of this approach, without giving any mathematical details.

The traditional sampling scheme depicted in Fig.~\ref{fig:sampath1} only accounts for the $Z$-sector of the phase space, the sector of the partition function. It only samples configurations with closed paths. Here the meaning of configuration includes both shape and connectivity of paths. The special feature of the WA is that it also explores the $G$-sector, the sector of the one-particle Matsubara Green function. This sector is represented by configurations with one open path, called the worm. There are seven types of configuration sampling moves in WA, six of which are in pairs: Open/Close; Insert/Remove; Advance/Recede; Swap. The Open move is to remove a segment of beads in a closed configuration.
It is one way to generate a worm. 
The Close move is to generate a segment of beads to connect the two dangling ends of an open configuration. 
It is one way to remove a worm. 
The Insert move is to insert a segment of beads with two dangling ends, i.e., a worm, given a closed configuration as the background. 
The Remove move is to erase an existent worm by removing all the beads it contains. 
The Advance move is to add several beads to one end of the worm if there is one. 
The Recede move is to remove several beads from one end of the worm if there is one. 
The Swap move is to remove a segment of beads from a closed path in a worm configuration, 
and generate a segment of beads between one of the original dangling ends to one of the newly generated dangling ends. 
This move keeps the system in the $G$-sector and generates permutations. 
Diagonal properties are evaluated when the configuration is in the $Z$-sector, while off-diagonal ones in the $G$-sector.

\begin{figure}[h]
\centerline {\includegraphics[width=1.0\columnwidth]{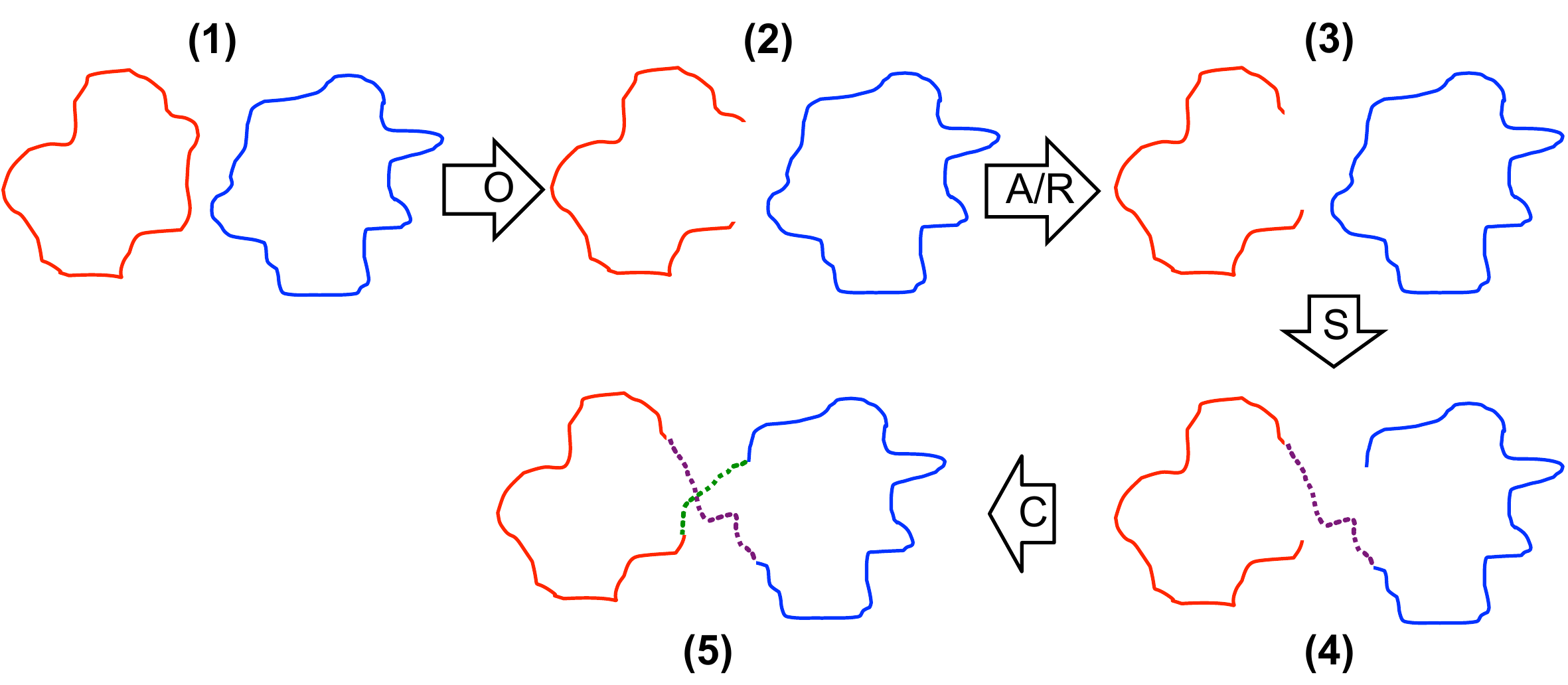}}
\caption{Sketch of the a sweep of WA sampling that starts from and returns to a closed configuration and realizes permutation. See the main text for discussion.} 
\label{fig:sampath2}
\end{figure}

A sketch of a typical sweep of the WA sampling steps is shown in Fig.~\ref{fig:sampath2}, 
where the moves are labeled by their first letters. 
This sequence of moves brings the configuration from the $Z$-sector to the $G$-sector and then back to the $Z$-sector, with a new permutation. 
The red and blue paths do not necessarily represent two particles. They can be paths that contain permutation cycles of several particles. 
The Insert/Erase pair are not included in the figure. 
The Insert move corresponds to the generation of configuration of the type shown in Fig.~\ref{fig:sampath2}(2) with only the blue path in configuration Fig.~\ref{fig:sampath2}(1), i.e., the red worm is inserted. 
The Erase move corresponds to removing the open paths in configuration Fig.~\ref{fig:sampath2}(2), (3), or (4). 
The Metropolis acceptance ratios of the Advance/Recede and Swap moves are strictly determined by the detailed balance of the Matsubara Green function and those of the Open/Close and Insert/Remove moves are also determined by the probabilities of being in the $Z$- and $G$-sectors. 
Obviously, the Swap (permutation) is just an intermediate step in exploring the $G$-sector and therefore, no permutation table is needed. 
Also, the sampling guided by the Matsubara Green function automatically creates Swap-favourable worm configurations and leads to a high acceptance rate of this move. 
Long permutation cycles can thus be efficiently sampled. 
Due to its high efficiency, the WA can be employed to simulate systems whose sizes are much larger than those that can be handled by the traditional PIMC approach.

It is finally worth noting that missing (gaining) some beads in a worm generated by an Open (Insert) move indicates that the number of particles is not conserved at the imaginary time slices corresponding to the worm section. Therefore, the WA is applicable for grand canonical ensemble simulations. 
As a matter of fact, it is the {\em first grand canonical QMC method with local updates to incorporate  full quantum statistics}~\cite{boninsegni_worm_2}. 
By choosing appropriate simulation parameters, especially the chemical potential, one can surely use this algorithm to perform a canonical ensemble simulation.

\subsection{Estimation of the effective moment of inertia} \label{subsec:Ieff_estim}

Since so far, the superfluidity of \phtwo~has only been predicted and observed to occur in clusters, 
we focus on the estimation of the effective moment of inertia and superfluid fraction for these finite size systems. 
A way to directly calculate $I^{\rm eff}$ for clusters is to use the area estimator proposed by Sindzingre \etal~in 1989~\cite{ceperley_area_estim}. 
The derivation of this estimator starts from discretizing the integral in Eq.~\ref{eqn:Ieff3} into a summation with finite increments in the variable $\tau$,
\begin{eqnarray}
I^{\rm eff}_n&=&\int_0^\beta \left< \hat{L}_n\left( \tau \right) \hat{L}_n \right> d\tau \nonumber\\
&=&\frac{1}{Z}\int_0^\beta {\rm Tr} e^{ -\left(\beta-\tau \right)\hat{H} }\hat{L}_n e^{ -\tau\hat{H} }\hat{L}_n d\tau \nonumber\\
&=&\frac{\Delta \tau}{Z}{\rm Tr} \sum_{k=1}^M  e^{ -\left(M-k \right)\Delta \tau }\hat{L}_n e^{ -k\Delta\tau\hat{H} }\hat{L}_n \nonumber\\
&=&\frac{\Delta \tau}{Z}{\rm Tr} \left(\sum_{k=1}^{M-1} e^{ -k\hat{H}\Delta \tau}\hat{L}_n e^{ -\left( M-k\right)\hat{H}\Delta \tau }\hat{L}_n +e^{ -M\Delta\tau\hat{H} }\hat{L}_n^2 \right) \nonumber\\
&=&\frac{\Delta \tau}{Z}{\rm Tr}\left(\hat{L}_n^2  e^{ -M\Delta\tau\hat{H} } \right) + \nonumber\\
&&\frac{\Delta\tau}{Z}\sum_{k=2}^M {\rm Tr} \left( \hat{L}_n e^{-\left( k-1 \right)\Delta\tau\hat{H} } \hat{L}_n e^{ -\left( M-k+1 \right)\Delta\tau\hat{H} } \right), \label{eqn:Ieff5}
\end{eqnarray}
where $M$ is a large integer with $M\Delta\tau=\beta$. Cyclic invariance of the trace operation has been employed to get to the last line of Eq.~\ref{eqn:Ieff5} and the reason for separating the summation into two terms will be clear below. 
For bosons, the summation over all  permutations is implicitly included in the ${\rm Tr}$ symbol. 
For ease of derivation below, we choose the $n$ direction to be along the $z$ axis of the frame of interest. 
The frame can be either the MFF or the SFF as pointed out in Sec.~\ref{sec:formu}\ref{subsec:SFFvsMFF}. 
Also, we use the conventional symbol $\tau$ for the small imaginary time interval to replace $\Delta\tau$. 
With the new symbols,
\begin{eqnarray}
I^{\rm eff}_z&=&\frac{\tau}{Z}{\rm Tr}\left(\hat{L}_z^2 e^{  -M\tau\hat{H} } \right) + \frac{\tau}{Z}\sum_{k=2}^M {\rm Tr} \left( \hat{L}_z e^{  -\left( k-1 \right)\tau\hat{H} } \hat{L}_z e^{ -\left( M-k+1 \right)\tau\hat{H} } \right). \label{eqn:Ieff6}
\end{eqnarray}
We first consider the system with only one type of point-like distinguishable particles and the generalization to more complicated cases will be discussed later.

Let us first look at the second term on the right-hand-side of Eq.~\ref{eqn:Ieff6}. 
For a summand in the summation over $k$, 
we factorize the exponentials into propagators with imaginary time interval $\tau$ and insert  $\int d {\bf R} \left| {\bf R} \right>\left< {\bf R} \right|$ -type resolutions of the identity between any adjacent time slices to have
\begin{eqnarray}
{\rm Tr}\left( \hat{L}_ze^{-(k-1)\hat{H}\tau}\hat{L}_ze^{-(M-(k-1))\hat{H}\tau} \right)&=&\int d{\bf R} d{\bf R}_1 \cdots d{\bf R}_{M-1} \nonumber\\
&&\left<{\bf R}\right|\hat{L}_ze^{-\tau\hat{H}}\left|{\bf R}_1\right>\left<{\bf R}_1\right|\cdots \nonumber\\
&&e^{-\tau\hat{H}}\left|{\bf R}_{k-1}\right>\left<{\bf R}_{k-1}\right|\hat{L}_ze^{-\tau\hat{H}}\left|{\bf R}_k\right>\nonumber\\
&&\left<{\bf R}_k\right|\cdots e^{-\tau\hat{H}}\left|{\bf R}_{M-1}\right>\nonumber\\
&&\left<{\bf R}_{M-1}\right|e^{-\tau\hat{H}}\left|{\bf R}\right>. \label{eqn:term_k}
\end{eqnarray}
The matrix elements of $\hat{L}_z$ can be expressed as
\begin{eqnarray}
\left<{\bf R}\right|\hat{L}_ze^{-\tau\hat{H}}\left|{\bf R}'\right>&=&(-i\hbar)\sum_{i=1}^N\left(x_i\frac{\partial}{\partial y_i}-y_i\frac{\partial}{\partial x_i} \right)\left<{\bf R}\right|e^{-\tau\hat{H}}\left|{\bf R}'\right>\nonumber\\
&=&(-i\hbar)\sum_{i=1}^N\left(x_i\frac{\partial}{\partial y_i}-y_i\frac{\partial}{\partial x_i} \right)\lambda_{\tau}^{-3N} \nonumber\\
&&\times e^{-\frac{\pi}{\lambda_\tau^2}({\bf R}-{\bf R}')^2}e^{-\tau V({\bf R}')}. \label{eqn:lzmat}
\end{eqnarray}
Apparently, the position representation of $\hat{L}_z$ and the Trotter factorization have been invoked. The action of $\hat{L}_{z}$ on the Gaussian operand yields 
\begin{eqnarray}
\left<{\bf R}\right|\hat{L}_ze^{-\tau\hat{H}_0}\left|{\bf R}'\right>&=&(-i\hbar)\left( -\frac{2\pi}{\lambda_\tau^2} \right)\sum_{i=1}^N\left( x_i\left(y_i-y_i' \right)-y_i\left(x_i-x_i' \right) \right) \nonumber\\
&&\times \lambda_{\tau}^{-3N}e^{-\frac{\pi}{\lambda_\tau^2}({\bf R}-{\bf R}')^2}e^{-\tau V({\bf R}')}\nonumber\\
&=&(-i\hbar)\frac{2\pi}{\lambda_\tau^2} \sum_{i=1}^N\left({\bf r}_i\times{\bf r}_i' \right)_z\left<{\bf R}\right|\hat{\rho}(\tau)\left|{\bf R}'\right>.
\end{eqnarray}
With this formula, the two terms involving $\hat{L}_z$ in Eq.~\ref{eqn:term_k} become
\begin{eqnarray}
\left<{\bf R}\right|\hat{L}_ze^{-\tau\hat{H}_0}\left|{\bf R}_1\right>&=&(-i\hbar)\frac{2\pi}{\lambda_\tau^2} \sum_{i=1}^N\left({\bf r}_i\times{\bf r}_{1,i} \right)_z\left<{\bf R}\right|\hat{\rho}(\tau)\left|{\bf R}_1\right>\nonumber\\
&=&(-i\hbar)\frac{4\pi}{\lambda_\tau^2}A_{1,z}\left<{\bf R}\right|\hat{\rho}(\tau)\left|{\bf R}_1\right>; \label{eqn:rlzr1}\\
\left<{\bf R}_{k-1}\right|\hat{L}_ze^{-\tau\hat{H}_0}\left|{\bf R}_k\right>&=&(-i\hbar)\frac{2\pi}{\lambda_\tau^2} \sum_{i=1}^N\left({\bf r}_{k-1,i}\times{\bf r}_{k,i} \right)_z\left<{\bf R}_{k-1}\right|\hat{\rho}(\tau)\left|{\bf R}_k\right> \nonumber\\
&=&(-i\hbar)\frac{4\pi}{\lambda_\tau^2}A_{k,z}\left<{\bf R}_{k-1}\right|\hat{\rho}(\tau)\left|{\bf R}_k\right>. \label{eqm:rklz}
\end{eqnarray}
The vectorial area between two adjacent slices
\begin{eqnarray}
{\bf A}_{k}&=&\sum_{i=1}^N\frac{1}{2}\left({\bf r}_{k-1,i}\times{\bf r}_{k,i} \right) \label{eqn:area_k}
\end{eqnarray}
has been used in these two equations. 
Obviously each summand in Eq.~\ref{eqn:area_k} is the area of the triangle made up of the two position vectors of the adjacent beads of one particle. 
$A_{k,z}$ is the projection of ${\bf A}_k$ along the $z$-axis. 
If $I^{\rm eff}$ is calculated in the SFF (response to an external field in Sec.~\ref{sec:formu}\ref{subsec:SFFvsMFF}), $
{\bf A}_k$ is projected onto the same axis independent of $k$. 
However, if $I^{\rm eff}$ is calculated in the MFF (response to molecular rotation in Sec.~\ref{sec:formu}\ref{subsec:SFFvsMFF}), 
the orientation of the $z$-axis varies with $k$. 
In the small $\tau$ limit, and for a heavy rotor, it is safe to assume that the $z$-axis orientation changes much less than the positions of the light bosons in the interval $\tau$ and 
therefore, one can project ${\bf A}_{k}$ along the $z$-axis in either slice $k-1$ or $k$ to calculate $A_{k,z}$.

Inserting Eqs.~\ref{eqn:rlzr1} and~\ref{eqm:rklz} into Eq.~\ref{eqn:term_k} and normalizing it with the partition function to get the ensemble average,  $\left< \right>$, we have
\begin{eqnarray}
\frac{1}{Z}{\rm Tr}\left( \hat{L}_ze^{-(k-1)\hat{H}\tau}\hat{L}_ze^{-(M-(k-1))\hat{H}\tau} \right)&=&(-i\hbar)^2\left( \frac{4\pi}{\lambda_\tau^2} \right)^2\left< A_{1,z}A_{k,z} \right>,
\end{eqnarray}
and the second term on the right-hand-side of Eq.~\ref{eqn:Ieff6} becomes
\begin{eqnarray}
\frac{\tau}{Z}\sum_{k=2}^M{\rm Tr}\left[ \hat{L}_ze^{-(k-1)\hat{H}\tau}\hat{L}_z 
e^{-(M-(k-1))\hat{H}\tau} \right]
\nonumber\\
=\tau(-i\hbar)^2\left(\frac{4\pi}{\lambda_\tau^2} \right)^2\sum_{k=2}^M\left<A_{1,z}A_{k,z} \right>\nonumber\\
=\tau(-i\hbar)^2\left(\frac{4\pi}{\lambda_\tau^2} \right)^2
\left( \left<A_{1,z}\sum_{k=1}^MA_{k,z} \right>-\left<A_{1,z}A_{1,z} \right> \right). \label{eqn:2nd_term}
\end{eqnarray}

A similar procedure can be applied to handle the first term on the right-hand-side of Eq.~\ref{eqn:Ieff6}.
\begin{eqnarray}
{\rm Tr}\left(\hat{L}_z^2e^{-\beta\hat{H}} \right)&=&\int d{\bf R} d{\bf R}_1\cdots d{\bf R}_{M-1}\left<{\bf R}\right|\hat{L}_z^2e^{-\tau\hat{H}}\left|{\bf R}_1\right>\nonumber\\
&&\times\left<{\bf R}_1\right|\cdots\left|{\bf R}_{M-1}\right>\left<{\bf R}_{M-1}\right|e^{-\tau\hat{H}}\left|{\bf R}\right>,\end{eqnarray}
and only the first matrix element contains $\hat{L}_z$. This matrix element is further derived as
\begin{eqnarray}
\left<{\bf R}\right|\hat{L}_z^2e^{-\tau\hat{H}}\left|{\bf R}_1\right>&=&(-i\hbar)^2\sum_{j=1}^N\left( x_j\frac{\partial}{\partial y_j} - y_j\frac{\partial}{\partial x_j} \right)\nonumber\\
&&\times \sum_{i=1}^N\left( x_i\frac{\partial}{\partial y_i} - y_i\frac{\partial}{\partial x_i} \right)\left<{\bf R}\right|e^{-\tau\hat{H}}\left|{\bf R}_1\right>\nonumber\\
&=&(-i\hbar)^2\left(-\frac{2\pi}{\lambda_\tau^2} \right)\sum_{j=1}^N\sum_{i=1}^N\left( x_j\frac{\partial}{\partial y_j} - y_j\frac{\partial}{\partial x_j} \right)\nonumber\\
&&\times\left( y_ix_{1,i}-x_iy_{1,i} \right)\left<{\bf R}\right|e^{-\tau\hat{H}}\left|{\bf R}_1\right>\nonumber\\
&=&(-i\hbar)^2\left(-\frac{2\pi}{\lambda_\tau^2} \right)\sum_{j=1}^N\sum_{i=1}^N\left\{ \delta_{ij}\left(x_jx_{1,i}+y_jy_{1,i} \right)+\right.\nonumber\\
&&\left. \left( y_ix_{1,i}-x_iy_{1,i} \right)\left( -\frac{2\pi}{\lambda_\tau^2}\right)\left(-x_jy_{1,j}+y_jx_{1,j} \right) \right\}\nonumber\\
&&\times\left<{\bf R}\right|e^{-\tau\hat{H}}\left|{\bf R}_1\right>\nonumber\\
&=&(-i\hbar)^2\left(-\frac{2\pi}{\lambda_\tau^2} \right)\left<{\bf R}\right|\rho(\tau)\left|{\bf R}_1\right>\nonumber\\
&&\times \left\{\sum_{i=1}^N \left(x_ix_{1,i}+y_iy_{1,i} \right)+ \right.\nonumber\\
&&\left.\left(-\frac{2\pi}{\lambda_\tau^2}\right)\sum_{j=1}^N\left(x_jy_{1,j}-y_jx_{1,j} \right)\sum_{i=1}^N\left(x_iy_{1,i}-y_ix_{1,i} \right)\right\}\nonumber\\
&=&(-i\hbar)^2\left(-\frac{2\pi}{\lambda_\tau^2} \right)\left<{\bf R}\right|\rho(\tau)\left|{\bf R}_1\right>\nonumber\\
&&\times\left\{ \sum_{i=1}^N \left(x_ix_{1,i}+y_iy_{1,i} \right)-\frac{8\pi}{\lambda_\tau^2}A_{1,z}^2 \right\}
\end{eqnarray}
Therefore, the first term on the right-hand-side of Eq.~\ref{eqn:Ieff6} can be expressed as an average  as
\begin{eqnarray}
\frac{\tau}{Z}{\rm Tr}\left(\hat{L}_z^2e^{-\beta\hat{H}} \right)&=&\tau(-i\hbar)^2\left( -\frac{2\pi}{\lambda_\tau^2} \right)\left< \sum_{i=1}^N \left(x_ix_{1,i}+y_iy_{1,i} \right) \right>\nonumber\\
&&+\tau(-i\hbar)^2\left( -\frac{4\pi}{\lambda_{\tau}^2} \right)^2\left<A_{1,z}^2 \right> \label{eqn:1st_term}
\end{eqnarray}
Adding Eqs.~\ref{eqn:1st_term} and~\ref{eqn:2nd_term} up and noticing that the second summands  cancel each other, we reach a programmable formula for $I^{\rm eff}$:
\begin{eqnarray}
I^{{\rm eff}}_z&=&\tau (-i\hbar)^2\left(-\frac{2\pi}{\lambda_\tau^2} \right)\left\{ \left< \sum_{i=1}^N \left(x_ix_{1,i}+y_iy_{1,i} \right) \right> \right.\nonumber\\
&&\left.+\left(-\frac{8\pi}{\lambda_\tau^2} \right)\left<A_{1,z}\sum_{k-1}^MA_{k,z} \right> \right\}\nonumber\\
&=&\tau (-i\hbar)^2\left(-\frac{2\pi}{\lambda_\tau^2} \right)\left\{ \left< \sum_{i=1}^N\left(\hat{z}\times{\bf r}_i \right)\cdot\left(\hat{z}\times{\bf r}_{1,i} \right) \right>\right.\nonumber\\
&&\left. +\left(-\frac{8\pi}{\lambda_\tau^2} \right)\left<A_{1,z}\sum_{k=1}^MA_{k,z} \right>\right\}. \label{eqn:Ieff7}
\end{eqnarray}

$I^{\rm eff}$ is evaluated only when a closed configuration is attained. Therefore, all beads should be considered equal and the averaged quantities with the specific subscript of ``1'' should be equal to  those with arbitrary $k$. This equivalence leads to a further average over imaginary time slices:
\begin{eqnarray}
\left< \sum_{i=1}^N\left(\hat{z}\times{\bf r}_i \right)\cdot\left(\hat{z}\times{\bf r}_{1,i} \right) \right>&=&\frac{1}{M}\sum_{k=1}^M\left< \sum_{i=1}^N\left(\hat{z}_{k-1}\times{\bf r}_{k-1,i} \right)\right.\nonumber\\
&&\left.\cdot\left(\hat{z}_{k-1}\times{\bf r}_{k,i} \right) \right>\nonumber\\
&=&\left<\frac{1}{M}\sum_{k=1}^M \sum_{i=1}^N\left(\hat{z}_{k-1}\times{\bf r}_{k-1,i} \right)\right.\nonumber\\
&&\left. \cdot\left(\hat{z}_{k-1}\times{\bf r}_{k,i} \right) \right>; \label{eqn:expec1}\\
\left<A_{1,z}\sum_{k=1}^MA_{k,z} \right>&=&\frac{1}{M}\sum_{k'=1}^M\left<A_{k',z}\sum_{k=1}^MA_{k,z} \right> \nonumber\\
&=&\frac{1}{M}\left<\sum_{k'=1}^MA_{k',z}\sum_{k=1}^MA_{k,z} \right>\nonumber\\
&=&\frac{1}{M}\left< \left(\sum_{k=1}^MA_{k,z}\right)^2 \right>\nonumber\\
&=&\frac{1}{M}\left< A_{z}^2 \right>, \label{eqn:expec2}
\end{eqnarray}
where the definition of sum-path-area
\begin{eqnarray}
{\bf A}=\sum_{k=1}^M{\bf A}_k
\end{eqnarray}
is used to obtain the last equality. 
Obviously, ${\bf A}$ is the total vectorial area of all the paths for a closed configuration.

Substituting Eqs.~\ref{eqn:expec1} and~\ref{eqn:expec2} in Eq.~\ref{eqn:Ieff7} and expressing $\lambda_\tau$ with Eq.~\ref{eqn:thermal_wl} leads to the final formula for $I^{\rm eff}$:
\begin{eqnarray}
I^{\rm eff}_z&=&\left<\frac{m}{M}\sum_{k=1}^M \sum_{i=1}^N\left(\hat{z}_{k-1}\times{\bf r}_{k-1,i} \right)\cdot\left(\hat{z}_{k-1}\times{\bf r}_{k,i} \right) \right>-\frac{4m^2}{\hbar^2\beta}\left< A_z^2 \right>\nonumber\\
&=&\left< I^{{\rm cl}}_z\right>-\frac{4m^2}{\hbar^2\beta}\left< A_z^2 \right>, \label{eqn:Ieff8}
\end{eqnarray} 
which is used in actual simulations. Now it becomes clear that the classical moment of inertia comes from the first term while the quantum reduction comes from the second term in the last equality of Eq.~\ref{eqn:Ieff5}, and this is why we separate the summation over $k$ into two parts there. The derivation of Eq.~\ref{eqn:Ieff8} is very lengthy and it has never been given in any publication. Usually, authors jump directly from Eq.~\ref{eqn:Ieff5} to Eq.~\ref{eqn:Ieff8}. We consider it necessary to include the derivation in this report to help readers who may be lost in the long march between the two equations.

Derived for a system of pure boltzmannons, Eq.~\ref{eqn:Ieff8} is directly applicable for bosons if the permutation sampling is involved in the simulation. It can also be used for systems with dopants to calculate $I^{\rm eff}$ in the MFF but as mentioned above, the $z$-axis changes orientation in the SFF during the simulation, as a result of the dopant rotation. To calculate the off-diagonal effective moments of inertia, e.g., $I^{\rm eff}_{xy}$, the corresponding angular momentum components, e.g., ${\hat{L}_x}$ and ${\hat{L}_y}$, will need to be employed in Eq.~\ref{eqn:Ieff6} and the final formula has the form 
\begin{eqnarray}
I^{eff}_{xy} &= &-\frac{m}{M}\left< \sum_{k=1}^M \sum_{i=1}^N y_{k-1,i}x_{k,i} \right> - \frac{4m^2}{\hbar^2\beta}\left< A_y A_x \right> \\
&=& \left< I^{cl}_{xy} \right> - \frac{4m^2}{\hbar^2\beta}\left< A_y A_x \right>.
\end{eqnarray}

So far there has been no report of effective moments of inertia for  systems composed of two superfluid species, e.g., both $^4$He and \phtwo. Here we provide a formula for this quantity without detailed derivation:
\begin{eqnarray}
I^{\rm eff, He/H_2}_z&=&\left<I^{\rm cl,H_2}_z \right>+\left<I^{\rm cl,He}_z \right>\nonumber \\
&&-\frac{4}{\hbar^2\beta}\left[ m_{\rm H_2}^2\left< A_{z,{\rm H_2}}^2 \right>+m_{\rm He}^2\left< A_{z,{\rm He}}^2 \right>\right.\nonumber\\
&&\left.+2m_{\rm He}m_{\rm H_2}\left< A_{z,{\rm He}}A_{z,{\rm H_2}} \right> \right].
\end{eqnarray}
A procedure similar to that used for obtaining Eq.~\ref{eqn:Ieff8} has been used here but with $\hat{L}_z=\hat{L}_z^{\rm H_2}+\hat{L}_z^{\rm He}$. One should notice that there is a mixed term
\begin{eqnarray}
2m_{\rm He}m_{\rm H_2}\left< A_{z,{\rm He}}A_{z,{\rm H_2}} \right> 
\end{eqnarray}
in the square bracket and it determines that the effective moment of inertia (superfluid response) of the composite system is not just a simple addition of the two components. 
This interference between $^4$He and \phtwo~in the total superfluid response has never been investigated 
(all studies of the mixed systems containing both $^4$He and \phtwo~introduced in Sec.~\ref{sec:results} assume additive superfluidity and omit the interference)
and we hope that our formula here will ignite studies on this effect.

The quantum reduction
\begin{eqnarray}
-\frac{4m^2}{\hbar^2\beta}\left< A_z^2 \right>
\end{eqnarray}
in Eq.~\ref{eqn:Ieff8} underlies the non-classical inertial response to the rotation of a probe. 
Consequently, the superfluid fraction along the $n$-axis is defined to be~\cite{ceperley_area_estim, sindzingre_pH2_superfluid,ceperley_rmp_1995}
\begin{eqnarray}
f_s^n&=&\frac{I^{\rm cl}_n-I^{\rm eff}_n}{I^{\rm cl}_n} \nonumber\\
&=&\frac{4m^2\left< A^2_n\right>}{\beta\hbar^2\left< I^{\rm cl}_n\right>}. \label{eqn:fs1}
\end{eqnarray}
The correlation between sum-path-area and superfluidity is evident, coining the name of area estimator of this method. Figs.~\ref{fig:sampath1} and~\ref{fig:sampath2} clearly demonstrate that bosonic exchange results in large path areas and explain the large superfluid response of bosonic systems like $^4$He and \phtwo. In Sec.~\ref{sec:results} below, all superfluid fractions from PIMC simulations are calculated using Eq.~\ref{eqn:fs1} unless further specified.

There have been several attempts to derive local superfluid estimator based on decomposing Eq.~\ref{eqn:fs1} into local contributions. Draeger and Ceperley proposed a local superfluid density defined as~\cite{rho_fs_draeger}
\begin{eqnarray}
\left.\rho_{s}\left(\bf r \right)\right|_{{n}}&=&\frac{4m^2N\left< A_n\left(\bf r \right)A_n \right>}{\beta\hbar^2\left< I^{\rm cl}_n\right>}, \label{eqn:N-norm-rhos}
\end{eqnarray}
where
\begin{eqnarray}
{\bf A}\left({\bf r} \right)&=&\frac{1}{2}\sum_{i=1}^N\sum_{k=1}^M\left( {\bf r}_{k-1,i} \times {\bf r}_{k,i}\right)\delta\left({\bf r}-{\bf r}_{k-1,i} \right).
\end{eqnarray}
Integrating this $\left.\rho_{s}\left(\bf r \right)\right|_{{n}}$ over the space results in the effective number of superfluid particles, $Nf_s^n$, which is not necessarily an integer. Eq.~\ref{eqn:N-norm-rhos} is then called $N$-normalized estimator of local superfluidity~\cite{rho_fs_kwon}. In 2006, Kwon \etal~proposed another estimator that also includes the decomposition of the classical moment of inertia~\cite{rho_fs_kwon}:
\begin{eqnarray}
\left.\rho_{s}\left(\bf r \right)\right|_{{n}}&=&\frac{4m\left< A_n\left(\bf r \right)A_n \right>}{\beta\hbar^2 r_{\perp}^2},
\end{eqnarray}
where $r_{\perp}$ is the distance from the $n$-axis. This density satisfies
\begin{eqnarray}
m\int \left.\rho_{s}\left(\bf r \right)\right|_{{n}} r_{\perp}^2 d{\bf r} = I^{\rm eff}_n=f_s^n I^{\rm cl}_n \label{eqn:I-norm-rhos}
\end{eqnarray}
and it is called $I$-normalized estimator of local superfluidity. Apparently, both local superfluid densities are anisotropic as they have a dependence on $n$. Also, decomposing one of the path area into local contribution while keeping the other intact in the angle brackets in Eq.~\ref{eqn:fs1} does not have a clear physical meaning. Therefore, the local superfluid densities should be considered as qualitative descriptions only. 
In the end, superfluidity is not a local effect.

Using Eq.~\ref{eqn:fs1} to measure superfluidity for large-scale systems like a beaker of $^4$He liquid in the macroscopic Andronikashvili experiment (Fig.~\ref{fig:bucket}(a)) is accurate. Because when there is no exchange, the path area is of the order of the thermal wavelength, which is far smaller than the size of the system and $\frac{4m^2}{\hbar^2\beta}\left<A^2_n\right>$ is negligible compared to the classical moment of inertia. Therefore for macroscopic systems, significant quantum reduction can only stem from exchange and the connection between superfluid response and bosonic exchange is clear. However, for finite size systems, other factors may contribute to the quantum reduction:
\begin{enumerate}
\item {\em Breakdown of  linear response theory.} The area estimator is derived from Eq.~\ref{eqn:Ieff3}, which results from  linear response theory. Therefore, any molecular dopant that does not satisfy the three requirements to be a good superfluid probe proposed at the end of Sec.~\ref{sec:formu}\ref{subsec:SFFvsMFF} will overestimate the quantum reduction in the calculation of the MFF response.

\item {\em Finite size effects.} For nano-scale clusters, the size of the system is not overwhelmingly larger than the path area, resulting in some background quantum reduction and superfluid fraction. This effect is present for both the MFF and SFF responses.

\end{enumerate}

The rotor-surrounding decouplings induced by the two factors above exist even when the dopant is surrounded by boltzmannons and therefore, they should not be considered as superfluid effects. 
But these decouplings are mixed with the coupling induced by bosonic exchange and contribute to $\left< A_n^2 \right>$ in Eq.~\ref{eqn:fs1}, making the superfluid fraction defined in the equation less accurate. 
A superfluid fraction without these background decouplings is desired. 
We propose an exchange (X) superfluid fraction to meet this need. 
The exchange superfluid fraction is defined as
\begin{eqnarray}
f_s^{n,X}&=&\frac{I_n^{\rm eff,BO}-I_n^{\rm eff,BE}}{I_n^{\rm eff,BO}}\nonumber\\
&=&1-\frac{I_n^{\rm eff,BE}}{I_n^{\rm eff,BO}},
\end{eqnarray}
where $I_n^{\rm eff,BO}$ is the effective moment of inertia from a simulation treating bosons as boltzmannons and $I_n^{\rm eff,BE}$ from treating bosons as bosons. 
Evidently, the background decouplings that affect both $I_n^{\rm eff,BO}$ and $I_n^{\rm eff,BE}$ are removed by the subtraction in the numerator in the first line or by the division in the second line of the equation. 
The exchange superfluid fraction only measures the superfluid response arising from exchange. 
An example is given in Sec.~\ref{sec:results}\ref{subsec:nonlinear} to demonstrate the usefulness of this new concept.

\section{Illustrative results} \label{sec:results}

In this section we discuss theoretical studies pertaining to superfluid \phtwo~systems. 
The discussion is divided into three parts. First, we look at theoretical works on pure \hydrogen~clusters in absence of a dopant probe molecule.
These systems include mixed clusters of \phtwo~and other \hydrogen~isotopologues. 
Without a molecular dopant, the calculated superfluid information is not measurable by a spectroscopic Andronikashvili experiment. 
So far theoretical simulation is still the only reliable tool to study their possible superfluidity and the calculated superfluid response for these systems corresponds to the response to an external field (SFF response, Sec.~\ref{sec:formu}\ref{subsec:SFFvsMFF}). 
We focus here  on the superfluidity of the \hydrogen~clusters. 
For other thermal effects in small \phtwo~clusters, e.g., excitation spectra, cluster abundance, etc., readers should refer to a recent brief review written by Navarro and Guardiola~\cite{navaro_ph2_review}. 
Second, we look at studies of \phtwo~clusters with a linear molecular dopant. 
Comparison between the theoretically predicted and experimentally measured superfluid responses of these systems is of extreme importance in judging the appropriateness of the theoretical methods and interpreting experimental results. 
Superfluid responses to both molecular rotation and an hypothetical external field (both MFF and SFF responses, Sec.~\ref{sec:formu}\ref{subsec:SFFvsMFF}) can be calculated. 
At last, we introduce studies of \phtwo~clusters with non-linear molecular dopants. 
Studies in each category will be introduced in a chronological sequence. 
Besides clusters, \phtwo~systems with reduced dimension, e.g. 2-D film, surface, 2-D crytalline matrix, and in metastable glassy phase are also objects of looking for superfluidity~\cite{wagner_h2_surface,ceperley_pimc_he_h2,wagner_melting_surface,ceperley_condition,gordillo_superfluid_h2film,boninsegni_ph2_layer_li,boninsegni_ph2_2d_matrix,boninsegni_ph2_c60,osychenko_h2_glassy}. 
Most of these studies involve periodic boundary condition and they are beyond the scope of the present report.
Pioneering studies~\cite{whaley_h2_clusters_structure,scharf_pimc_ph2_clusters,scharf_ph2_od2_isotope,mcmahon_quantum_liquid,whaley_vmc_dmc_h2,buch_pd2_od2_mixed,cheng_quantum_liquid} on nano-scale hydrogen systems before the dawn of microscopic superfluidity provided very useful background knowledge for the subsequent investigation on \phtwo~superfluidity. However, they are not covered in this work in the interest of length.

\subsection{\hydrogen~clusters without molecular rotor dopant} \label{subsec:h2_nodopant}

In 1999, Gordillo conducted a PIMC simulation of  $^4$He/\phtwo~binary clusters~\cite{gordillo_4He/H2}. 
The author varied the number of \phtwo~and $^4$He to investigate the structure and superfluidity dependence on the composition. 
When there is only one \phtwo, because of its large quantum delocalization, it tends to stay away from the centre of the cluster~\cite{whaley_1h2inhe}. 
On the contrary, Gordillo found that when there are more \phtwo particles, the strong \phtwo-\phtwo~attraction overcomes the quantum delocalization and the \phtwo~particles are located  at the core of the cluster. 
Those form a sub-cluster with a structure similar to that of pure \phtwo~ clusters as the \phtwo ~ particles are not surrounded by $^4$He.  
The $^4$He atoms compress the sub-cluster, making it more solid-like and less superfluid. 
This compression and superfluid reduction increase with the number of surrounding $^4$He until the added $^4$He are too far away from the \phtwo~sub-cluster. 
It is of interest to compare spectroscopic Andronikashvili experiments for the same \phtwo~cluster with and without a helium droplet and see whether this predicted superfluid reduction is reflected in the $B_{\rm eff}$ constants of the rotor. 
A technical challenge for this comparison is to have temperatures that are close enough in the two cases  to have a sensible comparison. This is not a problem for theoretical simulations. However, so far, we have not seen any theoretical study that makes such a comparison.

The reason why \phtwo~clusters can be liquid-like is because each \phtwo~has fewer neighbours than in the bulk phase and the effective attraction between \phtwo~is therefore weaker. 
Following this logic, Gordillo and Ceperley used the PIMC method to study 2-D (\phtwo)$_N$ clusters~\cite{gordillo_2d_ph2_cluster}. 
The authors considered several clusters whose sizes ranged from $N=6$ to $61$. They expected these finite size clusters with low dimension to be more liquid-like and their purpose was to investigate the relation between liquidity and cluster size, and the relation between superfluidity and cluster structure. 
Constrained on a 2-D surface, each \phtwo~can have at most six neighbours and the authors found substantial superfluid fraction only for clusters with two shells or when the third shell is not completely developed. 
By inspecting the diffusivity of \phtwo, which is quantified by the Lindemann ratio, 
they found that the two inner shells are frozen when the third shell is completed, while the outer shell retains some mobility. 
This combination of frozen core and mobile outer shell continues as the clusters become larger. Even the largest cluster ($N=61$) has a Lindemann ratio more than twice  that of the 2-D solid limit.

In 2006, Mezzacapo and Boninsegni carried out a PIMC study of (\phtwo)$_N$ clusters, with $N$ up to $40$ and in the temperature range of $0.5 \le T \le 4.0$~K~\cite{Boninsegni_pH2_melting}. 
This was the first theoretical study of (\phtwo)$_N$ clusters in 3-D space since the pioneering work of Sindzingre \etal~\cite{sindzingre_pH2_superfluid}. 
In the first paragraph of the paper, the authors summarized two questions regarding microscopic superfluidity to be addressed by theoretical studies: {\em What is the smallest finite size system for which superfluidity can be observed?; Which condensed matter systems, besides helium, can display this phenomenon, if not in the bulk at least in sufficiently small clusters?} 
The purpose of the present report is about the answer to the second question. We also propose a new answer for the first question which will be discussed in Sec.~\ref{sec:results}\ref{subsec:nonlinear}. 
In their study, the authors focused on the relation between superfluidity, structure, and quantum melting of the clusters. 
They employed the worm algorithm discussed in Sec.~\ref{sec:algor}\ref{subsec:worm} to account for \phtwo~exchange. 
They found that with $N < 22$, (\phtwo)$_N$ clusters are liquid-like and with superfluid fraction close to unity at $T=1.0$~K. 
For clusters with $22 \le N \le 30$ and at the same temperature, their superfluid fractions are generally lower and show a clear dependence on $N$. 
$f_s$ can change dramatically by adding only one \phtwo, 
and the authors attributed this phenomenon to the {\em alternating liquid-like (superfluid) and solid-like (insulating) characters of the clusters}, i.e., the evolution from liquid-like to solid-like structure is not a continuous process. 
This argument is supported by the \phtwo~radial distributions. For example, $f_s$ changes from 0.8 to 0.1 and to 0.25 for $N$ changes from 25 to 26 and to 27. 
These fractions are correlated with the flat liquid-like radial distribution of (\phtwo)$_{25}$, solid-like distribution with pronounced peaks of (\phtwo)$_{26}$, and the intermediate type distribution of (\phtwo)$_{27}$. The more delocalization in a liquid-like structure favours  larger degrees of overlap between particles and higher probability of the consequent exchange, resulting in enhanced superfluidity. 
For clusters with $N > 30$, superfluidity is largely suppressed.

\begin{figure}[h] 
\centerline {\includegraphics[width=0.7\columnwidth]{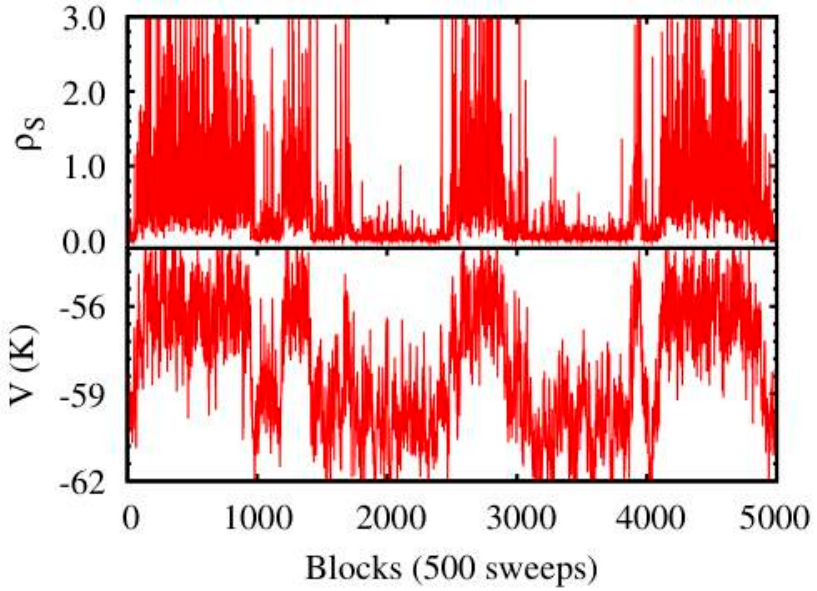}}
\caption{Superfluid fraction (upper panel) and potential energy per \phtwo~(lower panel) observed in each block of a PIMC simulation for the (\phtwo)$_{23}$ cluster at $T=1$~K. This figure is taken from Fig. 4 of Ref.~\cite{Boninsegni_pH2_melting}. Reproduced with the permission from~\cite{Boninsegni_pH2_melting}. Copyright \copyright~2006 The American Physical Society.} 
\label{fig:coexist}
\end{figure}

Another interesting phenomenon they observed is the coexistence of superfluid and non-superfluid phases within the same cluster at the same temperature. 
This phase coexistence is clearly shown in Fig.~\ref{fig:coexist}, in which the superfluid fraction and potential per \phtwo~observed at each block of a PIMC simulation for (\phtwo)$_{23}$ at 1~K are plotted. 
It is evident that there are two phases switching back and forth in the simulation, one with high superfluid fraction (averaged to 1) and higher potential, and the other with null superfluid fraction and lower potential. 
The potential profile indicates that the superfluid phase is liquid-like while the other is solid-like. 
The averaged superfluid fraction for this cluster was reported to be 0.5, meaning  equal probabilities of the two phases at this temperature. 
As temperature decreases, the superfluid phase becomes more pronounced and the cluster melts. 
This solid to liquid phase transition as temperature decreases is termed ``quantum melting'' by the authors. 
It is induced by the zero-point motion of the \phtwo~molecules and the exchange effects that give more mobility to the molecules.
This quantum delocalization of \phtwo~at low temperature was confirmed by another PIMC study by Warnecke~\etal~\cite{warnecke_ph2_clusters}

\begin{figure}[h] 
\centerline {\includegraphics[width=0.7\columnwidth]{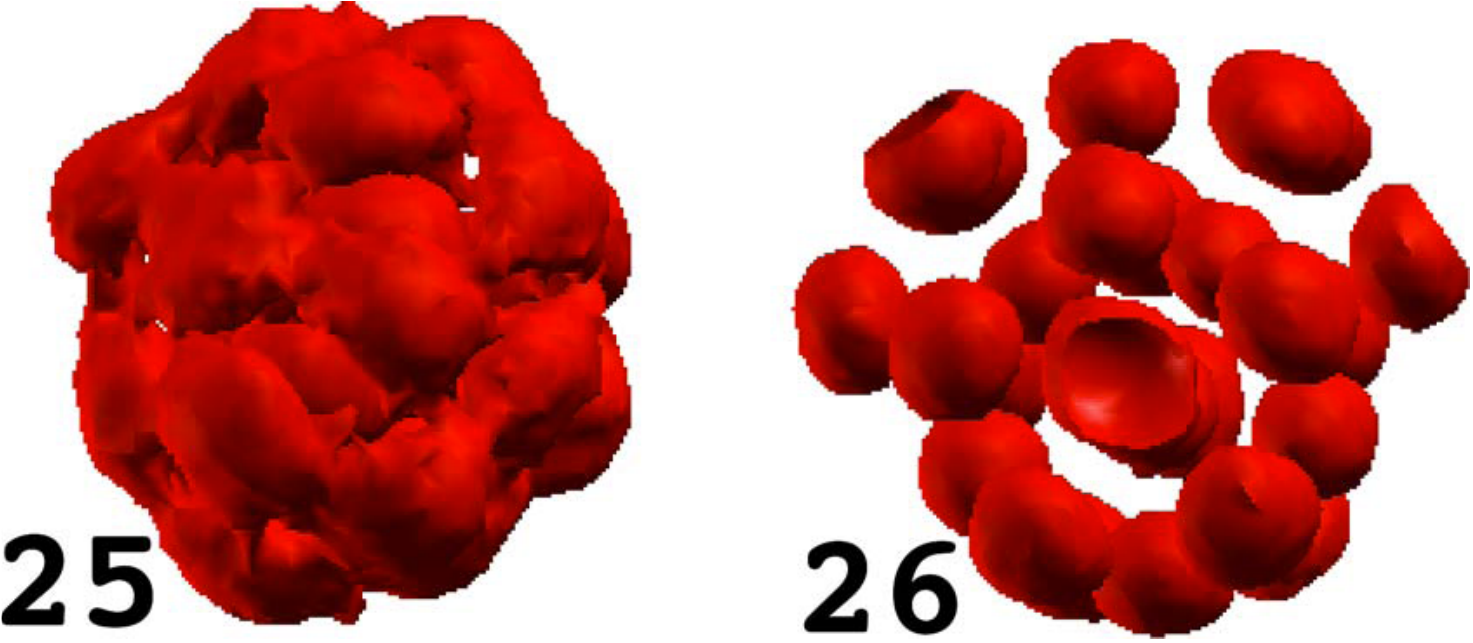}}
\caption{\phtwo~density iso-surfaces of (\phtwo)$_{25}$ and (\phtwo)$_{26}$ clusters at $T=1$~K. This figure is taken from Fig. 5 of Ref.~\cite{quantum_melting}. Reproduced with the permission from~\cite{quantum_melting}. Copyright \copyright~2007 The American Physical Society.} 
\label{fig:25vs26}
\end{figure}

In a follow-up study~\cite{quantum_melting}, Mezzacapo and Boninsegni investigated the energetics of the (\phtwo)$_N$ clusters discussed in the two paragraphs above. Moreover, they studied superfluid behaviour of (\odtwo)$_N$ clusters with $3\le N\le 20$. They pointed out that compared to helium, hydrogen has the advantage of having more isotopologues and therefore, one can investigate the mass effect on superfluidity. In this work, \odtwo~were treated as pure substance with zero nuclear spin. Therefore, the aforementioned spin mixture of \odtwo (Sec.~\ref{sec:exp}\ref{subsec:droplet_exp}) is omitted and the molecules can be treated as indistinguishable bosons. 
The energetics of (\phtwo)$_N$ confirms the magic size of $N=13$~\cite{guardiola_h2_cluster,cuervo2006}, which corresponds to a local maximum in chemical potential. 
The authors attributed this magic size to the completion of the first shell, rather than the formation of a solid-like structure as explained in Ref.~\cite{guardiola_h2_cluster}. On the contrary, another magic size cluster, (\phtwo)$_{26}$, does exhibit solid-like character, as illustrated by its density iso-surface in Fig.~\ref{fig:25vs26}, in comparison with (\phtwo)$_{25}$. The figure also illustrates the proposed alternation of liquid-like and solid-like structure by adding only one \phtwo. The solid-like structure explains the low superfluidity of (\phtwo)$_{26}$. 
The authors also pointed out that the superfluidity increase does not always come with a solid-liquid transition. 
For example, both (\phtwo)$_{18}$ and (\phtwo)$_{23}$ become more superfluid as $T$ decreases from 2 to 0.75~K but the former maintains the same liquid-like structure while the latter undergoes a solid-liquid structure change. Such oscillations between solid and liquid-like structures persist down to $T=0$ according to the PIGS calculations reported in Ref. \cite{cuervo2008}.

As to the (\odtwo)$_N$ clusters, their energetics points to magic sizes of $N=13$ and $19$. It is evident that \hydrogen~clusters, both \phtwo~and \odtwo, grow with the icosahedral-derived structure. 
This topology of cluster growth is clearly illustrated in Fig.~16 of Ref.~\cite{quantum_melting}.
It is noteworthy that (\hydrogen)$_N$~clusters with normal \hydrogen~(not considering the exchange effect of the intramolecular H atoms) also grow with the icosahedral pattern.~\cite{martinez_h2_clusters}
The superfluidity of the (\odtwo)$_N$ clusters is generally lower than the (\phtwo)$_N$, 
consistent with their larger molecular mass and being less quantum. 
The authors found that at the low temperature of $T=0.5$~K, (\odtwo)$_N$ clusters with $N\le 14$ possess significant superfluidity, but not the larger ones. 
Down to $T=0.5$~K, the phase coexistence and quantum melting were not observed for the (\odtwo)$_N$ clusters yet. The heavier mass of \odtwo~favours solid-like structure of the clusters.

In 2007, Khairallah \etal~published a study of (\phtwo)$_N$ clusters with $N\le 40$ and $0.5 \le T \le 4.5$~K~\cite{magicnumber_pH2}. 
They looked into the interplay between magic sizes and superfluidity of the clusters. 
As in the previously introduced study, magic sizes are determined by their larger chemical potentials. 
The authors found liquid-like structures and significant superfluidity for clusters with $N < 26$ at $T\le 1.5$~K. 
The superfluidity of the magic size clusters (\phtwo)$_{13,19,23}$ is generally lower than that of the others 
but still substantial. 
For larger size magic clusters (\phtwo)$_{26,29,32,34,37}$, superfluidity is largely quenched at temperatures down to 0.5~K. 
For the clusters between those large magic sizes, pronounced superfluidity is observed at $T=0.5$~K. 
The authors proposed that the additional \phtwo~are loosely bound and  explore more surface structures. 
These freely moving \phtwo~mediate long permutation cycles that are prohibited by the strong localization in the magic size clusters and increase  superfluidity. Based on this model, they drew the following conclusion: superfluidity of large  clusters mainly comes from their surfaces, while their central cores have solid-like structures.  We will see below that this conclusion was challenged by subsequent studies.

Mezzacapo and Boninsegni pursued their  study of \phtwo~clusters with isotopic dopants~\cite{mezzacapo_isotope_h2}. 
They studied clusters with a total number of molecules greater than 15 and  $T=0.5$ and $1$~K, with isotopic dopants consisting of \odtwo~and \ohtwo. 
They focused on how the isotopic dopants affect the structure and superfluidity of the clusters. 
It was found that the presence of \odtwo~dopants, even just one or two, greatly solidifies the clusters and reduces their superfluidity. 
This is especially true for clusters with more than 22 molecules. 
On the contrary, doping \ohtwo~in \phtwo~clusters have  lesser effect. 
Through studying the radial distributions, the authors revealed that the \odtwo~dopants tend to stay in the core of the clusters while the \ohtwo~stay on the surface. 
This is because of the larger mass of \odtwo~and higher localization. 
With a localized core of \odtwo~dopants, the \phtwo~are less mobile and tend to solidify. 
Furthermore, the central \odtwo~disconnect long permutation cycles of \phtwo~to the greatest extent, and therefore suppress their possible quantum melting and diminish their superfluidity. 
Such reduction in superfluid fraction are absent for \ohtwo~dopants for which the liquid-like structure and superfluidity of  \phtwo~ are largely retained. 
Although the authors predicted that the \ohtwo~doped \phtwo~clusters are significantly superfluid, 
we need to point out that their treatment of \ohtwo~is inaccurate. 
They treated \ohtwo~particles as point-like just as they did for \phtwo.
The same interaction potential was used and the point like \ohtwo~particles were not allowed to exchange due to their distinguishalbility.
However, \ohtwo~molecules have a $j=1$ degenerate ground rotational state, and they look more like dumbbells as the atomic $p$ orbitals as opposed to  spheres. 
This geometrical consideration gives a permanent quadrupole moment to the \ohtwo~molecules and the \ohtwo-\ohtwo~and \ohtwo-\phtwo~interactions should be stronger than that of the \phtwo-\phtwo~pair. 
These stronger interactions may force the \ohtwo~dopants to sit in the centre of the clusters and exert similar effects as the \odtwo.
Actually, there has been evidence of this central localization of \ohtwo~dopants when they are mixed with \phtwo~molecules.~\cite{akimov_oh2_in_ph2}.
An even more serious problem is that with a degenerate ground state, the rotation of \ohtwo~molecules in the field of other molecules is intrinsically non-adiabatic, i.e., it involves multiple potential energy surfaces which are coupled to each other through kinetic operators~\cite{tully_perspective}. 
The angular momenta of the \ohtwo~also need to be explicitly coupled to contribute to the total angular momentum of the whole cluster. These non-adiabatic effects were omitted in the oversimplified model.  A more advanced model to describe \ohtwo~molecules should be used in future studies.

Choo and Kwon also studied isotopically doped \phtwo~clusters~\cite{choo_d2h2_clusters}. 
They employed the PIMC method to simulate \odtwo(\phtwo)$_N$ clusters with $N=13$ and $18$ at $0.625 \le T \le 5$~K. 
As observed by Mezzacapo and Boninsegni, the authors found that the \odtwo~dopant is surrounded by \phtwo~in both clusters. 
The \odtwo~is located at the centre of mass of the \odtwo(\phtwo)$_{13}$ cluster for the whole temperature range under consideration, with a flatter radial distribution at higher temperature, a result of  thermal fluctuations. 
For \odtwo(\phtwo)$_{18}$, however, the \odtwo~is located at about 1.7~\AA~away from the cluster centre at $T<2.0$~K, and as $T$ increases, the dopant has its averaged position at the centre. 
This is also due to the larger thermal fluctuation of \odtwo~at higher temperature and higher propensity 
to move around inside the cluster. 
The thermal fluctuations of \odtwo~at high temperature affect the \phtwo~distribution in both clusters as \phtwo~particles can occupy the vacancy left by the thermally mobile \odtwo. 
The authors compared superfluid fractions of \odtwo(\phtwo)$_{13}$ and (\phtwo)$_{13}$ as functions of temperature and found similar behaviour. 
They then concluded that the superfluidity suppressions from having a magic size of 13 and doping an \odtwo~in (\phtwo)$_{13}$ are similar. The \odtwo(\phtwo)$_{18}$ cluster, having both a magic size and a dopant, is less superfluid than the (\phtwo)$_{18}$ cluster.

Within the same year Choo and Kwon published another PIMC study on \odtwo(\phtwo)$_N$ clusters with $10 \le N \le 19$ at $T=1.6$~K~\cite{choo_d2_h2_2}. 
Both the chemical potential and energy per \phtwo~profile show that $N=12$ and $18$ are magic size clusters that are more tightly bound compared to their neighbours. 
The two magic clusters have lower superfluid fractions compared to the others. These two findings are consistent with what was found for the pure (\phtwo)$_{13,19}$ clusters by Khairallah \etal~\cite{magicnumber_pH2}. 
The radial distributions of \odtwo~and \phtwo~indicate that the dopant is surrounded by \phtwo, and as $N$ increases to $18$, the dopant is not located at the centre of the cluster any more. 
This is consistent with a configuration where the \odtwo~dopant occupies one of the two equivalent central sites in a double-icosahedron. 
The authors calculated the $I$-normalized superfluid density (Eq.~\ref{eqn:I-norm-rhos}) for \odtwo(\phtwo)$_{15}$ and found that the radial distributions of the superfluid and total densities have similar shell structure, i.e., the superfluidity of the cluster is uniformly distributed.

\begin{figure}[h] 
\centerline {\includegraphics[width=0.5\columnwidth]{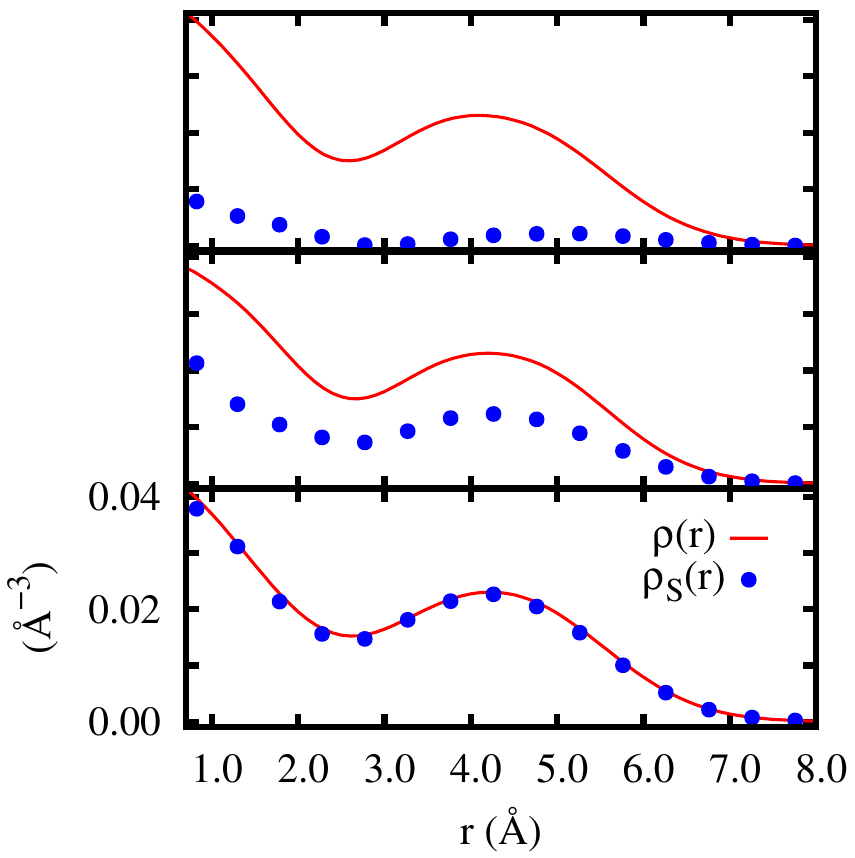}}
\caption{Radial profiles of density and superfluid density of (\phtwo)$_{18}$ at $T=3$, $2$, and $1$~K in the order of upper to lower panel. This figure are taken from Fig. 1 of Ref.~\cite{local_sup_boninsegni}. Reproduced with the permission from~\cite{local_sup_boninsegni}. Copyright \copyright~2008 The American Physical Society.} 
\label{fig:rhos_radial}
\end{figure}

The aforementioned conclusion of Khairallah \etal~\cite{magicnumber_pH2}~that the superfluid response of larger clusters mainly comes from their surfaces does not agree with the findings of the Boninsegni group and the Kwon group presented above. 
To obtain a clearer picture of the superfluidity distribution in \phtwo~clusters, Mezzacapo and Boninsegni studied the $I$-normalized superfluid densities for (\phtwo)$_N$ clusters with $N$ up to 27~\cite{local_sup_boninsegni}. 
Their main finding is illustrated in Fig.~\ref{fig:rhos_radial}.
For clusters with  a liquid-like structure such as (\phtwo)$_{18}$ shown in Fig.~\ref{fig:rhos_radial}(a), when $T$ decreases, the increased superfluid fraction is uniformly distributed in the whole cluster, including the most inner region.
The authors then challenged the conclusion of Khairallah \etal~and concluded that the superfluidity of \phtwo~clusters stems from long permutation cycles that involve \phtwo~over the whole cluster, regardless of their positions. 
This leads to a uniform superfluid response of the clusters. 

Mezzacapo and Boninsegni published a technical study in 2009 to investigate how simulation results  depend on the potential models in use~\cite{mezzacapo_ph2_models}. 
They calculated energies and superfluid fractions for (\phtwo)$_N$ clusters with $N$ up to 40, using three commonly used \phtwo-\phtwo~potentials: the Silvera-Goldman~\cite{sg_h2_pot}, the Buck~\cite{buck_h2_pot}, and the Lennard-Jones~\cite{magicnumber_pH2,lj_h2_pot} potentials. 
They found a strong relation between the detailed values of the properties and the potential. 
However, the general trend of changes with respect to temperature is preserved. 
This study reinforces the conclusions made by the two authors in their series of studies cited above, and those studies are summarized in recent review chapters~\cite{mezzacapo_review_h2_sf,boninsegni_melting_review,alonso_hydrogen_review}.

The Kwon group further studied  \odtwo~doped \phtwo~clusters with a larger range of compositions~\cite{shim_h2_d2_sf}. 
They used the PIMC method to simulate (\odtwo)$_M$(\phtwo)$_N$ clusters with $1\le M \le 5$ and $10 \le N \le 20$. 
The \odtwo~are all located at the centre of the clusters, as previously observed
The presence of \odtwo~does not modify the structure of the clusters and the magic sizes are preserved, consistent with an icosahedral growth pattern. 
Again, they found that \odtwo~suppresses \phtwo~superfluidity and the extent of this suppression is correlated with the number of dopants. 
To challenge the conclusion of Khairallah \etal~\cite{magicnumber_pH2}~about surface superfluidity, they calculated the $I$-normalized superfluid density of the clusters and concluded that the \phtwo~superfluidity is uniformly distributed except near the \odtwo, with no surface enhancement.

Sevryuk \etal\cite{why_superfluid_pH2} attempted to explain the superfluidity of \phtwo~clusters with the Quantum Theorem of Corresponding States.~\cite{dyugaev_corresponding1,dyugaev_corresponding2} 
This method was originally proposed by de Boer and Blaisse~\cite{deboer_qtcs_1,deboer_qtcs_2} to explain the relation between thermodynamic properties and quantum effects of light systems. 
Sevryuk \etal~studied two representative clusters, (\phtwo)$_{13,26}$, at $T=0.5$ and $1.5$~K, the former of which is low enough for both clusters to be superfluid. (\phtwo)$_{13}$ is representative of clusters with magic numbers and (\phtwo)$_{26}$ represents those with a quick drop of superfluidity after one \phtwo~is added. 
The main goal was to resolve the apparent contradiction between the superfluidity and solid-like structure of \phtwo~clusters, especially those with magic sizes. 
In this study, the quantumness of \phtwo~is represented by a dimensionless wavelength
\begin{eqnarray}
\Lambda^*&=&\frac{\hbar}{r_0\sqrt{m\epsilon}},
\end{eqnarray}
where $r_0$ and $\epsilon$ are the conventional parameters of the \phtwo-\phtwo~Lennard-Jones (12,6) potential and $m$ is the mass of the particle. 
Obviously, the larger $\Lambda^*$ is, the more quantum the particle is.
$\Lambda^*$ is called de Boer parameter. 
The authors tuned the quantumness of \phtwo~by modifying the interaction strength $\epsilon$.

Sevryuk \etal~found that the inclusion of exchange stabilizes the clusters when the \phtwo~quantumness increases. This is because when particles are quantum enough to have large overlaps, the inclusion of exchange forces clusters to only occupy the bosonic states and the probability of occupying the ground state is increased. As $\Lambda^*$ increases further to about 0.57, the phenomenon of quantum unbinding for both clusters emerges. 
When the particles are spatially fluctuating with a large enough thermal wavelength, the potential cannot bind them and the clusters dissociate. 
The \phtwo~radial distribution dependence on $\Lambda^*$ suggests that the smaller $\epsilon$ leads to overlap between shells and inter-shell exchange is possible. The authors called this disappearance of structure due to decreasing interaction of strength ``potential melting'', in order to differentiate from the aforementioned ``quantum melting'' induced by temperature reduction and increased thermal fluctuations. 
They also noticed that the density of the clusters decreases linearly with $\Lambda^*$. 
This is due to the larger quantum delocalization associated with weaker interactions. 
The authors looked into the dependence of the superfluid fractions of the clusters on $\Lambda^*$ and $T$, and they found that the quantumness of \phtwo~is on the borderline between a non-superfluid solid and a superfluid liquid. 
This explains the switching back and forth between the two phases of \phtwo~clusters with changes in temperature and size. This also resolves the contradiction between the solid-like rigidity and superfluidity of the magic size clusters.

\begin{figure}[h] 
\centerline {\includegraphics[width=0.7\columnwidth]{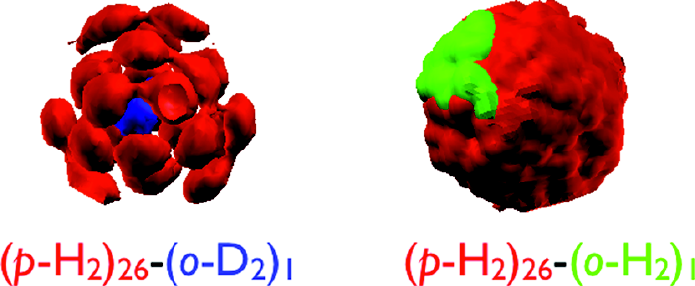}}
\caption{Particle densities of clusters (\phtwo)$_{26}$-\odtwo~(left) and (\phtwo)$_{26}$-\ohtwo~(right). \phtwo are represented by red colour, \odtwo~by blue, and \ohtwo~by green. This figure is taken from Fig. 6 of Ref.~\cite{mezzacapo_supersolid}. Reproduced with the permission from~\cite{mezzacapo_supersolid}. Copyright \copyright~2011 American Chemical Society.} 
\label{fig:od2_vs_oh2}
\end{figure}

In 2011, Mezzacapo and Boninsegni published a PIMC study of three clusters, 
(\phtwo)$_{25,26,27}$ at $0.125 \le T \le 2$~K~\cite{mezzacapo_supersolid}. 
These clusters, identified  in their previous research as~{\em quantum melters}, 
exhibit a remarkable variation in superfluidity as \phtwo~are sequentially added. 
They found that both (\phtwo)$_{25}$ and (\phtwo)$_{27}$ undergo a similar quantum melting process as $T$ decreases, i.e., solidephase $\rightarrow$ phase coexistence $\rightarrow$ liquid phase, and their increased superfluidity is ``strictly related'' to this structural transformation. 
(\phtwo)$_{26}$, however, maintains its solid-like structure in the temperature range of interest and its superfluid fraction saturates to 1.0 at $T=0.125$~K. 
Based on this finding, the authors tend to suggest that (\phtwo)$_{26}$ is an example of a supersolid, a long sought-after elusive state in low temperature physics that exhibits both solid character and superfluidity~\cite{chester_supersolid,Leggett_supersolid,kim_chan_supersolid,kim_chan_supersolid_he,leggett_defy_supersolid,ceperley_defies_supersolid,rittner_he_supersolid,rittner_supersolid_he_2,graves_he_supersolid,kondo_supersolid,penzev_he_supersolid,chan_supersolidity_2008}. 
They also carried out a similar simulation as in Ref.~\cite{mezzacapo_isotope_h2} for isotopically substituted (\phtwo)$_{25,26,27}$ clusters with one \odtwo~or \ohtwo~molecule. 
Similar structures were found for the substituted (\phtwo)$_{25,27}$ clusters, i.e., the massive \odtwo~is located at the centre of the clusters while the light \ohtwo~stays on the surface, 
and the \odtwo~doped clusters tend to solidify and become less superfluid. 
These structural characters are illustrated in Fig.~\ref{fig:od2_vs_oh2} taking the substituted (\phtwo)$_{27}$ clusters as examples. 
The substituted (\phtwo)$_{26}$ clusters with either \odtwo~or \ohtwo~molecules share a similar structure and the dopant is mainly located in the outer shell. 
The two substituted clusters undergo a similar decrease in superfluid fraction from $f_s=1.0$ to 0.2. 
The authors attributed the similarity between (\phtwo)$_{25}$-\odtwo~and (\phtwo)$_{25}$-\ohtwo~to the solid structure of (\phtwo)$_{26}$ and the dopant can only replace a \phtwo~at specific sites. 
This finding further substantiates their conclusion that (\phtwo)$_{26}$ is both solid and superfluid (supersolid).
In this work, the two authors employed the same treatment for \ohtwo~as in Ref.~\cite{mezzacapo_isotope_h2}. Therefore, the problems we point out above in the discussion of Ref.~\cite{mezzacapo_isotope_h2} also apply to this later work and their results for \ohtwo~substituted clusters contain an approximation that needs to be further investigated. 
For a more extensive discussion of supersolids, the interested readers can refer to the two reviews by Boninsegni and Prokof'ev~\cite{massimo_superdolid_rmp} and Boninsegni~\cite{boninsegni_review_supersolid}.
One, however, should note that the observation of supersolid substance substance remains elusive to this day
~\cite{chan_supersolid_absence}.

\subsection{\phtwo~clusters with a linear dopant molecule} \label{subsec:linear_dopant}

Before introducing any theoretical studies with a dopant in \phtwo~clusters, we would like to clarify 
that the presence of dopant affects the structure and superfluidity of the clusters. 
Mazzarello and Levi pointed out that a foreign molecule influences the solidification 
of \phtwo~clusters through the following four aspects~\cite{h2_solidify_levi,h2_solidify_dopant}: 
(1) it occupies some space now made inaccessible to \phtwo; 
(2) it breaks the isotropy of  space and introduces a non-spherical core for the solid; 
(3) it provides a template for layer-by-layer growth of the solid; 
(4) it facilitates solidification through its stronger interaction with \phtwo~compared to the \phtwo-\phtwo~interaction. 
Among them the first favours the liquid while the other can induce solidification, 
especially the last one which is dominant. 
Therefore, doping a \phtwo~cluster with a molecule would more or less suppress its superfluidity and it is this suppressed superfluid response that is probed in the studies discussed below. 
The reduction in superfluidity from a dopant and a helium surrounding (see the first paragraph of Sec.~\ref{sec:results}\ref{subsec:h2_nodopant}) are reflected by the low temperature onset of superfluidity   (between 0.15 to 0.38~K) for  OCS(\phtwo)$_{14,15,16}$ clusters embedded in helium droplets. 
Theoretical simulations predicted a higher superfluidity onset temperature (between 1 to 2~K) for (\phtwo)$_{13,18}$ clusters~\cite{sindzingre_pH2_superfluid}, indicating their more pronounced superfluidity. 
It is possible to employ PIMC simulations to separate the reduction in superfluidity  into  contributions from the helium droplet and the dopant. This would be a very interesting subject for future study.

In 2002, Kwon \etal~published the first theoretical study of a linear molecule in a \phtwo~cluster~\cite{kwon_OCS_pH2}. They carried out PIMC simulations for the OCS(\phtwo)$_{17}$ cluster with a fixed (non-rotating and non-translating) OCS. 
For this system, the MFF and SFF superfluid responses are identical. 
They found that seventeen \phtwo~form a complete solvation shell around OCS and the \phtwo~distribution can be separated into four rings around the OCS axis, with four, six, six, and one molecules. 
This pronounced localization is missing in OCS(He)$_N$ clusters~\cite{kwon_superfluid_helium}. 
It stems from the strong \phtwo-\phtwo~and OCS-\phtwo~interactions. 
The anisotropic superfluid response was obtained. The superfluid fraction parallel the OCS axis ($f_{s,\parallel}$) is significantly larger than that perpendicular to the axis ($f_{s,\perp}$) in a temperature range from 0.15 to 2.5~K. Furthermore, $f_{s,\parallel}$ rises  sharply to unity in the $0.15\le T \le 0.3$~K range, while $f_{s,\perp}$ rises  to less than 0.2. 
This anisotropic superfluid response is a natural result of the anisotropic interaction potential between hydrogen and OCS. 
The increase of $f_{s,\parallel}$ appears to be a two-stage process: it increases slowly from about 0.1 to about 0.15 as $T$ decreases from 2.5 to 1.2~K and remains steady until the aforementioned sharp rise at $T=0.3$~K. 
The authors explained this behaviour as follows: 
in the temperature range of the first slow rise-up, only intra-ring \phtwo~exchange can occur and at the lower temperature of the steep rise, inter-ring exchanges appear. 
This explanation is consistent with the behaviour of $f_{s,\perp}$, 
which is associated with  inter-ring exchanges and remains close to zero until the sharp rise of $f_{s,\parallel}$. Also, the local exchange density analysis (Fig. 3 of Ref.~\cite{kwon_OCS_pH2}) shows inter-ring exchange at $T=0.156$~K, but not at $T=0.375$~K. 
The sharp increase of $f_{s,\parallel}$ at $0.15\le T \le 0.3$~K is consistent with the disappearance of the Q-branch for the OCS(\phtwo)$_{14,15,16}$ clusters in the same temperature range~\cite{grebenev_OCS_pH2} and therefore, provides further evidence that what Grebenev \etal~observed is a superfluid phenomenon.

One shortcoming of the above study is that a fixed OCS dopant was used. 
One year after that publication, the same research group published a study on how the rotation of dopants (including OCS) affect the $^4$He distribution ~\cite{patel_doprot_he}. 
They found that even for the heaviest rotors, the solvating $^4$He cannot completely adiabatically follow its rotation and the $^4$He distribution is different between a rotating and a fixed dopant. 
Also in 2003, Kwon and Whaley performed a PIMC simulation for a OCS-\phtwo~dimer embedded in a cluster of 63 $^4$He, again with a non-rotating OCS~\cite{kwon_h2-ocs_in_he63}. 
This work was motivated by the experiment of Ref. \cite{grebenev_OCS_pH2}.
They found that their calculated moments of inertia of the OCS-\phtwo~complex are very close to those from experiment and exact bound states calculation, which involve OCS rotation. 
They then concluded that the \phtwo~molecule is rigidly coupled to the OCS rotation and this finding seems to support their use of a fixed OCS to study \phtwo~superfluid clusters. 
However, superfluidity may have a subtle dependence on localization. 
This is especially true for the calculation of $f_{s,\perp}$ because the OCS rotation may induce smearing of the \phtwo~rings and let them overlap with each other more easily, facilitating inter-ring exchange. 
A detailed study of the effect of a rotating OCS on the superfluidity of the surrounding \phtwo~cluster is needed.

Kwon and Whaley also studied OCS(\phtwo)$_{5,6,17}$ clusters with a fixed rotor and compared their structures with and without a $^4$He surrounding ~\cite{kwon_ocs_6ph2_39he}. 
They found that the structures of OCS(\phtwo)$_{5,17}$ are not significantly affected by the presence of the $^4$He outer shell. 
However, OCS(\phtwo)$_{6}$ has all 6 \phtwo~at the global minimum of the OCS-\phtwo~potential, i.e., forming a six-membered \phtwo~ring around the waist of OCS when the cluster is surrounded by  $^4$He. 
This is very different from the structure of an isolated OCS(\phtwo)$_{6}$ cluster, which has a five-membered ring at the global minimum and one \phtwo~at the oxygen side and close to the ring. 
The influence of the $^4$He surrounding (analogous to the helium droplet of actual experiments~\cite{grebenev_pH2_5ring,vilesov_ocs_ph2_he}) on the structure and superfluid response of the cluster is evident. 
This six-membered ring is a manifestation of the aforementioned effect of the helium on the structure of the interior \phtwo~cluster\cite{gordillo_4He/H2}. 
Six \phtwo~are squeezed by the $^4$He into a ring and the \phtwo-\phtwo~distance is shorter than 3.74~\AA, the minimum potential distance between two \phtwo. 
Both OCS(\phtwo)$_{5}$ and OCS(\phtwo)$_{6}$ exhibit no Q-branch when they are embedded in helium droplets~\cite{grebenev_pH2_5ring}. These similar superfluid phenomena should be related to their single ring structure and the associated intra-ring exchanges.

Paesani \etal~employed the DMC method to study OCS(\phtwo)$_N$ clusters with $N=1-8$~\cite{paesani_ocs_ph2_1-8}. 
They included OCS rotation and carried out rigid-body DMC to obtain the ground state wave functions of the clusters. 
They also calculated excited energies for the states with excitation of OCS rotation by using the POITSE method. 
The authors found that all the clusters have a fairly rigid structure, except OCS(\phtwo)$_{3,4}$. 
Those two clusters have floppy angular distribution of \phtwo~in the global minimum potential ring around OCS and they attributed this floppiness to the coupling between the breathing mode of the incomplete \phtwo~ring and the rotation of OCS. 
When the ring is complete at $N=5$, it becomes rigid and the breathing mode has a much smaller amplitude, 
reducing the coupling. 
The interaction between the primary five-membered ring and the additional \phtwo~particles that form the secondary ring around OCS confers rigidity to OCS(\phtwo)$_{6,7,8}$~clusters. 
The authors then concluded that the clusters with $N\ge 5$ can be sufficiently described with a symmetric top model. 
They fitted their excitation energies to the symmetric top rotational energy levels to obtain $B_{\rm eff}$, 
the effective rotational constant of OCS when its rotation is hindered by the surrounding \phtwo. 
They found that the $B_{\rm eff}$ monotonically decreases as $N$ increases, 
indicating that the \phtwo~superfluid response to the end-over-end rotation of OCS is not significant enough for these clusters. 
The authors also employed the clamped coordinate quasiadiabatic DMC (ccQA-DMC) method~\cite{quack_ccqa} to calculate the effective moment of inertia around the OCS axis (reported as $A$ in Table III of Ref.~\cite{paesani_ocs_ph2_1-8}). 
The $A_{\rm eff}$ have finite values and decrease monotonically with $N$. 
This is inconsistent with the experimental observation~\cite{tang_h2_ocs_2nd} 
that these clusters have no Q-branch in their spectra and that their $A_{\rm eff}$ should be infinitely large. 
Therefore, the ccQA-DMC model may not be an effective method to describe this superfluid response. 
In the end, this method is based on an assumption of \phtwo~rigidly following the OCS rotation, and this is, of course, not the case for superfluid clusters.

A better description of the superfluidity of OCS(\phtwo)$_5$ was given by Kwon and Whaley in their PIMC study, again with a fixed rotor~\cite{kwon_ocs_ph2_five}. 
They employed the area estimator to calculate the superfluid response of the five-membered \phtwo~ring around the OCS axis and obtained essentially 1.0 superfluid fraction for $T < 1$~K, consistent with the disappearance of the Q-branch of the same cluster when it is embedded in a helium droplet. 
Since OCS(\phtwo)$_5$ exhibits both crystalline structure, demonstrated as the highly peaked azimuthal pair angle distribution of \phtwo~in the ring, and high superfluid response, they considered this cluster as an example of a supersolid. 
However, whether the highly peaked pair angle distribution reflects a solid-like structure is questionable. 
This is because the distribution reflects the \phtwo~pair angle within the same imaginary time slices, i.e., it is a dynamical distribution, not a static one.
Another example of the difference between static and dynamic structures of nano-scale \hydrogen~clusters can be seen in Ref.~\cite{rabani_ph2_delocalization}.

In 2005, Moroni \etal~published a combined theoretical and experimental study on CO(\phtwo)$_N$ clusters with $N=1-17$~\cite{mroni_co_ph2_ir}. 
The IR spectra of these clusters were measured and assigned with the assistance of their RQMC simulations. 
They found that 12 \phtwo~form the first solvation shell around this smaller rotor, and they specifically investigated two series of $R(0)$ transitions: the $a$-type that is associated with the end-over-end rotation of the whole cluster and the $b$-type of the CO rotation. Both series of transition energies exhibit turnarounds as $N$ increases. They did not associate this phenomenon to \phtwo~superfluidity, as they mainly focused on the structure of the clusters and how it affects their spectra. 
Nevertheless, this study inspired two follow-up studies, which will be discussed below.

In the same year, Baroni and Moroni published another RQMC study to look into the special CO(\phtwo)$_{12}$~\cite{moroni_pH2_melting}. 
This time they were concentrated on the melting of this cluster. 
They ingeniously proposed a quantity to describe the rigidity of a finite size system, 
the multipole imaginary time correlation function:
\begin{eqnarray}
c_l\left( t\right)&=&\frac{\sum_{m=-l}^l\left< {Q^l_m}^*\left(\tau\right)\bar{Q}^l_m\left( \tau+t \right) \right>}{\sum_{m=-l}^l \left<{Q^l_m}^*\left(\tau\right)Q^l_m\left(\tau \right)\right>}.
\end{eqnarray}
$\left\{Q^l_m\right\}$ are the spherical multipole moments of mass density distribution around the centre of mass of the whole system and the overhead bar indicates that the multipole has been rotated backward by a rotation of the whole cluster in the imaginary time evolution. 
This backward rotation is to minimize the effect of cluster rotation on blurring the shape of the cluster. 
The faster the decay of $c_l\left( t\right)$, the less rigid the cluster is. 
The authors found that the $c_6\left(t \right)$ of CO(\phtwo)$_{12}$ becomes steady with a substantial value. 
This clearly indicates a rigid icosahedral structure of the cluster. 
On the other hand, (\phtwo)$_{13}$, with the central CO being replaced by a \phtwo, exhibits a quick and complete decay of $c_6\left(t \right)$ to zero, indicating that it has a molten structure with icosahedral character. 
This molten cluster is consistent with its predicted high superfluidity at low temperature~\cite{sindzingre_pH2_superfluid}. Comparatively, CO(\phtwo)$_{12}$ should be less superfluid. With an additional \phtwo, CO(\phtwo)$_{13}$ also has a quickly and completely decaying $c_6\left(t \right)$. 
The authors attributed this melting and the possible superfluidity to the addition of \phtwo~which can mediate and facilitate large-scale \phtwo~permutations. Those are greatly hindered by the central CO in CO(\phtwo)$_{12}$.
The \phtwo~densities of the three clusters are illustrated in Fig.~\ref{fig:coh2_comp} and clearly show the relative rigidities.

Paesani \etal~further studied OCS(\phtwo)$_N$ clusters with the same simulation methods as in Ref.~\cite{paesani_ocs_ph2_1-8} but with $N=9-17$, completing the first solvation shell~\cite{paesani_OCS_pH2}. 
This time they observed a turnaround of $B_{\rm eff}$ at $N=14$, indicating a superfluid decoupling from the end-over-end rotation of OCS. 
They employed Eq.~\ref{eqn:N-norm-rhos} to calculate local superfluid density of \phtwo. 
They observed a turnaround in $B_{\rm eff}$  consistent with the maximum local superfluid density perpendicular to the OCS axis of OCS(\phtwo)$_{14}$.
For larger clusters, e.g. OCS(\phtwo)$_{16}$, the perpendicular superfluid density is reduced, 
maybe because of their more stable and localized structure as a shell is completed. 
The superfluid responses parallel to the OCS axis of clusters with $10\le N\le 17$ are all close to unity, 
consistent with high local parallel superfluid densities. 
This theoretical study predicted the absence of a Q-branch in spectra of those clusters, 
similar to their analogues embedded in helium droplet.

Kwon and Whaley performed a PIMC study of both OCS(\phtwo)$_5$ and OCS(\odtwo)$_5$ clusters and compared their different superfluid responses~\cite{kwon_ocs_five_h2}. 
In that study, they ignored the possible spin mixture of the \odtwo~particles(Sec.~\ref{sec:exp}\ref{subsec:droplet_exp}) and treated the species as indistinguishable. 
Both clusters have the same five-membered ring structure around the OCS axis and OCS(\odtwo)$_5$ has even a more peaked azimuthal angular pair distribution, as a result of the lesser quantum fluctuation of the heavier \odtwo. 
In the  $0.3$ to $4$~K temperature range, this structure persists for both clusters. 
Both clusters have negligible perpendicular superfluid response.  
The parallel superfluid fraction of OCS(\phtwo)$_5$ saturates to unity at $T=1$~K while that of OCS(\odtwo)$_5$ rises more slowly and reaches 0.8 at $T=0.62$~K. 
This different behaviour is also related to the heavier mass of \odtwo. 
The authors compared the single-stage increase of the parallel superfluid fractions of the two clusters in this work and the aforementioned two-step rise for OCS(\phtwo)$_{17}$~\cite{kwon_OCS_pH2}. 
They concluded that this is because the complete parallel superfluid response in OCS(\phtwo)$_{17}$ requires inter-ring \phtwo~exchanges.
This is not needed in OCS(\phtwo)$_5$ and OCS(\odtwo)$_5$. 
The increase of the parallel superfluid fractions correlate with the decrease of the Q-branch intensities of the spectra of the two clusters as $T$ decreases. 
But for OCS(\odtwo)$_5$, the Q-branch intensity does not decay to zero. 
The authors described this phenomenon as evidence of spin mixture of $I=2$ and $I=0$ \odtwo~in the actual experiment. 
With regards to the experimental observation that OCS(\phtwo)$_5$ displays no Q-branch at both $T=0.15$ and $0.38$~K, while OCS(\phtwo)$_N$ with $N\ge 11$ only lack Q-branches at  lower temperature, the authors attributed this phenomenon to the smaller dimension of the OCS(\phtwo)$_5$ (effectively 1-D cluster) and the consequent weaker localization of \phtwo. 
The authors found that five-particle cyclic permutations are the only probable exchange cycles because they are compatible with the five-membered ring structure.

In 2006, Piccarreta and Gianturco studied OCS(\phtwo)$_N$ clusters with $N$ up to 30~\cite{piccarreta_ocs_h2}. 
They carried out bound-state calculations to obtain energies and wave functions of the ground and low-lying excited states of the OCS-\phtwo~dimer and performed DMC simulations for clusters with larger sizes. 
OCS rotation was included in their simulations. 
Their most interesting finding is that OCS(\phtwo)$_N$ clusters with $N$ up to 12 are well described by a Hartree product picture, i.e., the additional \phtwo~occupy orbitals which resemble the bound states of the OCS-\phtwo~dimer. 
This is strong evidence that the OCS-\phtwo~interaction dominates for those clusters. 
For $N > 12$, the \phtwo-\phtwo~interaction (correlation) starts to take over and the Hartree picture breaks down. 
For $N > 20$, the chemical potential saturates at a value close to the \phtwo-\phtwo~potential minimum, 
indicating the filling of an outer shell.

In 2010, Li \etal~published a combined PIMC theoretical and IR experimental study of CO$_2$(\phtwo)$_N$ clusters with $1\le N \le 18$ at $T=0.5$~K~\cite{huili_prl}. 
Unlike the PIMC studies cited above, the rotation of the dopant is included in order to have more reliable results and conclusions. 
A similar five-membered \phtwo~ring as in OCS(\phtwo)$_N$ clusters is found to be around the C atom and the first solvation shell is completed at $N=17$. 
The authors used both the area estimator and the imaginary time orientational correlation function~\cite{blinov_RCF} to calculate $B_{\rm eff}$ as a function of $N$ and the calculated results are in very good agreement with the experimental results. 
The $B_{\rm eff}$ has a first turnaround at $N=8$, where the overlap between the central five-membered ring and the ring beside it starts to emerge. 
The $B_{\rm eff}$ reaches a maximum at $N=12$, corresponding to $f_{s,\perp}=0.8$. 
The \phtwo~density of this cluster (Fig.~\ref{fig:co2_h2}(a), (b), and (c)) shows a liquid-like structure and pronounced overlap between the central and two side-rings. 
The fractional occupations of the rings, $5\frac{1}{2}$ for the central one and $3\frac{1}{4}$ for each side-ring, 
further confirm the fluidity of the \phtwo. 
The $\frac{1}{2}$ additional \phtwo~in the central ring can be viewed as a result of the \phtwo~flow expanding the ring in order to go through it. 
On the other hand, the superfluidity of CO$_2$(\phtwo)$_{17}$ is largely quenched by its solid-like double-icosahedral structure (Fig.~\ref{fig:co2_h2}(d), (e), and (f)). 
The $B_{\rm eff}$ of this cluster is close to its $B_{\rm cl}$. 
The three rings of the cluster have integer occupations of five \phtwo, lacking any sign of flow. 
From the experimental $B_{\rm eff}$ and calculated $B_{\rm cl}$, the authors provided the first direct measurement of $f_s$ in a molecular system. 
The measured $f_s$ values are consistent with the calculated ones (see Fig. 3(b) of the reference), validating the usefulness of PIMC and area estimator in the search for a molecular \phtwo~superfluid.

A new theoretical breakthrough was made by de Lara-Castells and Mitrushchenkovin in 2011~\cite{ph2_ocs_ci}. 
They adapted their full-configuration-interaction nuclear-orbital (FCI-NO) method, 
which resembles the traditional CI method of quantum electronic structure theory~\cite{jungwirth_fcino} and was originally developed to handle the Fermionic $^3$He clusters~\cite{lara_fcino1,lara_fcino2,lara_fcino3,lara_fcino4,lara_fcino5}, to bosonic systems and applied it to investigate CO$_2$(\phtwo)$_N$ clusters. 
The authors adapted the Born-Oppenheimer approximation of quantum chemistry calculations to their simulation and treated CO$_2$ as a fixed molecule that only provides an external potential for \phtwo. 
They focused on  clusters with $N\le5$, i.e., clusters with \phtwo~on its central ring. 
They found that for $N=4$ and $5$, the first excited state of the \phtwo~rotation about the CO$_2$ axis has axis-projected angular momentum of $\pm N $. 
This is exactly what was proposed by Grebenev \etal~when they tried to explain the absence of a Q-branch for the OCS(\phtwo)$_5$ cluster embedded in helium droplet~\cite{grebenev_pH2_5ring}. 
Their explanation was finally confirmed and elucidated by the work of de Lara-Castells and Mitrushchenkovin, whose calculation was almost completely {\em ab initio}. de Lara-Castells and Mitrushchenkovin also revealed that the bosonic symmetry of \phtwo~is not a sufficient condition for the special excitation pattern for their axial rotation. 
The \phtwo-\phtwo~bending mode will need to be delocalizing enough to let the \phtwo ~explore the whole ring, i.e., to have a distribution similar to the rigid \phtwo~ring proposed by Grebenev \etal. 
This study greatly deepens our understanding of the disappearance of the Q-branch in some linear molecule-doped \phtwo~clusters and at the end of the paper, the authors planned to employ the same method to study clusters with multiple \phtwo~rings to explain the perpendicular superfluid response. 
The success of this proposed approach will be another feat in the field. 
Although not specifically mentioned in that work, we note that the similar collective rotational excitations of CO$_2$(\phtwo)$_4$ and CO$_2$(\phtwo)$_5$ suggest that  superfluidity may occur with as few as only $4$ \phtwo~ particles. 
This conjecture is further confirmed by our work on asymmetric top-doped \phtwo~clusters discussed below.

Following the two aforementioned pioneering studies of CO(\phtwo)$_N$ clusters by the Moroni group~\cite{mroni_co_ph2_ir,moroni_pH2_melting}, Raston \etal~conducted a combined theoretical and experimental study of CO(\phtwo)$_N$ clusters with $N\le 14$ and published their work in 2012~\cite{raston_coh2_superfluid}. 
They measured the MW spectra of the clusters and extracted superfluid information from the $a$- and $b$-type $R_1\left(0 \right)$ transitions and compared with the results of PIMC simulations. 
CO is a rotor that violates two of the three guidelines of a good probe summarized in Sec.~\ref{sec:formu}\ref{subsec:SFFvsMFF}: 
it is a fairly fast rotor with rotational constant about 2~\wno~and its interaction with \phtwo~is weakly anisotropic. 
The CO-\phtwo~binding energy is also fairly weak but strong enough to allow CO to sit at the centre of the clusters. 
The authors chose to use this probe because they wanted to maximally reduce the \phtwo~localization effect from the dopant. 
The cost is that the rotational spectra of the clusters have two types of transitions ($a$- and $b$-types), 
which cannot be interpreted as arising from an effective rotors with one clear $B_{\rm eff}$.
This more complex situation required additional theoretical developments.

One of us (PNR) proposed a new method to extract $I^{\rm eff}$ by associating the second order Stark shift formula to the transition frequencies and spectral weights of the MW spectra. 
The Stark shift energy for the ground state of the cluster, derived from  second order perturbation theory, is
\begin{eqnarray}
\Delta E_0^2=-\sum_n\frac{\left| \left<0 \left|\cos \alpha \right| n\right>\right|^2\mu_z^2E_z^2}{E_n-E_0},
\end{eqnarray}
and for a rigid rotor, this shift is related to the rotational constant $B$ and moment of inertia $I$ as~\cite{zare_1988}
\begin{eqnarray}
\Delta E_0^2=-\frac{\mu_z^2E_z^2}{6B}=\frac{\mu_z^2E_z^2}{3\hbar^2}I.
\end{eqnarray}
Assuming the rigidity of the CO(\phtwo)$_N$ cluster and equating the two above equations, one has
\begin{eqnarray}
I^{\rm eff,cluster}&=&\sum_n\frac{3\left| \left<0 \left|\cos \alpha \right| n\right>\right|^2\hbar^2}{E_n-E_0}, \label{eqn:Ieff_stark}
\end{eqnarray}
of which the denominator is related to the experimental transition frequency and the matrix element square is the normalized spectral weight of a transition. 
Both of these are obtainable from the spectra and simulations. 
Only two rotational transitions, the $a$-type end-over-end cluster rotation and $b$-type CO hindered rotation, contribute to the summation over $n$ and both $I^{\rm eff}$ and $B_{\rm eff}$ were calculated using Eq.~\ref{eqn:Ieff_stark}, an average of the two contributions.

As shown in Fig.~\ref{fig:coh2_beff}, the averaged experimental $B_{\rm eff}$, $I^{\rm eff}$, and $f_s$ (they are $f_{s,\perp}$) are in excellent agreement with those calculated with the area estimator (two-fluid) and validate the accuracy of Eq.~\ref{eqn:Ieff_stark}. 
The agreement also indicates that a ``bad'' probe in an experimental sense may not be so bad in a theoretical sense, 
since the linear response theory behind the area estimator still holds for CO. 
This fact corroborates the accuracy of the calculated results for larger sized clusters which were not identified in the spectra. 
The comparison between the effective and classical quantities in Fig.~\ref{fig:coh2_beff}(a) and (b) indicates a significant decoupling between the rotor and the surrounding. 
Moreover, the decoupling does not die-off as the clusters grow, in contrast to the case of CO$_2$(\phtwo)$_N$ as shown in Fig.~\ref{fig:coh2_beff}(c). The authors concluded that the \phtwo~superfluid response to the probe is persistent with such a light molecule.

\subsection{\phtwo~clusters with a non-linear dopant molecule} \label{subsec:nonlinear}

The first theoretical  simulation of non-linear dopants in  \phtwo~clusters was conducted by Mak \etal~in 2005~\cite{mak_ch4_h2}. They studied CH$_4$(\phtwo)$_{12,14,16}$ clusters using PIMC simulations, and focused on their structure and superfluidity as functions of temperature. 
The authors chose CH$_4$ as the dopant because of its weak interaction with \phtwo, which is only about twice of the \phtwo-\phtwo~interaction, and they hoped that this weak interaction would minimally suppress the \phtwo~superfluidity. 
However, choosing this dopant is in contradiction to the guidelines we summarize in Sec.~\ref{sec:formu}\ref{subsec:SFFvsMFF} and it would be quite unlikely to measure and confirm their calculated results in a spectroscopic Andronikashvili experiment with the same dopant. Furthermore, they employed an isotropic CH$_4$-\phtwo~potential. 
Therefore, the dopant was treated as a point-like particle and the superfluidity discussed therein is actually the SFF response. 
They found that the presence of the dopant has negligible effect on the superfluidity of the clusters, which have a superfluid transition temperature close to that of the pure clusters~\cite{sindzingre_pH2_superfluid}, about 1~K. 
The \phtwo~radial distributions of all three clusters exhibit only one peak, meaning that with $N\le 16$ only a single solvation shell is formed around the CH$_4$. 
The radial distributions do not depend on temperature in the range of $0.5\le T\le 2$~K considered in their work.
The \phtwo~angular distributions are also independent of temperature in that range. 
The different temperature dependences of superfluidity and structure of the clusters suggests little correlation between them. 
The angular distribution of CH$_4$(\phtwo)$_{12}$ shows evidence of an icosahedral structure. 
They observed a correlation between quantum dispersion of \phtwo~and exchange probability as $T$ decreases and found that the quantum dispersion (and exchange) is favoured along the wetting surface of the dopant, because of the weaker \phtwo-\phtwo~potential barrier. 
The most probable permutation cycles of \phtwo~for the three clusters include from 4 to 7 molecules, and they adopt ring-like structures around CH$_4$.

The first actual theoretical study on \phtwo~clusters doped with a non-linear dopant molecule (also the first on superfluid clusters doped with an asymmetric top) was conducted by us and recently published~\cite{tobywater}. 
We employed the newly derived Noya propagator (Eq.~\ref{eqn:noya_rho}) and a newly developed accurate and efficient \water-\phtwo~potential with an Adiabatic-Hindered-Rotor approximation~\cite{zeng_ahr_pot} to perform a PIMC simulation for \pwater(\phtwo)$_N$ clusters with $N$ up to 20. 
We found similar magic cluster sizes as the pure \phtwo~clusters and concluded that the \pwater(\phtwo)$_N$ clusters have a similar icosahedron-derived structure with the dopant replacing one of the central \phtwo, as shown in Fig.~\ref{fig:ph2oph2}(b). 
The calculated superfluid fractions using the area estimator in the \water~MFF are generally greater than unity, indicating a breakdown of  linear response theory. 
In contrast to the aforementioned CO probe used in Ref.~\cite{raston_coh2_superfluid}, \pwater~is really a {\em bad} probe. 
It exhibits the same imaginary time orientational correlation function~\cite{blinov_RCF} as if it were a free rotor when it is doped in all different sizes of \phtwo~clusters and the same correlation function is obtained when the \phtwo~are treated as boltzmannons. Therefore, the perfect decoupling between the dopant and the solvent is not connected to \phtwo~superfluidity. 
Rather, it originates from the fairly isotropic \water-\phtwo~potential and the fact that \pwater~is a fast rotor, a much faster one than CO. 
Fig.~\ref{fig:ph2oph2}(a) shows that treating \phtwo~as a point-like particle by spherically averaging the \water-\phtwo~potential leads to a similar chemical potential profile to that of the full treatment. 
This means that \pwater~is diabatically sitting in its spherical ground rotational state and explains the negligible coupling between its rotation and  \phtwo.

By calculating the SFF superfluid fractions of the clusters, we found that the presence of \pwater~substantially quenches their superfluidity. This is because the dopant disconnects the long permutation cycles that pass through the centre of the clusters. 
In the paper we thoroughly discuss the requirements of a good superfluid probe and propose the three guidelines summarized in Sec.~\ref{sec:formu}\ref{subsec:SFFvsMFF}. A technical breakthrough of this work is that we combined the worm algorithm for bosonic exchange and asymmetric top rotation in a PIMC program for the first time.
This allows one to simulate superfluid clusters doped with any type of dopant.

We recently conducted another study with an asymmetric top dopant, namely the  \sotwo~molecule.~\cite{zeng_so2ph2} 
We simulated  \sotwo(\phtwo)$_N$ clusters with $N \le 17$ at $T=0.37$~K and found that the first solvation shell is complete at this upper limit. \sotwo~satisfies all three guidelines for a good probe and it is meaningful to calculate its MFF superfluid fractions. 
Fig.~\ref{fig:so2_fs} shows the MFF superfluid fractions as functions of $N$, orientation, and permutation symmetry. 
The superfluid fractions vary differently or even oppositely along different principal axes in the MFF and this is due to the lower symmetry of the asymmetric top. Their dependence on permutation symmetry indicates that the rotor-solvent decoupling is a true superfluid effect. 
We also proposed the exchange superfluid fraction (see the end of Sec.~\ref{sec:algor}\ref{subsec:Ieff_estim}) in this work and it provides a more precise measurement of superfluidity. 
For example, $f_s^{c,{\rm X}}$ is zero for \sotwo(\phtwo)$_1$, which should not have any superfluidity, but the regular $f_s^{c, {\rm B.E}}$ fails to catch it, for the \phtwo~is distributed around the $c$-axis and has fairly large projected area along the axis. 
A very striking finding of the research is that \sotwo(\phtwo)$_4$ has substantial superfluid response along the $a$-axis (Fig.~\ref{fig:so2_fs}(a)). For the first time we predict that cluster as small as with only four bosons can be superfluid. 
A crucial finding is that the averaged MFF superfluid fraction agrees very well with the SFF value.
(See panels (c) and (d) of Fig. 4 in Ref. \cite{zeng_so2ph2}).
Therefore, \sotwo~is such a good probe that it probes the genuine superfluidity of the clusters. This finding corroborates the other conclusions we made. 
Note that there have been very few publications on the simulation superfluid clusters doped with non-linear molecules. 
In the future, more studies are needed to explore this mostly untouched field and enrich our understanding on microscopic superfluids.

\section{Future challenges} \label{sec:challenge}

Despite the great advances made towards the theoretical simulation of \phtwo~superfluids since the beginning of this century, 
many challenges remain. 
One of these has been mentioned in Sec.~\ref{sec:results}\ref{subsec:h2_nodopant} and it is the treatment of dopants with strong non-adiabatic character, e.g., \ohtwo. 
Recent progress in non-adiabatic path-integral theory~\cite{schmidt_na_pi,ananth_semi_na,krishna_na_pi,hanasaki_na_pi,ananth_na_continuous,huo_na_pi,ananth_flux} may be an avenue to tackle this problem. 
This non-trivial problem requires further formal developments. We expect to see increased future activities in this direction.

Another challenge is the system size problem. 
Great insight could be deduced if one could simulate larger clusters up to the droplet limit and see how the hindered rotation of a probe evolves with the growth in size. 
More efficient computational tools are necessary to achieve such an objective.
In 2008, Markland and Manolopoulos proposed a method called the ring polymer contraction scheme to make path-integral simulations much faster~\cite{markland_contraction,markland_contraction2}. 
The Trotter approximation (Eq.~\ref{eqn:trotter_prim}) requires smaller $\tau$ (more beads) if the potential varies rapidly with the particle positions, which is the case for the short-range portion of the potential but not for the long-range one. 
Therefore, it was proposed to contract a polymer (path) with a larger number of beads to a double with fewer beads used for the long-range potential evaluation. Substantial time saving was observed in their simulation and this technique is very encouraging for the future studies of large size clusters. On the other hand, constructing PIMC actions that are beyond the Trotter approximation such as the pair product approximation \cite{ceperley_rmp_1995} will also be very helpful in reducing the number of imaginary time slices and facilitate large cluster studies.
It should be noted that ten years ago, standard PIMC method could be used to simulate a system with 500 $^4$He atoms.~\cite{rho_fs_draeger} With the latest methodological advances in the last decade, it is not far-fetched to simulate a system with thousands of particles, i.e., the size of an actual nano-droplet.

The next challenge worth noting is the possibility of carrying out real-time dynamics simulations for \phtwo~superfluids. 
The theoretical methods that have been used so far in this field, PIMC, DMC, and RQMC, are all imaginary time simulation methods. 
They are not based on the real time Schr\"odinger equation and therefore, cannot provide the real time information directly, e.g., spectral line shapes. 
Whether the line shape~\cite{ravi_ch4_he,raston_hocl_he} or other real time dynamics properties reveal superfluid information is
  unknown and deserves more exploration. 
 The ring polymer molecular dynamics~\cite{craig_rpmd1,braams_rpmd_shorttime,habershon_rpmd_correlation,habershon_rpmd_arpc} proposed and developed by Manolopoulos \etal~can be used to calculate real time correlation function for large-scale quantum systems in an $NVE$ ensemble. A hybrid PIMC/RPMD scheme may solve the problem: PIMC is used to generate path configurations with all possible permutations in an $NVT$ ensemble and RPMD is used to evolve a given configuration in the real time and calculate correlation functions of interesting properties, e.g., dipole-dipole correlation of the dopant. Thermally averaged real time information can be obtained through this scheme.
A closely related approach is centroid molecular dynamics (CMD) \cite{cao94II,voth96}. The method allows the calculation of time correlation functions based on the centroid of a Feynman path. The formalism has been extended to account for Bose-Einstein and Fermi-Dirac statistics  \cite{roy99I,roy99II}. An operator formulation of CMD also exists \cite{jang99I,jang99II} and has been useful
in dealing with non-linear operators \cite{reichman2000} and in the development of an operator formulation for bosons and fermions \cite{nb2001I,nb2001II,moffatt2004}.
The use of a centroid permutation potential has been propose to tackle the problem of exchange sampling \cite{kinugawa99,kinugawa2001,nb2002I}.
One can also construct a pseudopotential along with bosonic exchange to speedup calculations \cite{nb2002II}.
It is important to note that centroid structural properties  differ from their real-space counterpart and that a connection has been proposed \cite{blinov:3759}.
An issue associated with centroid based methods is the loss of quantum coherence inherent to the dynamical approximation.
A promising approach is the Bose-Einstein semiclassical formulation of Nakayama and Makri  \cite{Nakayama22032005}.
In order to study bosonic clusters doped with rigid molecules, rigid-body semiclassical techniques \cite{bilkiss_trimers_jcp_123_84103,bilkiss_excitedtrimers_jcp_126_024111,bilkiss_fullSCwater_jcp_127_054105,bilkiss_TA_SCIVR_H2O_jcp_127_144306} would be required.

Another challenge is to investigate the superfluid effect on the off-diagonal moments of inertia. All the theoretical works reported so far only considered the effective diagonal moments of inertia of the dopants, e.g., $I^{\rm eff}_{\rm perp}$ and $I^{\rm eff}_{\rm parallel}$ of a linear molecule. This is because the symmetry of the direct products of the angular momentum components determines that the off-diagonal moments of inertia are zero for all the dopants studied thus far. For a linear molecule, none of the direct products of the irreducible representations (irreps) of $\hat{L}_i$ and $\hat{L}_j$ with $i\ne j$ belongs to the totally symmetric irrep, nullifying the matrix element $\left<\hat{L}_i\left(\tau \right)\hat{L}_j \right>$ and $I^{\rm eff}_{ij}$. In making this statement, the totally symmetric property of
$e^{-\tau \hat{H} }$
has been employed. For example, in the MFF of a linear molecule with $z$-axis coincident with the molecular axis, $\hat{L}_x\hat{L}_y$ belongs to the $\Delta$ and $\Sigma^-$ irreps and $\hat{L}_x\hat{L}_z$ and $\hat{L}_y\hat{L}_z$ belong to the $\Pi$ irrep. Only the diagonal products $\hat{L}_x\hat{L}_x$, $\hat{L}_y\hat{L}_y$, and $\hat{L}_z\hat{L}_z$ contain the totally symmetric $\Sigma^+$ irrep, giving nonzero diagonal effective moments of inertia. Similarly, for \water~and \sotwo~molecules with $C_{2v}$ symmetry, $\hat{L}_x\hat{L}_y$, $\hat{L}_x\hat{L}_z$, and $\hat{L}_y\hat{L}_z$ in their MFF belong to $A_2$, $B_1$, and $B_2$ irreps respectively, if the $z$-axis is taken to be along the $C_2$ axis. Only the diagonal products belong to the totally symmetric $A_1$ irrep, rendering the non-zero moments of inertia. We have not seen any similar symmetry analysis in all the publications so far. 
The authors of those works may have considered this symmetry argument as obvious enough not to discuss for linear molecules, and for asymmetric tops, we are the only group that have contributed to date. But in the future, if one considers to use asymmetric top dopants with lower symmetry, e.g., $C_s$ or $C_1$, the symmetry consideration is no more trivial and the nonzero off-diagonal effective moments of inertia may reflect superfluidity of the surrounding bosons. The presence of off-diagonal effective moments of inertia will certainly reorient the principal axes of the effective rotor and redistribute the electric dipole moment along those axes, affecting its spectrum. More studies shall be dedicated to this untouched field.

The last challenge mentioned here is to simulate clusters with multiple dopants. It would be interesting to study clusters with dopant combinations, which include linear-non-linear, multi linear, multi non-linear, linear-isotopologues, non-linear-isotopologues, etc. 
Due to the present limit of simulation programs, no such studies have ever been reported. 
However, the generalization of the present computer programs to handle those cases is straightforward and we expect to see the first study of this kind in the near future.

\section{Summary and outlook} \label{sec:summary}

In this report, we have reviewed theoretical and experimental advances towards the elucidation of the properties of molecular scale superfluid response with a special emphasis on \phtwo, the only known molecular superfluid.
We have emphasized the difference between molecular scale superfluidity and an actual molecular superfluid such as \phtwo.
This active field of study continues to generate new insights into the nature of low temperature quantum clusters.
Research in superfluid response is well established for bulk systems and after several years of experimental and theoretical efforts,
the phenomenon has been shown to also occur at the microscopic scale in droplets containing thousands of helium atoms. 
This finding is rooted in the microscopic analogue of the famous Andronikashvili experiment where a molecular rotor is doped inside a microscopic superfluid system. 
The tell-tale sign of superfluid response is the resolved rotational structure in the spectrum of the dopant, a feature associated with rotational coherence albeit with renormalized effective moments of inertia. 
The smaller rotational constant (larger moment of inertia) can be attributed to the adiabatic following of the normal component of the microscopic fluid relative to the rotation of the rotor.
The sharp rotational lines and associated rotational coherence are due to the decoupling between the superfluid and the normal components of the fluid. 
Besides $^4$He, \phtwo~is a promising molecular superfluid candidate because of its light mass and and bosonic character. 
The stronger \phtwo-\phtwo~interaction, when compared to that of a pair of helium atoms, somewhat reduces quantumness and has so far prevented the realization of a \phtwo~bulk superfluid phase. 
The discovery of microscopic superfluid helium droplets however suggests the possible observation of a \phtwo~superfluid in the form of clusters. 
Because of experimental limitations, theoretical efforts have been playing a special role in the search for a \phtwo~superfluid. 
Such a \phtwo~superfluid was first predicted based on PIMC simulations of pure \phtwo~clusters. 
Almost a decade later, that work motivated a follow-up experiment that provided indirect evidence from a microscopic Andronikashvili measurement. 
Since then, theory and experiment have played complementary roles in the study of \phtwo~superfluidity.
Many-body simulation approaches such as quantum Monte Carlo methods are the tools of choice because superfluidity is a quantum phenomenon involving multiple bosons. 
The PIMC technique is the most widely used method to simulate microscopic superfluids. We have seen that the non-classical moment of inertia and its relation to superfluid response is key in our understanding of rotational decoupling observed experimentally. 
We have thoroughly discussed the differences between the superfluid response to a hypothetical external field that rotates infinitesimally slowly and the response to a quantum rotor. We learned that the former reveals the genuine space-fixed frame superfluidity while the latter is actually measured in the spectroscopic Andronikashvili experiment. Based on this difference, we summarize three guidelines for a rotor to be a good dopant to probe the genuine superfluid response.
We have discussed  algorithms for the sampling of rigid rotors in PIMC simulations, especially a newly proposed method for sampling asymmetric tops. 
We also presented the worm algorithm, an efficient approach for the sampling of bosonic exchange.
The method has  recently been applied in simulations of microscopic superfluids.
We have provided a detailed derivation of the area estimator that is used to calculate effective moments of inertia and superfluid fractions in both the rotor-fixed  frame and the space-fixed frame.
We provided a thorough coverage of the theoretical studies on \phtwo~clusters since 2000, when the first experimental evidence of \phtwo~superfluidity was reported. Simulations of \hydrogen~clusters without any dopant, with a linear dopant, and with a non-linear dopant are discussed. 
We point out some of the shortcomings of the simulation methods and comment on the possible avenues for improvement. 
We connect results from different studies to obtain a better and deeper understanding of \phtwo~superfluidity. 
In summary, the study of \phtwo~superfluidity has opened a wealth of questions and is setting the bar for efficient quantum simulation techniques. 
Even though substantial progress has been made in terms of our theoretical understanding and the development of practical simulation tools, many unanswered questions remain. Further work is indeed required to tackle challenges such as the realtime dynamics and the nature of the quantum entanglement (see Ref. \cite{herdman2013particle} for promising new QMC approach) between a dopant and its superfluid environment.

\section*{Acknowledgements}
We thank N. Blinov, M. Boninsegni, M. Gingras, G. Guillon, W. J\"ager, R.J. Le Roy, H. Li,  A.R.W. McKellar, R. Melko, M. Nooijen, J. P. Toennies, A. Vilesov, and  Y. Xu for stimulating discussions.
This research has been supported by the Natural Sciences and Engineering Research Council of Canada (NSERC) and the Canada Foundation for Innovation (CFI). TZ thanks NSERC (PDF-403739-2011) and the Ministry of Research and Innovation of Ontario for postdoctoral fellowships.

\section*{Appendix: Real Basis of Non-Linear Rotor} \label{app:real_basis}

Here we provide an example of real basis in the representation of Euler angles for the non-linear rotor. Only integral $j$, $k$, and $m$ are needed for the consideration of rigid-body rotation. These functions can be obtained through a unitary transformation from the Wigner functions, $\left\{\left<{\bf \Omega}\right.\left|jkm\right>\right\}$, the most commonly used basis functions for the non-linear rotor. We first define three real functions:
\begin{eqnarray}
\Theta^{j,0}\left({\bf \Omega}\right)&=&\sqrt{\frac{2j+1}{8\pi^2}}d^j_{00}\left(\theta\right);\\
\Theta^{j,c}_{mk}\left({\bf \Omega}\right)&=&\sqrt{\frac{2j+1}{4\pi^2}}d^j_{mk}\left(\theta\right)\cos\left(m\phi+k\chi \right);\\
\Theta^{j,s}_{mk}\left({\bf \Omega}\right)&=&\sqrt{\frac{2j+1}{4\pi^2}}d^j_{mk}\left(\theta\right)\sin\left(m\phi+k\chi \right),
\end{eqnarray}
where the superscript $s$ and $c$ denote the $\sin$ and $\cos$ factors in the expression while $0$ for no such factors.

It can be easily shown that for $k=m=0$,
\begin{eqnarray}
\left| \Theta^{j,0} \right>&=&\left|jkm \right>,
\end{eqnarray}
or else
\begin{eqnarray}
\left| \Theta^{j,c}_{mk} \right>&=&\frac{1}{\sqrt{2}}\left( \left|jkm \right>+\left|j -k -m \right> \right) \; {\rm if} \; m-k \; {\rm even};\\
\left| \Theta^{j,c}_{mk} \right>&=&\frac{1}{\sqrt{2}}\left( \left|jkm \right>-\left|j -k -m \right> \right) \; {\rm if} \; m-k \; {\rm odd};\\
\left| \Theta^{j,s}_{mk} \right>&=&\frac{1}{i\sqrt{2}}\left( \left|jkm \right>-\left|j -k -m \right> \right) \; {\rm if} \; m-k \; {\rm even};\\
\left| \Theta^{j,s}_{mk} \right>&=&\frac{1}{i\sqrt{2}}\left( \left|jkm \right>+\left|j -k -m \right> \right) \; {\rm if} \; m-k \; {\rm odd}.
\end{eqnarray}
All the linear combinations in the above five equalities are unitary transformations and therefore, the complex $\left\{\left|jkm \right>\right\}$ Wigner basis can be transformed to a real basis. If all the $\left| \Theta^{j,0} \right>$, $\left| \Theta^{j,c}_{mk} \right>$ and $\left| \Theta^{j,s}_{mk} \right>$ bases for a $j$ value are included, the number of bases is larger than the correct dimension of $\left(2j+1 \right)^2$. This means there is some overlap among the $\left\{ \left| \Theta^j \right>\right\}$ states. By inspection, we find the following selection of states in the basis can solve the problem: for $k=1$ to $j$ and $m=-j$ to $j$, both $\left| \Theta^{j,c}_{mk} \right>$ and $\left| \Theta^{j,s}_{mk} \right>$ are included; for $k=0$ and $m=1$ to $j$, both $\left| \Theta^{j,c}_{mk} \right>$ and $\left| \Theta^{j,s}_{mk} \right>$ are included; for $k=0$ and $m=0$, $\left| \Theta^{j,0} \right>$ is included. The dimension of this real basis is the correct $\left(2j+1 \right)^2$.

Since orthogonal transformation of a real basis will lead to another real basis expanding the same space, there are more than one choice of the real basis for the non-linear rotor. The example basis shown here is just to confirm the possibility of making the basis real, and we have no intention to derive the best real basis for the non-linear rotor.

\section*{Abbreviations}

B.E.: Bose-Einstein; Boltz: Boltzmann; ccQA-DMC: clamped coordinate quasiadiabatic diffusion Monte Carlo; CMD: centroid molecular dynamics; DMC: diffusion Monte Carlo; FCI-NO: full-configuration-interaction nuclear-orbital; $f_n$: normal fraction; $f_s$: superfluid fraction; HENDI: helium nano-droplet isolation; IR: infrared; irreps: irreducible representation; LePIGS: Langevin equation Path integral Ground State ; \ohtwo: {\it ortho}-hydrogen; \owater: {\it ortho}-water; \phtwo: {\it para}-hydrogen; PI: path-integral; PIMC: path-integral Monte Carlo; PIGS: path-integral ground state; \pwater: {\it para}-water; POITSE: projection operator imaginary time spectral evolution; MFF: molecule-fixed frame; MW: microwave; $NVE$ ensemble: micro-canonical ensembles with a fixed number of particles ($N$), a fixed volume ($V$), and a fixed energy ($E$); $NVT$ ensemble: canonical ensembles with a fixed number of particles ($N$), a fixed volume ($V$), and a fixed temperature ($T$); RBDMC: rigid body diffusion Monte Carlo; RFF: rotor-fixed frame; RPMD: ring polymer molecular dynamics; RQMC: Reptation Quantum Monte Carlo; SFF: space-fixed frame; WA: worm algorithm; X: exchange.

\section*{References}

\providecommand{\newblock}{}

\clearpage

\clearpage

\begin{figure}[h]
\centerline {\includegraphics[width=0.5\columnwidth]{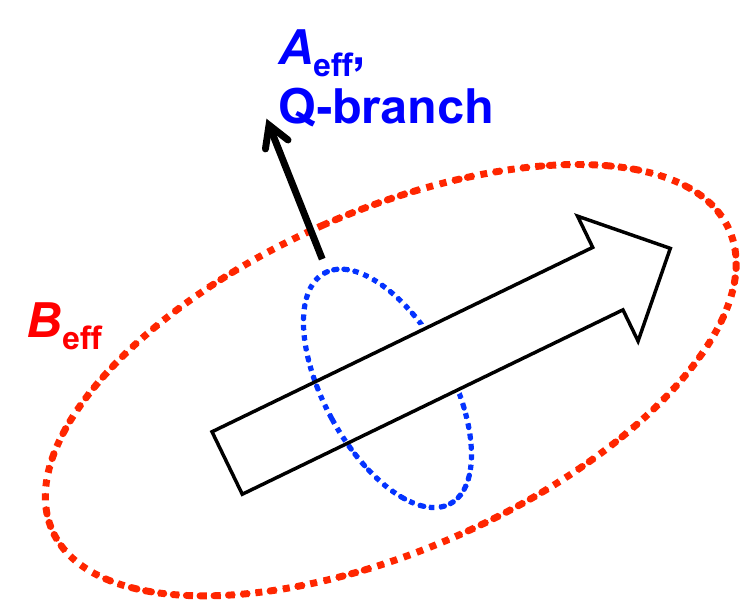}}
\caption{Sketch of parallel and perpendicular superfluid responses to the rotation of a linear rotor. The rotor is represented by a hollow arrow. The parallel response is represented by a blue dashed circle around the waist of the rotor, indicating a response of the surrounding bosons to an infinitesimally slow rotation about the molecular axis of the rotor. The perpendicular response is represented by a red dashed circle wrapping the whole rotor, indicating a response to the end-over-end rotation of the rotor. The quantities and phenomenon that are affected by the respective responses are labelled by the respective colours.} 
\label{fig:sf_paral_perp}
\end{figure}

\begin{figure}[h] 
\centerline {\includegraphics[width=1.0\columnwidth]{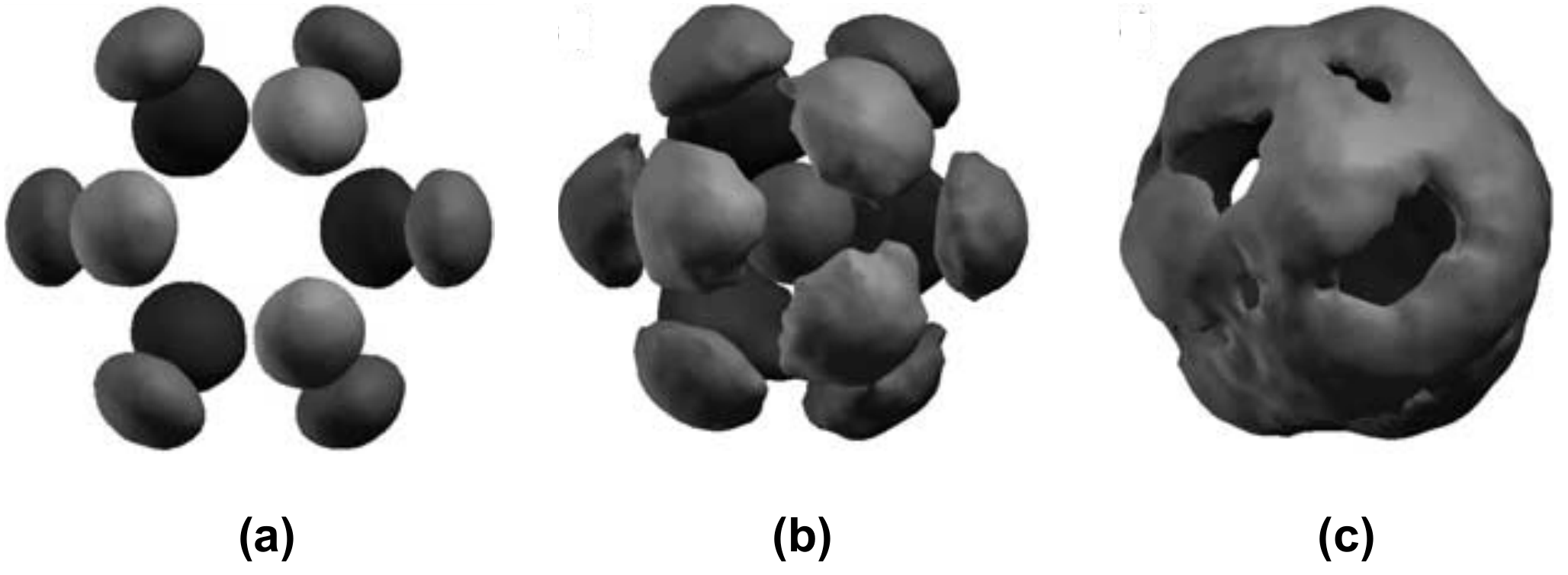}}
\caption{\phtwo~density in (a) CO(\phtwo)$_{12}$, (b) (\phtwo)$_{13}$, and (c) CO(\phtwo)$_{13}$ clusters. CO is located at the centre of the clusters and is not shown. The three images are taken from Figs. 3, 4, and 5 of Ref.~\cite{moroni_pH2_melting}. Reproduced with the permission from~\cite{moroni_pH2_melting}. Copyright \copyright~2005 WILEY-VCH Verlag GmbH \& Co. KGaA, Weinheim.} 
\label{fig:coh2_comp}
\end{figure}

\clearpage

\begin{figure}[h] 
\centerline {\includegraphics[width=0.8\columnwidth]{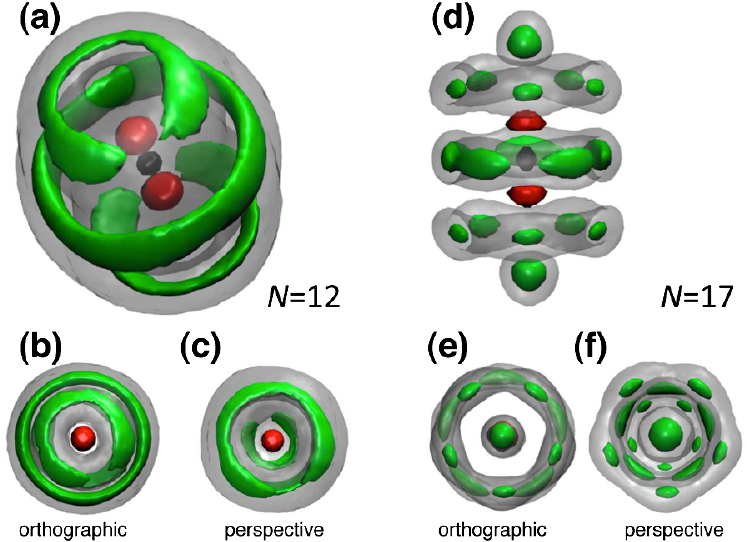}}
\caption{\phtwo~densities for CO$_2$(\phtwo)$_{12}$ (Panels (a), (b), and (c)) and CO$_2$(\phtwo)$_{17}$ (Panels (d), (e), and (f)). The green (dark) and gray (light) colours represent high and low densities respectively. This figure is taken from Fig. 4 of Ref.~\cite{huili_prl}. Reproduced with the permission from~\cite{huili_prl}. Copyright \copyright~2010 The American Physical Society.}
\label{fig:co2_h2}
\end{figure}

\begin{figure}[h] 
\centerline {\includegraphics[width=0.7\columnwidth]{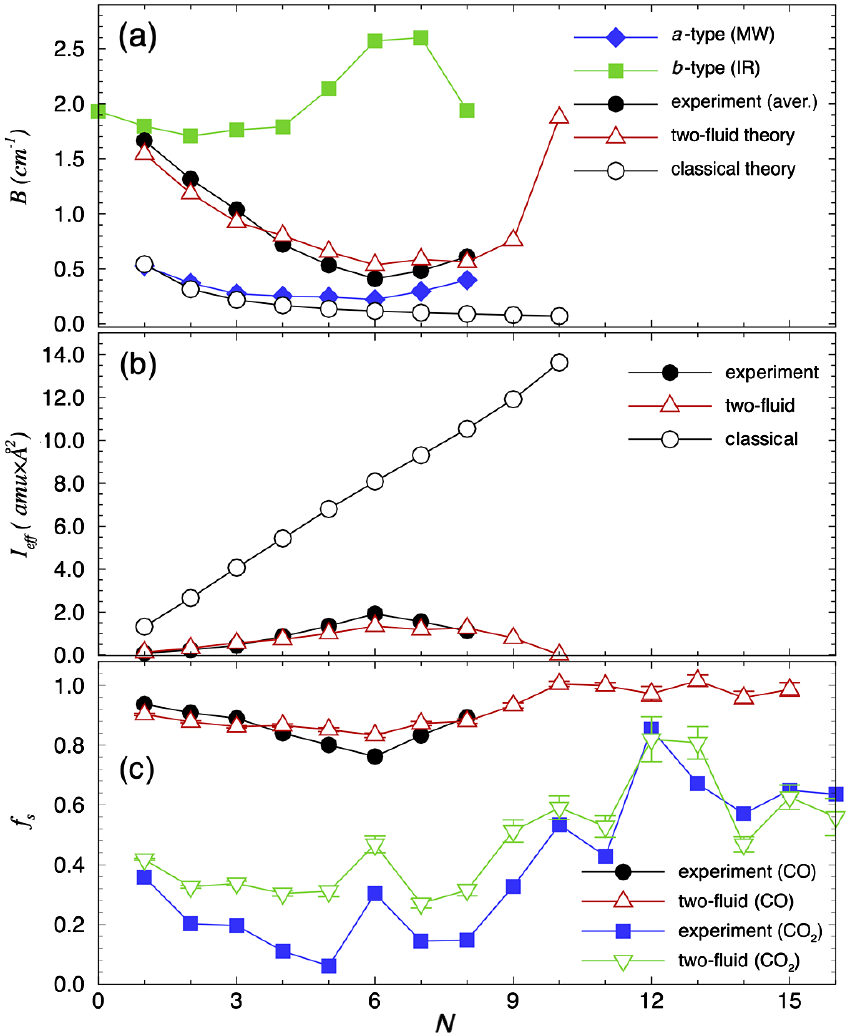}}
\caption{(a) Effective rotational constants of CO(\phtwo)$_N$ from different experimental measured transitions ($a$- and $b$-types), their averaged using Eq.~\ref{eqn:Ieff_stark}, and theoretical calculations using two-fluid model and classical mass distribution. (b) Effective moment of inertias of CO(\phtwo)$_N$ from experimental average (Eq.~\ref{eqn:Ieff_stark}) and theoretical calculations using two-fluid model and classical mass distribution. (c) Experimental and superfluid superfluid fractions of CO(\phtwo)$_N$ and CO$_2$(\phtwo)$_N$. This figure is taken from Fig. 2 of Ref.~\cite{raston_coh2_superfluid}. Reproduced with the permission from~\cite{raston_coh2_superfluid}. Copyright \copyright~2012 The American Physical Society.}
\label{fig:coh2_beff}
\end{figure}

\begin{figure}[h] 
\centerline {\includegraphics[width=0.7\columnwidth]{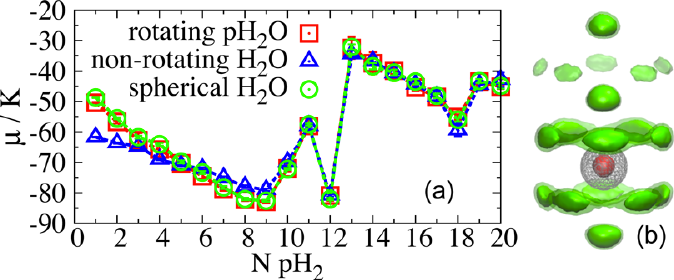}}
\caption{(a) Chemical potential ($\mu$) as a function of $N$ for \pwater(\phtwo)$_N$ with thee treatments of \pwater. (b) Density iso-surfaces of \pwater(\phtwo)$_{18}$: \phtwo~in green O in red, and H in gray wire. The solid and transparent green iso-surfaces represent high and low densities of \phtwo~respectively. This figure is taken from Fig. 1 of Ref.~\cite{tobywater}. Reproduced with the permission from~\cite{tobywater}. Copyright \copyright~2013 American Chemical Society.}
\label{fig:ph2oph2}
\end{figure}

\begin{figure}[h]
\centerline {\includegraphics[width=0.6\columnwidth]{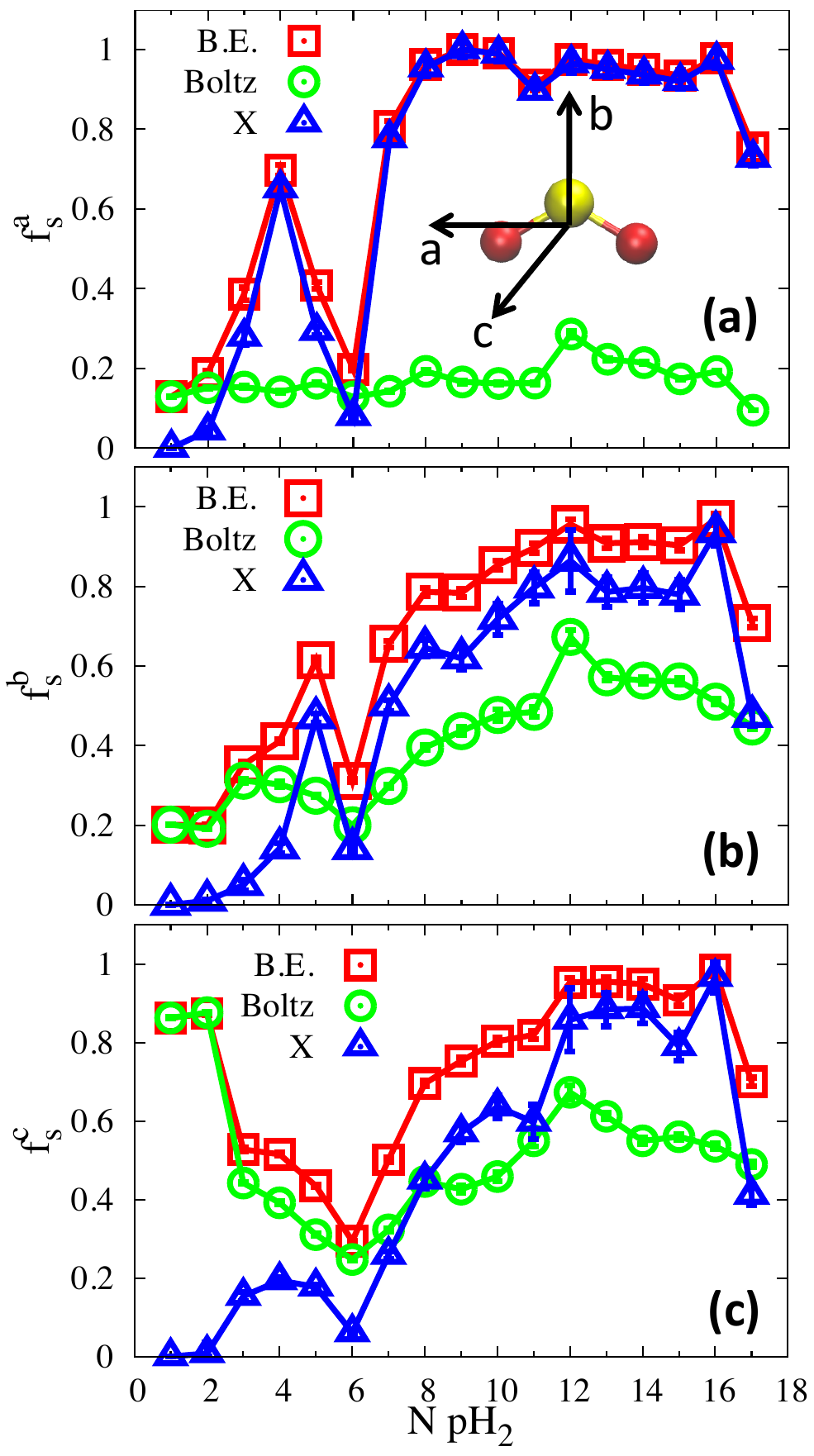}}
\caption{Superfluid fractions along the three principal axes of \sotwo~as functions of number of \phtwo: (a) $a$-axis; (b) $b$-axis; (c) $c$-axis. ``B.E.'' and ``Boltz'' denote $f_s$ from calculations treating \phtwo as bosons and boltzmannons. ``X'' denotes the exchange $f_s$. This figure is taken from Fig. 1 of Ref.~\cite{zeng_so2ph2}. Reproduced with the permission from~\cite{zeng_so2ph2}. Copyright \copyright~2013 American Chemical Society.}
\label{fig:so2_fs}
\end{figure}

\end{document}